\documentclass[12pt,a4paper,oneside,english,headsepline]{MastersDoctoralThesis}
\usepackage[latin1]{inputenc}
\usepackage{amsmath}
\usepackage{amsfonts}
\usepackage{physics}
\usepackage[11pt]{moresize}
\usepackage{color,soul}
\usepackage{amssymb}
\usepackage{makeidx}
\usepackage{subcaption}
\usepackage{graphicx}
\usepackage{pifont}
\usepackage{hyperref}
\begin{document}
	\pagestyle{headings}
	\def\baselinestretch{1.2}
\hoffset=-1.0 true cm
\voffset=-2 true cm
\topmargin=1.0cm
\thispagestyle{empty}
\def\thefootnote{\fnsymbol{footnote}}

\thicklines
\begin{picture}(370,60)(0,0)
\setlength{\unitlength}{1pt}
\put(40,53){\line(2,3){15}}
\put(40,53){\line(5,6){19}}
\put(40,53){\line(1,1){27}}
\put(40,53){\line(6,5){33}}
\put(40,53){\line(3,2){25}}
\put(40,53){\line(2,1){19}}
\put(40,53){\line(5,-6){17}}
\put(40,53){\line(1,-1){22}}
\put(40,53){\line(6,-5){30}}
\put(40,53){\line(3,-2){22}}
\put(40,53){\line(-2,1){15}}
\put(40,53){\line(-3,1){23}}
\put(40,53){\line(-4,1){26}}
\put(40,53){\line(-6,1){36}}
\put(40,53){\line(-1,0){40}}
\put(40,53){\line(-6,-1){32}}
\put(40,53){\line(-3,-1){20}}
\put(40,53){\line(-2,-1){10}}
\put(75,45){\Huge \bf IFT}
\put(180,56){\small \bf Instituto de F\'\i sica Te\'orica}
\put(165,42){\small \bf Universidade Estadual Paulista} 
\put(-25,2){\line(1,0){433}}
\put(-25,-2){\line(1,0){433}}
\end{picture}  


\vskip .3cm
\noindent
{DISSERTA\c C\~AO DE MESTRADO}
\hfill    IFT--D.001/19\\


\vspace{3cm}
\begin{center}
{\LARGE \bf Quantum Field Theory at high multiplicity: The Higgsplosion mechanism.}

\vspace{1.2cm}
Carlos Henrique Costa Duarte de Lima
\end{center}

\vskip 3cm
\hfill Advisor
\vskip 0.4cm

\hfill {\em Eduardo Pont\'on}
\vskip 1.5cm
\hfill 
\vskip 0.2cm

\hfill 
\vskip 0cm
\hfill 
\vskip 2cm
\vfill
\begin{center}
July, 2019
\end{center}

\newpage

\pagenumbering{roman}

\clearpage
\begin{center}
    \thispagestyle{empty}
    \vspace*{\fill}
 To the memory of Eduardo Pont\'on
    \vspace*{\fill}
\end{center}
\clearpage

\begin{center}
{\Large \bf Acknowledgments}
\end{center}
\vskip 2.0cm
Firstly, I would like to express my sincere gratitude to my advisor Prof. Eduardo Pont\'on, for the continuous support of my MSc study and related research, for his patience, motivation, and immense knowledge. His guidance helped me in all the time of research and writing of this thesis. I could not have imagined having a better advisor and mentor for my MSc study.

I would like to say a massive thank you to Prof. Ricardo D'Elia Matheus for the fantastic courses that I attended, and the continuous help throughout my formation in the Institute of Theoretical Physics (IFT).

My thanks also go out to the support I received from Prof. Jos\'e Helay\"el Neto and Prof. Sebasti\~ao Alves Dias from Brazilian Center for Research in Physics (CBPF). Having always the door open for me, and giving me opportunities to improve as a physicist. 

I gratefully acknowledge the funding received towards my MSc from CAPES.

I am indebted to all my friends at IFT. The uncountable coffee hours discussions, and who were always so helpful in numerous ways. Special thanks to Victor, Dean, Maria Gabriela, Guilherme, Andrei, Renato, Vinicius, Lucas, Matheus, Felippe and Andr\'e.

Some special words of gratitude go to my friends from UERJ, who have always been a major source of support: Leonardo, Jo\~ao Gabriel, Lucas Braga, Lucas Toledo and Nathan.

 I must express my very profound gratitude to my parents for providing me with unfailing support and continuous encouragement throughout my years of study and through the process of researching and writing this thesis. This accomplishment would not have been possible without them. Thank you.

And finally to Camila Fernandes, who has been by my side throughout this MSc and all this trajectory, living every single minute of it, and without whom, I would not have had the courage to embark on this journey in the first place. I can only be grateful for your support for all that time and more to come.

\newpage

\begin{center}
{\Large \bf Resumo}
\end{center}
\vskip 2.0cm


O presente trabalho busca entender o que acontece com uma teoria Qu\^antica de Campos quando estamos em um regime de alta multiplicidade. A motiva\c c\~ao para esta busca \'e em grande parte vinda de um novo (2017) proposto mecanismo que ocorreria em teorias escalares neste regime: \textit{Higgsplosion}. Ser\'a revisado o que se conhece at\'e ent\~ao dos calculos perturbativos e alguns outros resultados vindo de aproxima\c c\~oes semiclassicas. Por fim,  ser\'a estudado qual as consequencias desse mecanismo para uma teoria escalar e se pode haver contribui\c c\~oes para o Modelo Padr\~ao. O foco deste trabalho \'e entender se esse mecanismo realmente pode acontecer em uma teoria de campos usual, essa pergunta ser\'a respondida no regime perturbativo pois uma resposta mais geral ainda \'e desconhecida. Adiconalmente, uma nova poss\'ivel interpreta\c c\~ao do mecanismo de \textit{Higgsplosion} \'e proposta e discutida.


\vskip 1.0cm
\noindent
{\bf Palavras Chaves}: Alta Mutiplicidade; Higgsplosion; Teoria Qu\^antica de Campos
\vskip 0.5cm
\noindent
{\bf \'Areas do conhecimento}: Ci\^encias Exatas e da Terra; F\'isica Te\'orica

\newpage

\begin{center}
{\Large \bf Abstract}
\end{center}
\vskip 2.0cm

The current work seeks to understand what happens to a Quantum Field Theory when we are in the high multiplicity regime. The motivation for this study comes from a newly (2017) proposed a mechanism that would happen in scalar theories in this limit, the Higgsplosion. We review what it is known so far about the perturbative results in this regime and some other results coming from different approaches. We study the consequences of this mechanism for a normal scalar theory and if it can happen in the Standard Model. The goal is to understand if this mechanism can really happen in usual field theory, this question will be answered in the perturbative regime because a more general solution is still unknown. Aditionally, a new possible interpretation for the Higgsplosion mechanism is proposed and discussed.
\vskip 0.5cm
\noindent
{\bf Areas}: Natural Sciences; Theoretical Physics.


\vfill \eject

	\tableofcontents
	\mainmatter

\chapter{Introduction}  \label{c1}

\pagenumbering{arabic}

The discovery of the Higgs Boson~\cite{higgs1} opened a new era for particle physics~\cite{higgs2}. The last fundamental piece necessary for the Standard Model (SM) to work as intended. Since its discovery, we entered a new phase of precision measurement and confirmation of the Standard Model. Despite its great achievements, it is known that the Standard Model cannot be the full history. That is motivated by the lack of understanding of observational results that comes from other sources, such as, Dark Matter~\cite{dm1,dm2} and Dark Energy~\cite{DE}, which are not accounted for by the Standard Model. Even the lack of a theoretical understanding of what Quantum Gravity\cite{gravity1,gravity2,gravity3,gravity4} looks like can enter as evidence for the need of Beyond Standard Model Physics(BSM). These results indicate that maybe the Standard Model is not the end.

Even if one ignores these hints, some unsolved puzzles can be identified already within the Standard Model. One of them is the fine-tuning problem associated with the weak scale~\cite{nat1,nat2}. Naively it is expected that the squared mass parameter of a scalar particle should be of the order of the cutoff of the theory. In the Standard Model, the cutoff can be assumed as the Planck scale $\Lambda_{p}$, the scale where Quantum Gravity becomes relevant:
\begin{align}
\Lambda_{p} \approx 10^{19} \text{GeV} \, .
\end{align}
This expectation happens because there is no symmetry to protect the theory from receiving large contributions to the squared mass term. The story is different from a fermion particle, where the presence of chiral symmetry in the $m_{f} \rightarrow 0 $ limit shields the fermions from being quadratically sensitive to the Ultra-Violet (UV). This does not occur for the scalar particle without any additional symmetry. Thus, it is surprising that the measured Higgs mass $m_{h}\approx 125$~GeV~\cite{higgs1} and the absence of other states or signs of new physics, indicates the presence of a scalar much lighter than a cutoff. That means the occurrence of a fine-tuning of the contribution from BSM physics in such a way that the Higgs mass is small. The theory has large numbers  that conspire to give a small physical contribution:
\begin{align}
m_{h}^{2} \approx m_{0}^{2} + \delta m_{BSM}^{2} \, .
\end{align}
There are a few potential solutions to the fine-tuning problem. The most famous are Supersymmetry~\cite{susy1,susy2,susy3,susy4} and Composite Higgs~\cite{comp1,comp2,comp3,comp4,comp5}.

In this thesis, we review a new possible mechanism that can render the Higgs mass parameter small naturally and potentially make the Standard Model UV finite. This new mechanism is called Higgsplosion and was proposed in 2017 by Valentin V. Khoze and Michael Spannowsky~\cite{Khoze-higgsplosion,Khoze-jun-17}.  This thesis aims to understand the proposal in detail and learn more about Quantum Field Theory at high multiplicity as well as the applicability of ordinary perturbation theory in such a regime. The study of Higgsplosion is intimately related to the question of what happens to a Quantum Field Theory when we have high multiplicity processes.

Inside the Quantum Field Theory framework, people developed powerful tricks~\cite{Brown-nov-92,Voloshin-apr-93} that made possible to compute high multiplicity processes. In all of the computation, one finds the unusual feature that the leading order is already growing exponentially with the number of final states. At the time it was interpreted as implying that the perturbation theory is not valid in this regime~\cite{Goldberg-may-90}. The leading term would not be a good approximation, and any partial sum would not reproduce the correct answer. In this thesis, this claim is reviewed, so we can understand precisely how the perturbation theory works in such a regime of a Quantum Field Theory. The picture changed later when Son~\cite{Son-may-95} developed a semiclassical computation to obtain expressions for high multiplicity processes that, in principle, can be trusted in a fixed limit of the theory. The decay rate for a high multiplicity process obtained in~\cite{Son-may-95} had an exponential form, but in the region of applicability it does not have an exponential growth.

 The next breakthrough came when these semiclassical computations were generalized to the strong 't Hooft like coupling ($\lambda n$) regime for $\phi^{4}$ in the broken phase in (1+3)D. In such a limit, the same behavior of exponential growth of some objects at high multiplicity appears. This fact is what motivated Valentin V. Khoze and Michael Spannowsky to propose the Higgsplosion mechanism. It was not well understood what the limitations of their results were, and we discuss this in detail. These result gave strong evidence that Higgsplosion may happen at least in this model, which we discuss also. In the Higgsplosion mechanism, these results are used together with some basic Quantum Field Theory to show that unitarity is preserved even with this exponential growth, but with the price of the propagator vanishing exponentially fast at high energies. This exponential suppression renders loops finite, and the theory stays at a UV interacting fixed point. That is a strong claim, and the role of this thesis is to investigate this and understand better if Higgsplosion can happen in a Quantum Field Theory. If this is true, then there will be consequences to the Higgs sector of the Standard Model that could explain the fine-tuning problem and in some sense UV complete the whole theory. Even if it turns out that it does not apply to the Standard Model, it could be right in some limiting case of other models, and we can learn more about Quantum Field Theory in a different regime.

The structure of this thesis is the following. In chapter~\ref{c1}, we review the notation and tools that we use. In chapter~\ref{c2}, we calculate some high multiplicity amplitudes at the threshold (the limit where all outgoing particles are at rest) and explore beyond threshold amplitudes. At the end of chapter~\ref{c2}, we show some recent results that we use later to discuss the possibility of Higgsplosion. We choose to focus more on the perturbative approach to see if Higgsplosion happens in this regime, but these results coming from the semiclassical calculation are useful to understand the current state of the Higgsplosion Proposal. In chapter~\ref{c3}, we present the Higgsplosion mechanism itself and what it can bring to the table. After that, we discuss some problems with the claims of Higgsplosion, as well as the known criticism of it, and present some potential solutions. At the end of chapter~\ref{c3}, we present two toy models that are useful to understand the applicability of perturbation theory and a new proposed interpretation of the Higgsplosion mechanism. We try to point out which directions are worth exploring to settle the open questions that have been raised about this mechanism.

\section{Toolbox}

\subsection{Green Functions}
In this thesis, we use different types of $n$-point correlators, so it is worth defining the notation here. These correlators are used to construct physical amplitudes through the standard LSZ reduction formula~\cite{qft}. Knowing these $n$-point functions, we can construct any S-matrix element of the theory. In this point, there is no mention of perturbation theory aside from the assumption of asymptotic states that enters in LSZ\footnote{This excludes theories that we cannot separate the particles from the interaction, for instance, a confined system.}.

\begin{figure}[h!]
\centering
\includegraphics[width=8cm]{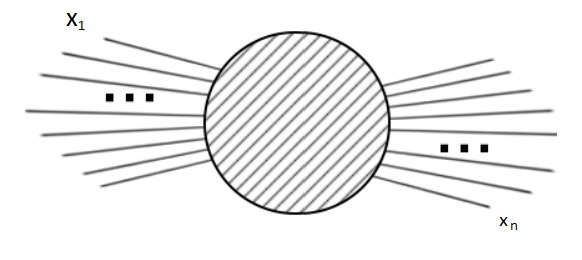}
\caption{Diagrammatic representation for n-point green functions inside perturbation theory.}
\label{fazer1}
\end{figure}

 First we define the $n$-point function as:
\begin{align} \label{green}
G^{(n)}(x_{1},\dots,x_{n}) \equiv \ev{T \left(\phi(x_{1})\dots \phi(x_{n}) \right)}{\Omega} \, ,
\end{align}
where $T$ is the time ordering operator. We can use a diagrammatic representation for this object inside perturbation theory of the form represented in Figure~\ref{fazer1}.

Given this Green function we can define its connected part diagrammatically were all external points are connected (interacting):
\begin{align}\label{cone}
G_{c}^{(n)}(x_{1},\dots,x_{n}) \equiv G^{(n)}(x_{1},\dots,x_{n}) - \text{disconnected parts} \, .
\end{align}
We will see later that we can define generating functional for these connected Green functions in such a way that we can work only using $G_{c}^{(n)}$. This definition is a generalization of the concept of cumulants in probability theory~\cite{cumu}. In particular, for $n=2$ (the propagator) all diagrams are connected:
\begin{align}
G_{c}^{(2)}(x_{1},x_{2}) = G^{(2)}(x_{1},x_{2}) \, .
\end{align}

\begin{figure}[h!]
\centering
\includegraphics[width=8cm]{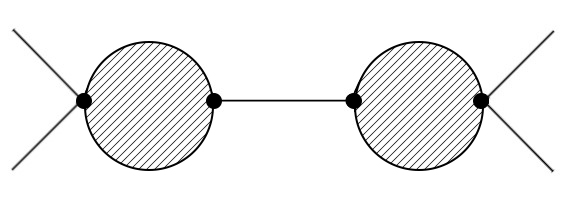}
\caption{Example of a process that is not 1PI.}
\label{not1}
\end{figure}

Finally, we can define a Green function that cannot be separated into subprocesses by cutting a single line. These are called one-particle irreducible Green functions(1PI): $G_{1PI}^{(n)}$. We will see how to obtain these objects using functional methods in  Section~\ref{sfun}. They are the fundamental blocks that we can use to construct arbitrary processes. For instance, the process represented in~\ref{not1} is not 1PI.

With these objects we can define its Fourier transform:
\begin{align}\label{furi}
(2\pi)^{4}\delta^{4}(p_{1}+\dots+ p_{n}) G_{c}^{(n)}(p_{1}, \dots, p_{n}) = \int \dd[4]{x_{1}}\dots \dd[4]{x_{n}} e^{-i(p_{1}\cdot x_{1}+\dots +p_{n}\cdot x_{n})}G_{c}^{(n)}(x_{1},\dots,x_{n}) \, .
\end{align}
Usually, we work in Fourier space and with the definition that $+p$ are entering momenta. It can be seen above in Eq~(\ref{furi}), that we have a total momentum conservation delta function. We can use these objects to compute off-shell amplitudes by picking any momentum in a physical amplitude and letting it be virtual. In other words, work only with the Green function in momentum space and ignore the overall conservation delta function. The full propagator in momentum space is then:
\begin{align} \label{ca}
G^{(2)}(p,-p) \equiv G^{(2)}(p) \, ,
\end{align}
 and with this we can define the amputated Green function that plays an important role in constructing amplitudes:
\begin{align}
G_{amp}^{(n)} (p_{1},\dots,p_{n}) \equiv \prod_{1}^{n} \left( G^{(2)}(p_{i}) \right)^{-1} G_{c}^{(n)}(p_{1},\dots,p_{n}) \, .
\end{align}

Diagramatically we are removing all the external legs of an amplitude using the full propagator. This can be used to generalize LSZ to off-shell amplitudes. We will not re-derive LSZ as this is standard textbook material~\cite{qft}. Nonetheless, to understand this last statement let us consider the LSZ reduction formula for a real scalar field:
\begin{align} \label{lsz}
i T_{n \rightarrow m} (p_{1},\dots,p_{n} ; p_{1}', \dots, p_{m}') = \lim_{p^{2}_{i} \rightarrow m^{2}}(2\pi)^{4} \delta^{4} \left( \sum_{1}^{n} p_{i} - \sum_{1}^{m} p_{i}' \right) \left( i\sqrt{Z_{\phi}} \right)^{-(n+m)} \times \\
\times (p_{1}^{2}-m^{2}) \dots (p_{m}'^{2}-m^{2}) G_{c}^{(n+m)}(p_{1},\dots,p_{n} ; p_{1}', \dots, p_{m}') \, , \nonumber
\end{align}
where $Z_{\phi}$ is the wave function normalization, and $T_{n \rightarrow m}$ is the transfer matrix elements that are related to the interacting part of the S-matrix: 
\begin{align}
\mathbb{S} =\mathbb{I} + i\mathbb{T} \, .
\end{align}
The object to the right of the delta function in Eq.~(\ref{lsz}) is the invariant amplitude $\mathcal{M}$ for this process:
\begin{align}
 T_{n \rightarrow m} (p_{1},\dots,p_{n} ; p_{1}', \dots, p_{m}') = (2\pi)^{4} \delta^{4} \left( \sum_{1}^{n} p_{i} - \sum_{1}^{m} p_{i}'\right)(-i\mathcal{M}(n \rightarrow m)) \, .
\end{align} 

The generalization goes as follows, instead of removing the propagator near the mass shell, we remove the full propagator and generate an off-shell amplitude. This amplitude became the physical amplitude when we put all particles on-shell. We can see that we are almost removing the inverse of the full propagator near the on-shell limit in Eq.~(\ref{lsz}), just changing the overall $Z_{\phi}$ factor:
\begin{align}
G^{(2)}(p) \overset{p^{2}\rightarrow m^{2}}{=} \frac{i Z_{\phi}}{p^{2}-m^{2}+i\epsilon} \, ,
\end{align}
this  $m$ is the physical mass, different than the bare mass $m_{0}$ that appear in the free propagator. The inverse is defined in such away that:
\begin{align}
G^{(2)} \cdot \left( G^{(2)} \right)^{-1} = -i \, .
\end{align}

The off-shell generalization is direct, we change these propagators near the on-shell limit to the full propagators and use the definition of the amputated Green function:
\begin{align}\label{genlsz}
-i\mathcal{M}(n \rightarrow m) = (i\sqrt{Z}_{\phi})^{(n+m)} G_{amp}^{(n+m)}(p_{1},\dots,p_{n} ; p_{1}', \dots, p_{m}') \, .
\end{align}
It is possible to see that the amputated Green function are the off-shell amplitudes aside from an overall normalization factor\footnote{If you work with a theory where $Z_{\phi}$ is zero up to some loop order, then the amputed Green function is the amplitude directly up to the same order.}. We can use this definition to work out a case that is used in this thesis, the $1 \to 1$ ``scattering":
\begin{align}
\mathcal{M}(1 \rightarrow 1) = - Z_{\phi} \left( G^{(2)} \right)^{-1} \, .
\end{align}
This case is interesting because we can use the Optical Theorem~\cite{qft} to relate the imaginary part of this amplitude to the total decay rate:
\begin{align}
\Im \left( \mathcal{M}(1 \rightarrow 1) \right) = m \Gamma_{total}(p) = -Z_{\phi} \Im \left( (G^{(2)}(p))^{-1}\right) \, ,
\end{align}
where the first equality comes from the Optical Theorem, while the second one comes from the $1 \to 1$ ``scattering" amplitude obtained trough generalized LSZ, Eq.~(\ref{genlsz}). Using the inverse of the full propagator (we will comment further on this form in the Section~\ref{secdy}):
\begin{align} \label{fullprop}
\left( G^{(2)} \right)^{-1} = -\left( p^{2}-m^{2} - \Sigma(p^{2}) \right) \, ,
\end{align}
we arrive at one of the most important relations that we will use in this thesis:
\begin{align}
\Gamma_{total}(p^{2}) = -\frac{Z_{\phi}}{m}\Im \left( \Sigma(p^{2}) \right) \, .
\end{align}

Thus, if the off-shell total decay rate grows exponentially, the imaginary part of $\Sigma(p^{2})$ grows as well. The physical total decay rate can be recovered by going to the mass shell. This feature of working with off-shell quantities let us gain more information about the theory in general, and it is a powerful tool in Quantum Field Theory. The exponential growth of the imaginary part of $\Sigma(p^{2})$ can in principle suppress the propagator, Eq.~(\ref{fullprop}), at a scale even when the real part of this function is well behaved (we cannot say much about its real part). Now, let us investigate further Eq.~(\ref{fullprop}) because this is a central point of this thesis.

\subsection{Dyson Resummation and the Full Propagator} \label{secdy}

Here we discuss important properties of the full propagator presented in Eq.~(\ref{fullprop}). Using the interacting part of the 1PI two-point function that we define as:
\begin{align} \label{1pid}
\Sigma (p^{2}) \, ,
\end{align}
we can recover all the information about the full propagator $G^{(2)}$. With $\Sigma(p^{2})$ we can re-construct the full propagator as a geometric sum of these graphs\footnote{It is interesting to note that Eq.~(\ref{umpa}) is not so straightfoward. It almost comes as a definition in Quantum Field Theory. This happens because we cannot reorganize terms in a divergent series, since summation is not infinitely associative and commutative. The 1PI organization of the perturbative series that appears in Quantum Field Theory is not immediate from perturbation theory alone. However, it is possible to justify this ordering using the definition of a Quantum Action as we show in Section~\ref{sfun}. Therefore, it is a consequence of the meaning of what a Quantum Field Theory is, and not something additional to that.}:
\begin{align}\label{umpa}
G^{(2)}= G_{0}^{(2)}+ G^{(2)}_{0}(-i\Sigma) G^{(2)}_{0} + G^{(2)}_{0}(-i\Sigma) G^{(2)}_{0}(-i\Sigma) G^{(2)}_{0} + \dots \, .
\end{align}

\begin{figure}[h!]
\centering
\includegraphics[width=15cm]{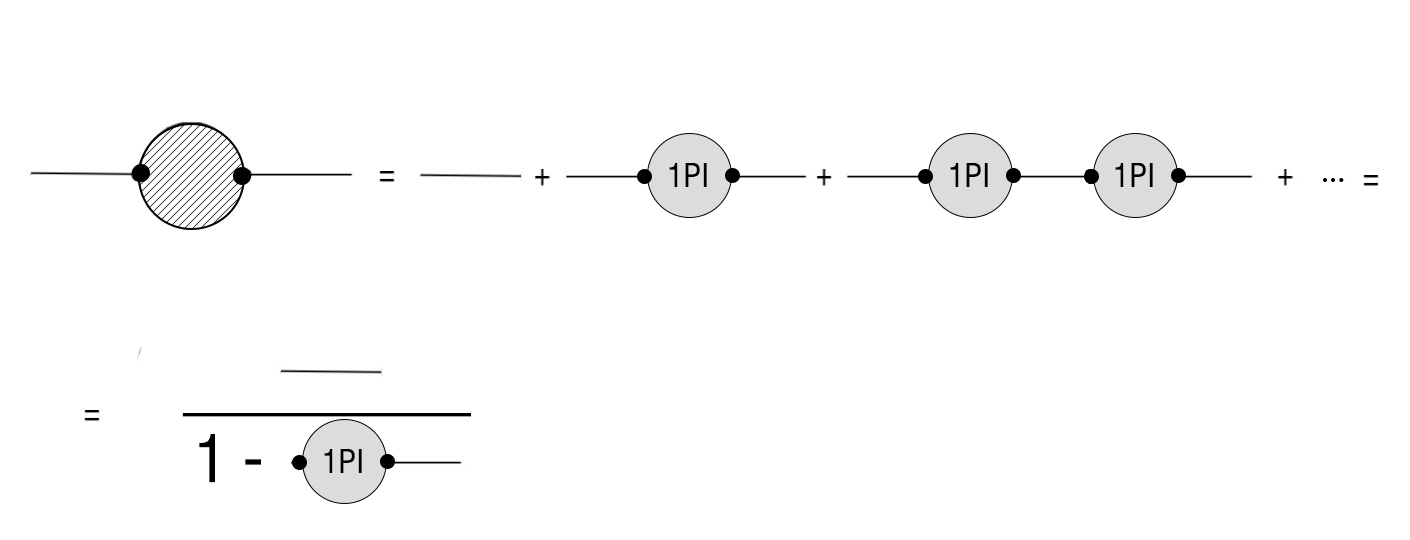}
\caption{Diagrammatic representation of the 1PI resummation.}
\label{diga}
\end{figure}
Diagrammatically this is represented in Figure~\ref{diga}. If we do this resummation we get the representation of the full propagator used before in Eq.~(\ref{fullprop}). Indeed, one has:
\begin{align}
G^{(2)}(p^{2})= \frac{G^{(2)}_{0}(p^{2})}{1+i\Sigma(p^{2}) G_{0}^{(2)}(p^{2})} \, ,
\end{align}
and using the usual free propagator:
\begin{align}
G^{(2)}_{0} = \frac{i}{p^{2}-m^{2}_{0}+i\epsilon} \, ,
\end{align}
we get the representation for the full propagator as:
\begin{align}\label{propa}
G^{(2)} = \frac{i}{p^{2}-m^{2}_{0}-\Sigma(p^{2})+i\epsilon} \, .
\end{align}

 It is important to note that the standard perturbation theory is inside $\Sigma(p^{2})$. We are free to do this resummation, and any non-perturbative effect is not lost but rendered inside the 1PI function $\Sigma(p^{2})$.  This resummation can be done even when $\Sigma(p^{2})$ is large because we can interpret this geometric series as a divergent series representation of the full propagator. Being a divergent series representation and having this geometric nature we can, in this case, find a region where the series converges.  For instance, in a given renormalization scheme, we can fix $\Sigma(m^{2})=0$, summing this series near the mass shell condition means that we are inside the convergence radius. After we resum this series, the expression can be expanded to the whole complex plane just like the regular geometric series for $x=2$ or any other complex value:
\begin{align}
1+2+4+8+ \dots = \frac{1}{1-2} = -1
\end{align} 
Typically in a divergent series, this is not the whole story, because of non-perturbative effects. For the expansion in $\Sigma(p^{2})$ given in Eq.~(\ref{umpa}), there is no effect of this kind. That does not mean that we solved the theory because we do not know how to calculate $\Sigma(p^{2})$ exactly. This result, Eq.~(\ref{propa}), is one of the few non-perturbative results in Quantum Field Theory that we currently have. With that information, we can guarantee the relation between the imaginary part of the 1PI function and the total decay rate as defined above, provided that the theory is unitary.  
This relation is re-derived without talking about this resummation (this is called Dyson Resummation in the literature) at the end of this section when we introduce functional methods. With this solved, we can start to investigate what else we can say about the full propagator, Eq.~(\ref{propa}).

\subsection{K\"all\'en-Lehmann Spectral Representation}

The object of interest here is the two-point function:
\begin{align} \label{eqalgo}
\ev{T \left( \phi(x)\phi(y) \right)}{\Omega} \, .
\end{align}

To explore this, we pick one time configuration and then in the end recover Eq.~(\ref{eqalgo}) constructing the time ordering. Choosing an ordering where $x^{0} < y^{0}$, we have:
\begin{align}
\ev{\phi(x)\phi(y)}{\Omega} \, .
\end{align}
Although we cannot compute this exactly in a interacting theory, we can extract much information from it. If we introduce a set of complete states between the operators and use translation invariance we can write:
\begin{align}
\ev{\phi(x)\phi(y)}{\Omega} = \sum_{n} \mel{\Omega}{\phi(0)}{n }\mel{n}{\phi(0)}{\Omega } e^{-i p_{n} \cdot (x-y)} \, ,
\end{align}
where $n$ runs over all states in the theory, discrete and continuous (the sum becomes an integral over the continuous states).  As expected this object depends only on the difference between the points $x$ and $y$. Now, we can introduce a delta function in a suggestive way to re-write this expression:
\begin{align}
\ev{\phi(x)\phi(y)}{\Omega} =  \int \frac{\dd[4]{p}}{(2\pi)^{4}} e^{-i p \cdot (x-y)} \left(  \sum_{n} (2\pi)^{4} \delta^{4}(p-p_{n}) \left| \mel{\Omega}{\phi(0)}{n }\right|^{2} \right) \, .
\end{align}

We can now define the spectral density:
\begin{align}
\tilde{\rho}(p) =  \sum_{n} (2\pi)^{4} \delta^{4}(p-p_{n}) \left| \mel{\Omega}{\phi(0)}{n } \right|^{2} \, ,
\end{align}
this measures the contribution to the two-point function of the states with momentum $p$. It receives contributions from bound states as well as multi-particle ones. This density is a Lorentz invariant object and vanishes when $p$ is not in the future lightcone~\cite{qft}. Using this we can write it as:
\begin{align}
\tilde{\rho}(p) = 2\pi \rho(p^{2}) \theta(p^{0}) \, .
\end{align}
Assuming there are no negative norm states it follows that the spectral density is positive semi-definite for all $p$ inside the lightcone:
\begin{align}
\rho(p^{2}) \geq 0  \, .
\end{align}
We can write the non-ordered two-point function with this spectral decomposition:
\begin{align}
\ev{\phi(x)\phi(y)}{\Omega}  =\int \frac{\dd[4]{p}}{(2\pi)^{3}} e^{-i p \cdot (x-y)} \rho(p^{2})\theta(p^{0}) \, ,
\end{align}
and using the propagator in position space:
\begin{align}
\Delta(x,y,m^{2}) = \int \frac{\dd[4]{p}}{(2\pi)^{3}} e^{-i p \cdot (x-y)}  \theta(p^{0})\delta(p^{2}-m^{2})
\end{align}
it is possible to write this ordered two-point function as:
\begin{align}
\ev{\phi(x)\phi(y)}{\Omega}  = \int_{0}^{\infty} \dd{s} \rho(s) \Delta(x,y,s) \, .
\end{align} 
To recover the time ordering in this two-point function we can use:
\begin{align}
\ev{T\left(\phi(x)\phi(y)\right)}{\Omega} = \theta(x^{0}-y^{0})\ev{\phi(x)\phi(y)}{\Omega} +  \theta(y^{0}-x^{0})\ev{\phi(y)\phi(x)}{\Omega}  
\end{align}
together with the following relation:
\begin{align}
e^{-iE_{p}(x^{0}-y^{0})} \theta(x^{0}-y^{0}) + e^{+iE_{p}(x^{0}-y^{0})}\theta(y^{0}-x^{0}) = \lim_{\epsilon \rightarrow 0 } \frac{- E_{p}}{\pi i} \int_{-\infty}^{\infty} \frac{\dd{E}}{E^{2}-E_{p}^{2}+i\epsilon}e^{i E (x^{0}-y^{0})} \, .
\end{align}
To construct the ordered two-point function Eq.~(\ref{eqalgo}), i.e. the Feynman propagator:
\begin{align}
\ev{T(\phi(x)\phi(y))}{\Omega} =  \int \frac{\dd[4]{p}}{(2\pi)^{3}} e^{i p \cdot (x-y)}  i\Delta_{F}(p^{2}) \, ,
\end{align}
were we have the full Feynman propagator in momentum space as\footnote{The $i$ was removed from the definition of this propagator to facilitate the analysis in the complex plane. Everything will be similar if we study $-i\Delta_{F}$ and kept the initial definition.}
\begin{align} \label{eq1}
\Delta_{F}(p^{2}) = \int_{0}^{\infty} \dd{s} \rho(s) \frac{1}{p^{2}-s+i\epsilon} \, .
\end{align}

\begin{figure}[h!]
\centering
\includegraphics[width=15cm]{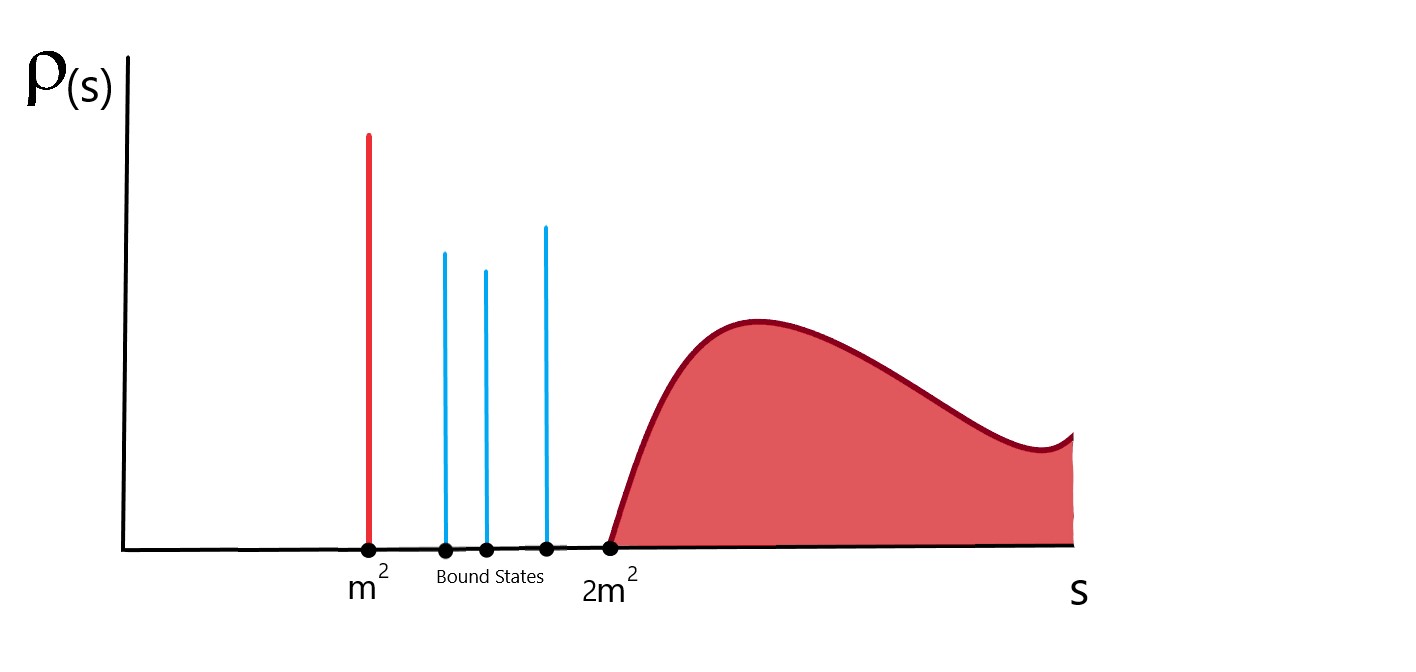}
\caption{Example of a spectral density. An usual theory has a single pole at its first excitation and then some bound states before having a continuum spectrum for the 2 particle states and above.}
\label{fig:spec}
\end{figure}

The Eq.~(\ref{eq1}) is known as K\"all\'en-Lehmann spectral representation of the full propagator. We did not use any information about the interaction or any expansion. This decomposition is a non-perturbative result. With this representation, we can derive a significant amount of information about the interacting theory. In an interacting theory, the spectral density will have singularities at locations of physical particles, illustrated in Figure~\ref{fig:spec}.

Because the spectral density is positive we can calculate the imaginary part using Cauchy theorem. Given the analytic structure of the propagator (simple poles for one-particle state and bound states, branch cuts for multi-particle states) we can write a contour integral representation where the contour is a circle around the real line where the cut lives:
\begin{align}
\Delta_{F}(p^{2}) = \frac{1}{2\pi i} \int_{\gamma} \dd{s} \frac{\Delta_{F}(s)}{p^{2}-s} \, .
\end{align}

This expression holds as long as $p^{2}$ is inside the contour and the contour path does not cross any singularity. Taking the radius of the contour $R \rightarrow \infty$ and assuming that $\rho(s) \rightarrow 0$ as $s \rightarrow \infty$ we can write:
\begin{align}
\Delta_{F}(p^{2})  = \frac{1}{2\pi i} \int_{s_{0}}^{\infty} \frac{\dd{s} \text{Disc}\left( \Delta_{F}(s)\right) }{p^{2}-s+i\epsilon} \, ,
\end{align}
were $s_{0}$ is the localization of the first singularity that we assume to be the single particle state, and the discontinuity of the function along the cut is equal to the imaginary part of it:
\begin{align}
\text{Disc}\left( \Delta_{F}(s) \right)  = 2i \Im \left( \Delta_{F}(s) \right) \, .
\end{align}
With this information, we can use the definition of the spectral decomposition Eq.~(\ref{eq1}) to write:
\begin{align} \label{eq4}
\rho(s) = -\frac{1}{\pi} \Im \left( \Delta_{F}(p^{2}) \right) \, .
\end{align}
This feature is a consequence that the propagator is real everywhere unless when the particle is on-shell.

 Another important feature is a constraint on the power with which the propagator can vanish at large momentum. Assuming that we can Wick rotate Eq.~(\ref{eq1}) (this is a non-trivial statement in an interacting theory) we get:
\begin{align}
p^{0} \rightarrow ip^{0}_{E} \quad \, , \quad  p^{2} \rightarrow - p^{2}_{E} \, .
\end{align}
This means that the modulo of the full propagator is:
\begin{align}
\left| \Delta_{F}(-p^{2}_{E}) \right| = \left| \int_{0}^{\infty} \dd{s} \rho(s) \frac{1}{p^{2}_{E}+s} \right| \, ,
\end{align}
this implies:
\begin{align}
\left| \Delta_{F}(-p^{2}_{E}) \right| \geqslant \left| \int_{0}^{s_{0}} \dd{s} \rho(s) \frac{1}{p^{2}_{E}+s} \right| \, ,
\end{align}
for any $s_{0}$ because the density is positive semi-definite. Taking the limit of Euclidean momentum going to infinity we arrive at some point in $p^{2}_{E}> s_{0}$ for a fixed $s_{0}$. Then:
\begin{align}
\lim_{p^{2}_{E} \rightarrow \infty} p^{2}_{E} \left| \Delta_{F}(-p^{2}_{E}) \right|  \geqslant  \lim_{p^{2}_{E} \rightarrow \infty} p^{2}_{E}  \left|    \int_{0}^{s_{0}} \dd{s} \rho(s) \frac{1}{p^{2}_{E}+s}  \right| \geqslant \lim_{p^{2}_{E} \rightarrow \infty} p^{2}_{E}  \left|    \int_{0}^{s_{0}} \dd{s} \rho(s) \frac{1}{p^{2}_{E}}  \right| = C 
\end{align}
for some finite positive number $C$:
\begin{align}
C = \int_{0}^{s_{0}} \dd{s} \rho(s)  \, .
\end{align}

Thus, with these assumptions, the propagator cannot fall faster than $1/p^{2}$ as $p^{2} \rightarrow \infty$. This does not mean that the propagator cannot ``look" as it is falling faster than $p^{-2}$ in intermediary regions (if we fix a $s_{0}$ such that $C=0$). This feature will be important in this thesis because it is a strong non-perturbative constraint to the two-point function. Possible loopholes of this are discussed in chapter~\ref{c3}. An important aspect of this derivation is that we can use any local operator to get similar non-perturbative information:
\begin{align}
\Delta^{O}(x) = \ev{O(x)O(0)}{\Omega}   \, ,
\end{align}
as longs as $O(x)$ is invariant under translations:
\begin{align} \label{eq2}
\Delta^{O}(x) = \int \frac{\dd{s}}{2\pi} \int \frac{\dd[3]{p}}{(2\pi)^{3}} \frac{\rho_{O}(s)}{2E_{s,p}} e^{-ip \cdot x}
\end{align}
where we define the spectral density of this operator as:
\begin{align}
\rho_{O}(s) = \sum_{n} \delta(s-p^{0})   |\mel{\Omega}{O(0)}{n }|^{2}  \, .
\end{align}

Now it is important to make some remarks. The first thing is to know that Eq.~(\ref{eq1}) and Eq.~(\ref{eq2}) may contain UV divergences. This is always the case for a 4-dimensional theory. This ultimately mean that Eq.~(\ref{eq1}) and Eq.~(\ref{eq2}) are ill-defined. The usual step to take is to regularize and renormalize these expressions. The nature of the distribution $\rho(s)$ dictates how we handle these divergences. If we assume $\rho(s)$ is a tempered distribution\footnote{ Space of tempered distributions or Schwartz space is the function space of all infinitely differentiable functions that are rapidly decreasing at infinity along with all partial derivatives.} then $\rho(s)$ must grow at $s \rightarrow \infty$ no faster than a polynomial. In this case we can re-arrange the expression by adding $m$ unknown coefficients to improve the behavior of the integral. These coefficients will be fixed by the renormalization condition:
\begin{align} \label{eq3}
\Delta_{F}(p^{2}) = P_{m-1}(p^{2}) + p^{2m} \int \dd{s} \rho(s) \frac{\Delta(p^{2},s)}{s^{m}}  \, ,
\end{align}
where $P_{m-1}$ is a polynomial of degree $m-1$.  We have to do as many subtractions as needed to render the integral finite.  The coefficients of these polynomials are fixed by experimental input. One way to justify Eq.~(\ref{eq3}) is to consider the case where we have logarithmic divergence. Doing one subtraction for the propagator, Eq.~(\ref{eq1}), we get:
\begin{align}
\Delta_{F}(p^{2})-\Delta_{F}(p_{0}^{2}) = \int_{0}^{\infty} \dd{s} \rho(s) \left( \frac{1}{p^{2}-s+i\epsilon} - \frac{1}{p^{2}_{0}-s+i\epsilon} \right)  \, ,
\end{align}
with the object on the right side now finite, and we can then define the once subtracted propagator as:
\begin{align}
\Delta_{F}(p^{2}) = \Delta_{F}(p_{0}^{2}) + (p^{2}_{0}-p^{2})\int_{0}^{\infty} \dd{s} \rho(s)  \frac{1}{(p^{2}-s+i\epsilon)(p^{2}_{0}-s+i\epsilon)}  \, .
\end{align}
Then, using one renormalization condition, we can fix the part depending on the arbitrary parameter $p_{0}$ and everything will be similar to doing renormalization in the usual way. In this language, regularizing the integral with a finite number of subtractions means that the theory is renormalizable. The story changes if we let $\rho(s)$ be a different kind of distribution~\cite{Khoze-sep-18,Jaffe-jan-67}, for instance, growing exponentially at large s. Then, what follows is that we need to do an inifinite number of subtractions. This can mean, in a worst case scenario, the necessity to fix an infinite number of constants using boundary or renormalization conditions. Theories with these kind of distributions normally are non-local, quasi-local or non-renormalizable. It can happen that after an infinite number of subtractions, only a finite number of constants need to be fixed in these three cases, but this is not a general feature. The behavior of $\rho(s)$ at infinity is fundamental to write the relation in Eq.~(\ref{eq4}). Doing this step with caution, we use the subtracted propagator where the new distributions vanish at large s.

Lastly, we can relate $Z_{\phi}$ with the spectral distribution if we remove the one-particle state from it. Using the fact that the field operators obey canonical commutation relations, this gives a constraint that one-particle plus multi-particle states should add up to one:
\begin{align}
1 = Z_{\phi}  + \int_{m^{2}}^{\infty} ds \eta(s)  \, ,
\end{align}
with $\eta(s)$ being the spectral density after the removal of the one-particle state. This fixes $Z_{\phi}$ to be a number smaller than 1:
\begin{align}
0 < Z_{\phi} < 1  \, .
\end{align}

The closer $Z_{\phi}$ is to zero, the more the multi-particle states dominate. In the limit of $Z_{\phi}=0$, we need to change the description of the theory because we do not have one-particle states anymore. That would appear as a reorganization of the degrees of freedom in the theory. With the previous understanding of the propagator, we need to introduce one more tool to start doing calculations.

\subsection{Functional Methods} \label{sfun}

Let us introduce important objects that we use throughout the thesis. The first one is the generating functional for a $n$-point Green function:
\begin{align}\label{ze}
Z[\rho] = \ev{T \left( e^{i\int \dd[4]{x} \rho(x)\phi(x)} \right) }{\Omega} = \braket{\Omega}_{\rho}  \, .
\end{align}
With this object, we can generate any Green function by differentiating with respect to the external source $\rho(x)$:
\begin{align}
(-i)^{n} \frac{\delta^{n}}{\delta \rho(x_{1}) \dots \delta \rho(x_{n})} Z[\rho] = \ev{T \left( \phi(x_{1})\dots \phi(x_{n})e^{i\int \dd[4]{x} \rho(x)\phi(x)} \right) }{\Omega}
\end{align}
in such a way that in the limit where the source goes to zero, we recover the $n$-point function Eq.~(\ref{green}). Given Eq.~(\ref{ze}), we can construct the generating functional for the connected $n$-point function defined in Eq.~(\ref{cone}) as:
\begin{align}
Z[\rho] = e^{iW[\rho]}  \, .
\end{align}
This means that:
\begin{align}
G_{c}^{(n)} = i (-i)^{n} \frac{\delta^{n}}{\delta \rho(x_{1}) \dots \delta \rho(x_{n})} W[\rho] \Bigg|_{\rho =0}  \, .
\end{align}
The last important definition is the quantum action, the generating functional for the 1PI Green function. To get this object we do a Legendre transform of $W[\rho]$:
\begin{align} \label{fdv}
\phi(x_{i}) \equiv \fdv{W}{\rho(x_{i})} = \ev{\phi(x_{i})}{\Omega}_{\rho}  \, ,
\end{align}
\begin{align}
\rho(x_{i}) \equiv - \fdv{\Gamma[\phi]}{\phi(x_{i})}  \, ,
\end{align}
\begin{align}
\Gamma[\phi] \equiv W[\rho] - \int \dd[4]{x} \rho(x)\phi(x)  \, ,
\end{align}
where we trade the $\rho$ dependence by a $\phi$ dependence. The $n$-point 1PI Green function is obtained by taking functional derivatives with respect to the field:
\begin{align}
G_{1PI}^{(n)} = \frac{\delta^{n}}{\delta \phi(x_{1}) \dots \delta \phi(x_{n})} \Gamma[\phi] \Bigg|_{\phi =0}  \, .
\end{align}

With these objects defined, we can start to analyze some relevant results. The first result is the importance of the expectation value of the field with the presence of a source. If we find a way to compute this in the presence of an arbitrary source, then we have all the information needed to recover the $n$-point functions:
\begin{align}
\ev{\phi(x_{i})}{\Omega}_{\rho}  \longleftrightarrow  \frac{\delta W[\rho]}{\delta \rho(x_{i})}  \, .
\end{align}
The only object that we cannot recover is the vacuum-vacuum amplitude $\bra{\Omega}\ket{\Omega}$, which is irrelevant when dealing with particle physics. It turns out that it is possible to find an equation for this object. It is precisely the expectation value of the classical equation of motion. That why Eq.~(\ref{fdv}) is called the classical field. It is not, in fact, all classical because non-linearity appears as $n$-point functions in this equation of $1$-point function. 

 The next result is about the full propagator and its relation to the 1PI two-point function.  Given the connected two-point function:
\begin{align}
G_{c}(x_{i},x_{2}) = -i \frac{\delta^{2}}{\delta \rho(x_{1})\delta\rho(x_{2})} W  \, ,
\end{align}
we can relate to the 1PI two point function using the identity:
\begin{align}
\fdv{\rho(x_{i})}{\rho(x_{j})} = \delta^{4}(x_{i}-x_{j})
= \fdv{\rho(x_{i})}{\phi(x_{k})} \fdv{\phi(x_{k})}{\rho(x_{j})} =  \\
= -  \left(\frac{\delta^{2}\Gamma}{\delta \phi(x_{i})\delta\phi(x_{k})} \right) \left( \frac{\delta^{2}W}{\delta \rho(x_{k})\delta\rho(x_{j})} \right)  \, .
\end{align}
This is just the inversion equation for the connected propagator written in terms of the 1PI propagator. This means that the 1PI propagator is the inverse of the connected propagator. In momentum space:
\begin{align}
G_{c}^{(2)}(p^{2}) = -\left( G^{(2)}_{1PI}(p^{2}) \right)^{-1}  \,  ,
\end{align}
this shows the overall consistency of Eq.~(\ref{fullprop}). 

The last thing that is worth pointing out about these objects is that the quantum action at tree level is the classical action, so we can use functional derivatives to derive the Feynman rules of any theory:
\begin{align}
\Gamma[\phi] = S[\phi] + \mathcal{O}(\hbar)  \, .
\end{align}

Usually, at this point, we would introduce a path integral representation for these generating functional and start calculating processes. However, here we go a different route because we are interested in high multiplicity amplitudes, and usually, Feynman diagrams do not help. Even at tree level, there are too many diagrams to count, and this method would not be useful. The method that we use is introduced in the next chapter. Now we are ready to start calculating some high multiplicity amplitudes and try to understand what is happening in this regime. After the exploration of these processes, we introduce the newly proposed mechanism of Higgsplosion and discuss the possibility of its occurrence in a scalar Quantum Field Theory.

\chapter{Perturbative Investigation of High Multiplicity Amplitudes} \label{c2}

\pagenumbering{arabic}

The focus of this chapter is the study of high multiplicity processes. The primary motivation for it comes from trying to understand the Higgsplosion proposal~\cite{Khoze-higgsplosion}. Nevertheless, this is not the only reason to look for these processes. We typically do not explore this regime in a Quantum Field Theory, and it is not clear what to expect. Maybe the particle interpretation of the field excitation ceases to be valid or useful in this regime. Because this is a complicated problem, first we do a perturbative investigation of this limit. The goal is to obtain enough information such that we can understand the applicability of perturbation theory in this regime. Ultimately this is answered at the end of chapter~\ref{c3}. 

In the presence of those perturbative results, we can start to explore different approaches for high multiplicity calculations. We do not work these additional results deeply because we chose to focus on the perturbative calculations. After we recover most of the essential results for high multiplicity scalar Quantum Field Theory, we start to work out the Higgsplosion framework.

\section{Tree Level Amplitude at Threshold}
\begin{figure}[h!]
\centering
\includegraphics[width=10cm]{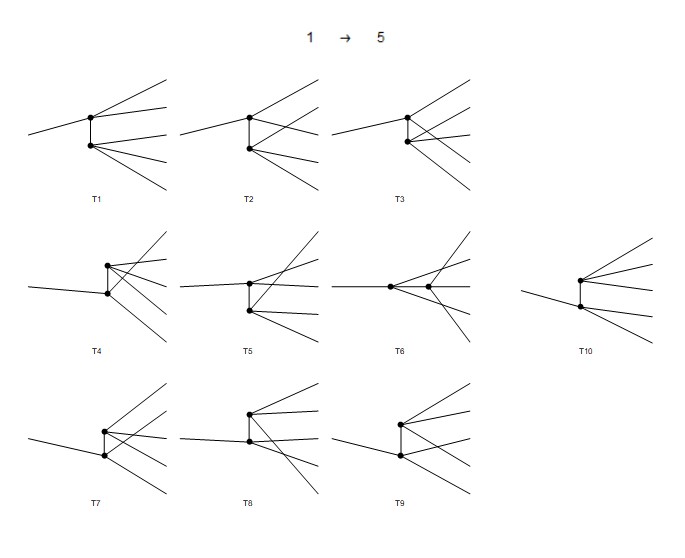}
\caption{Diagrams contributing for the $1 \to 5$ process in a $\phi^{4}$ theory in the unbroken phase. Generated with FeynArts~\cite{feyart}.}
\label{fig:1to5}
\end{figure}

\begin{figure}[h!]
\centering
\includegraphics[width=13cm]{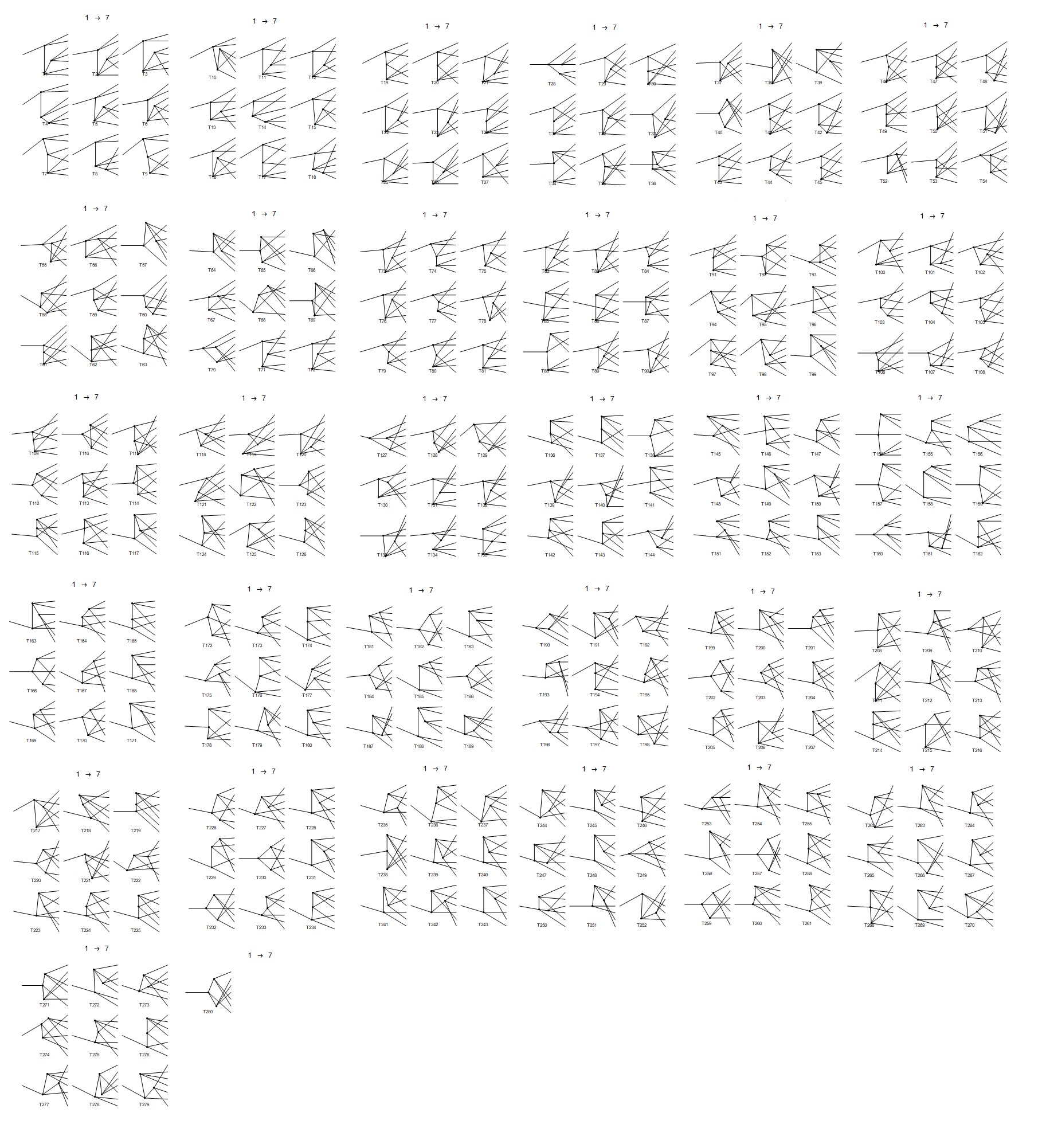}
\caption{Diagrams contributing for the $1 \to 7$ process in a $\phi^{4}$ theory in the unbroken phase. Generated with FeynArts~\cite{feyart}.}
\label{fig:1to7}
\end{figure}

We are interested in calculating the decay rate at high multiplicity of final states in a scalar theory. This could, in principle, be calculated with Feynman diagrams. However, the high number of final states makes this a tedious and challenging task. For example, if we are interested in $1 \to 5$ processes at tree level in an unbroken $\phi^{4}$ theory, we have only ten diagrams, showed in Figure~\ref{fig:1to5}. If we go for $1 \to 7$ processes, we get 280 diagrams, as it is represented in Figure~\ref{fig:1to7}. Going beyond nine particles in the final state, we rapidly pass the 1000 diagrams and becomes increasingly hard. This counting is only at tree level, adding quantum corrections creates more diagrams, and the Feynman diagrammatic approach becomes almost useless. 

 
\begin{figure}[h!]
\centering
\includegraphics[width=15cm]{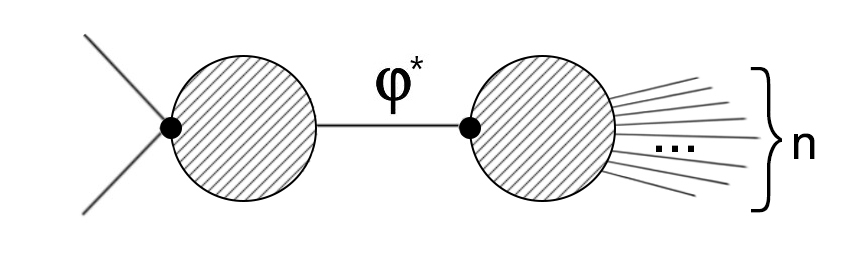}
\caption{Process where the off-shell amplitude can contribute.}
\label{higgspersion}
\end{figure}

 Here we use a different approach that was proposed by Brown~\cite{Brown-nov-92}. In this approach, we calculate the amplitude that enters in the decay rate taking advantage of the LSZ reduction formula, Eq.~(\ref{lsz}). The decay rate that we are interested in calculating is of the form:
\begin{align} \label{decay}
\Gamma_{1 \rightarrow n}(p^{2}) = \frac{1}{2m} \int \dd{\Pi_{n}} \abs{\mathcal{A}(1 \rightarrow n)}^{2} \, ,
\end{align}
where $\dd{\Pi_{n}}$ is the Lorentz invariant phase space factor, including the $1/n!$ factor since the end particles are identical. This process is a highly virtual one because there is only one scalar field, and it is stable. However, in the middle of a process, this could contribute as is represented in the Figure~\ref{higgspersion}.

The amplitude as written in Eq.~(\ref{decay}) is the one without amputating the incoming leg. The reduction formula for this case is:
\begin{align}
\mathcal{A}(1\rightarrow n)[x,p_{1},\dots,p_{n}] = (iZ_{\phi})^{-(n+1)/2} \lim_{p_{i}^{2}\rightarrow m^{2}} \int \dd[4]{x_{1}}\ldots \dd[4]{x_{n}} e^{-ip_{1} \cdot x_{1}}\ldots  \\
e^{-ip_{n}\cdot x_{n}}(p_{1}^{2}-m^{2})\ldots (p_{n}^{2}-m^{2}) \ev{T \left( \phi(x_{1})\ldots\phi(x_{n})\phi(x) \right) }{\Omega}  \nonumber \, .
\end{align}

Here we have to be careful because of the use of a nonstandard notation. The initial off-shell particle is in the position representation, and the rest are in the momentum representation. This amplitude has mixed momentum and position dependence. All of this dependence vanishes at threshold:
\begin{align}
\vec{p}_{i} = 0 \, , \quad \text{for all $i$} \, ,
\end{align}
where $p_{i}$ are the outgoing momenta in the amplitude. We calculate first these amplitudes in the threshold limit, and in the end, try to recover the momentum dependence. Another point to notice is that, if we want to relate this amplitude to the Feynman diagram computation, we need to amputate the virtual particle. The Feynman amplitude would be $\mathcal{M}$, and its relation to $\mathcal{A}$ is:
\begin{align}
\mathcal{M}(p,p_{i})= \left( p^{2}-m^{2}\right) \mathcal{A}(p,p_{i})  \, .
\end{align}

From now on, we can can ignore the $Z_{\phi}$ factor, because all the computations are done up to one-loop. The effects of the $Z_{\phi}$ enters only at higher loops for the class of theories that we work in this thesis. We can drop out also the overal phase factor for this process because we square this amplitude in the end. The interesting observation used by Brown is that we can re-write the $(n+1)$-point correlator in terms of functional derivatives using Eq.~(\ref{fdv}):
\begin{align}\label{tomano}
\ev{T \left( \phi(x_{1})\ldots\phi(x_{n})\phi(x) \right) }{\Omega} =\fdv{\rho(x_{1})}\ldots\fdv{\rho(x_{n})}\ev{\phi(x)}{\Omega}_{\rho} \Bigg|_{\rho = 0}  \, ,
\end{align}
where the expectation value is taken in presence of an arbitrary source. Using Eq.~(\ref{tomano}), we write the amplitude as:
\begin{align} \label{amp}
\mathcal{A}(1\rightarrow n) = \prod_{i=1}^{n} \left[ \lim_{p_{i}^{2} \rightarrow m^{2}} \int \dd[4]{x_{i}} e^{-ip_{i}x_{i}}(p_{i}^{2}-m^{2}) \fdv{\rho(x_{i})} \right] \ev{\phi(x)}{\Omega}_{\rho} \Bigg|_{\rho = 0}  \, .
\end{align}
It is possible to find a differential equation for $\ev{\phi(x)}{\Omega}_{\rho} \bigg|_{\rho = 0}$ that is just like the classical equation of motion, and then find an analytic solution for $\mathcal{A}(1\rightarrow n)$. This equation simplifies when we want to calculate only the tree-level contribution. Let us specialize for this case now, for the $\phi^{4}$ theory (any other interaction or kind of matter would follow a similar path).

\subsection{$\frac{\lambda\phi^{4}}{4!}$ in the Unbroken Phase}

It is known that the tree level approximation for the expectation value is just the classical solution in presence of an arbitrary source~\cite{Brown-nov-92}:
\begin{align}
\ev{\phi(x)}{\Omega}_{\rho} \rightarrow \phi_{cl}(x)[\rho]  \, .
\end{align}
This statement will be made a little more rigorous when we compute loop corrections at Section~\ref{loop}. Choosing the $\phi^{4}$ theory and using the usual particle physics normalization we get the equation of motion:
\begin{align} \label{eqmov}
(\Box + m^{2}) \phi_{cl}(x) + \frac{\lambda}{3!} \phi_{cl}^{3}(x)=\rho(x)  \, .
\end{align}

We want to find a solution of this equation to get the $\rho$ dependence of the field and then do $n$-functional derivatives to obtain the amplitude, Eq.~(\ref{amp}). Because this is a differential equation, we need boundary conditions. They are set by the Feynman prescription of the propagator that tell us how to project to the right vacuum.  We have transformed the problem of finding the tree level amplitude into solving a non-linear second order differential equation with an arbitrary source, which is still a difficult problem. The approach developed by Brown is to focus on the threshold limit, where the source can be taken to have a simple enough form that the equation can be solved, as we will show. Note that we want the dependence of the field with respect to the source, not the actual form of the field solution by itself. For this, we need to be careful because the source needs to be able to excite all modes of the field. Since we want the amplitude at threshold, this means that there is no spatial momentum in the final states. In that limit, the source and the field are homogeneous in space and depend only on time. Hence, the threshold limit simplifies the equation to only one dimension. 

The next step to find a solution for Eq.~(\ref{eqmov}) is to choose a simple exponential source:
\begin{align} \label{salsa}
\rho(t) = \rho_{0}e^{i\omega t}  \, .
\end{align}
The equation of motion then becames:
\begin{align} \label{socorro}
(\partial_{t}^{2}+m^{2})\phi_{tree}+\frac{\lambda}{3!}\phi_{tree}^{3}= \rho(t)  \, .
\end{align}
This is still a non-linear problem but now is solvable. We look for a solution in perturbation theory, using $\lambda$ as our deformation parameter that turns on and off the non-linearity. We first consider the free equation:
\begin{align}
(\partial_{t}^{2}+m^{2})\phi_{tree}^{0}=\rho_{0} e^{i\omega t} \, ,
\end{align}
whose solution is:
\begin{align} \label{zdete}
\phi_{tree}^{0} = - \frac{\rho_{0}}{\omega^{2}-m^{2}+i\epsilon} e^{i\omega t} = z(t)  \, .
\end{align}
Turning the coupling on generates a series of the form:
\begin{align} \label{pert}
\phi_{tree}= z(t) + \lambda \phi_{tree}^{(1)}+\lambda^{2} \phi_{tree}^{(2)} + \ldots  \, .
\end{align}
Plugging this in the equation of motion:
\begin{align} 
(\partial_{t}^{2}+m^{2})(z(t) + \lambda \phi_{tree}^{(1)}+\lambda^{2} \phi_{tree}^{(2)} + \ldots)+\frac{\lambda}{3!}(z(t) + \lambda \phi_{tree}^{(1)}+\lambda^{2} \phi_{tree}^{(2)} + \ldots)^{3}= \rho_{0}e^{i\omega t}  \, ,
\end{align}
shows that the $\phi_{tree}(t)$ can be written only in terms of $z(t)$, and the source dependence comes only from it, for instance the first term in the expansion is:
\begin{align}\label{fi}
\phi_{tree}^{(1)} = \lambda \frac{z_{0}(\rho_{0},\omega)}{3!(9\omega^{2}-m^{2})} e^{3i\omega t} = - \lambda \frac{z(t)^{3}}{3!(9\omega^{2}-m^{2})z_{0}(\rho_{0},m)^{2}}   \, .
\end{align}
The notation that we used is:
\begin{align}
z_{0}(\rho_{0},\omega) = \frac{\rho_{0}}{\omega^{2}-m^{2}+i\epsilon} \, ,
\end{align}
in such a say that the double limit of $\rho_{0} \rightarrow 0$ and $ \omega \rightarrow m$ this function becomes a constant $z_{0}$.

We can trade the functional derivative with respect to the source for ordinary $z(t)$ derivatives in the amplitude, Eq~(\ref{amp}):
\begin{align} \label{co}
\mathcal{A}(1\rightarrow n) = \prod_{i=1}^{n} \left[ \lim_{p_{i}^{2} \rightarrow m^{2}} \int \dd[4]{x_{i}} e^{-ip_{i}x_{i}}(p_{i}^{2}-m^{2}) \fdv{\rho(x_{i})}  \right] \ev{\phi(x)}{\Omega}_{\rho} \Bigg|_{\rho = 0} =
\end{align}
\begin{align} \nonumber
 = \prod_{i=1}^{n} \left[ \lim_{\omega \rightarrow m} \int \dd[4]{x_{i}} e^{-i\omega t_{i}}(\omega^{2}-m^{2}) \fdv{\rho(t_{i})}  \right] \ev{\phi(x)}{\Omega}_{\rho} \Bigg|_{\rho = 0}  \, .
\end{align}
Now we can use the dependence of the source from Eq.~(\ref{zdete}) and Eq.~(\ref{pert}):
\begin{align} \label{sour}
\frac{\delta \phi[z]}{\delta \rho(t_{i})} = \frac{\delta z[\rho]}{\delta \rho(t_{i})} \frac{\partial \phi(z)}{\partial z} = -\frac{1}{\omega^{2}-m^{2}} \delta(t-t_{i}) \delta^{3}(x_{i})\frac{\partial \phi(z)}{\partial z} \, .
\end{align}
Using Eq.~(\ref{sour}) in Eq.~(\ref{co}) we can see the simplification of the problem, after doing the delta integrations and ignoring the overall phase:
\begin{align} \label{amp1}
\mathcal{A}(1 \rightarrow n) = \left( \pdv{z} \right)^{n} \phi_{tree}[z(t)] \Bigg|_{z=0}  \, ,
\end{align}
where the on-shell condition corresponds to the limit $\omega \rightarrow m$ and $\rho \rightarrow 0$, in such a way that $z_{0}$ remains finite. The solution for $\phi_{tree}$ can be obtained order by order in $\lambda$ like the first term obtained in Eq~(\ref{fi}). One finds that the perturbative series Eq.~(\ref{pert}) can be ressumed:
\begin{align} \label{unbroken}
\phi_{tree}(t)= \frac{z(t)}{1-z^{2}\frac{\lambda}{48m^{2}}}  \, .
\end{align} 

It is easy to check that this is indeed a solution in a well-defined limit. Defining $\alpha = \frac{\lambda}{48m^{2}}$ we compute:
\begin{align}
\dot{\phi}_{tree} = \dot{z} \frac{(1+\alpha z^{2})}{(1- \alpha z^{2})^{2}}  \, ,
\end{align}
\begin{align}
\ddot{\phi}_{tree} =\frac{ \ddot{z}(1-\alpha z^{4}) + \dot{z}^{2}(6\alpha z +2 \alpha^{2} z^{3})}{(1-\alpha z^{2})^{3}}  \, ,
\end{align}
and using the definition of $z(t)$, Eq.~(\ref{zdete}) we have that $z \propto e^{i \omega t}$ so:
\begin{align}
\dot{\phi}_{tree} = i \omega z \frac{(1+\alpha z^{2})}{(1- \alpha z^{2})^{2}}  \, ,
\end{align}
\begin{align}
\ddot{\phi}_{tree} =-\omega^{2} \frac{(\alpha^{2}z^{5}+6\alpha z^{3}+ z)}{(1-\alpha z^{2})^{3}}  \, .
\end{align}
Putting this in the equation of motion, Eq.~(\ref{socorro}), and simplifying the denominator we get:
\begin{align} \label{direito}
\alpha^{2} z^{5}(m^{2}-\omega^{2}) + z^{3}(\frac{\lambda}{3!}-6\omega^{2}\alpha -2m^{2}\alpha) + z(m^{2}-\omega^{2})=-(\omega^{2}-m^{2})z(1-2\alpha z^{2}+\alpha^{2}z^{4})  \, .
\end{align}
We can see that this is a solution when $\omega \rightarrow m$ since the middle term is just the definition of $\alpha$. Here we can take $\rho \rightarrow 0 $ in such a way that $z(t)$ remains finite. Even though we arrived at a solution using perturbation theory, we have obtained a representation, Eq~(\ref{unbroken}) that can be trivially continued to the full complex $z(t)$ plane. An interesting thing comming from the form of the amplitude Eq.~(\ref{amp1}) is that we can work without the source if we change the boundary conditions. In the solution this can be seen trivially in the right side of the Eq.~(\ref{direito}), since the source contribution vanishes on-shell. This happens because we only want the solution on-shell, and before doing this limit the $\rho_{0} \rightarrow 0$ is similar to $z \rightarrow 0$. Then, we can solve without any source to find these amplitudes. Doing this changes the boundary conditions of the solution until we set $\rho=0$, because we are solving with a source at all times. This dictates that the solution, Eq.~(\ref{unbroken}), should vanish in positive Euclidean times:
\begin{align}
Im(t) \rightarrow\infty  \, .
\end{align}
This boundary condition remains for the broken case and loop corrections. Another interesting point is that we started with a real field but got a complex solution. This complexification is happening because of the source, Eq.~(\ref{salsa}). If we had chosen a real source, the solution would be real as well. However, we only use the source as a trick to get the scattering amplitude. It is arbitrary and can be chosen to be of this particular form.

With that in mind we can now find the decay amplitude, Eq.~(\ref{amp1}), by taking $n$-derivatives with respect to $z(t)$ in Eq.~(\ref{unbroken}). To facilitate this, we can use a series representation of this expression and pick up the nth-term in it:
\begin{align}
\frac{z}{1-\alpha z^{2}} = \sum_{0}^{\infty} z^{2n+1} \alpha^{n} \, .
\end{align}
The amplitude is then:
\begin{align}
\mathcal{A}(1 \rightarrow n) = \frac{n!}{2\alpha} [ (\alpha)^{(1+n)/2)} + (-\alpha)^{(1+n)/2}]   \, .
\end{align}

It is possible to draw some conclusions about this amplitude at tree level and threshold. First it is necessary to remember that this is a partially off-shell amplitude and because of that it is not a physical object by itself. Nevertheless we can use this amplitude to construct physical observables. This amplitude has the interesting features:
\begin{itemize}
\item It vanishes for even final states:
\begin{align}
\mathcal{A}(1 \rightarrow 2k) =0  \, .
\end{align}
\item For odd final states it has the factorial growth:
\begin{align}
\mathcal{A}(1 \rightarrow 2k+1) = (2k+1)! \left( \frac{\lambda}{48m^{2}} \right)^{k}  \, .
\end{align}
\end{itemize}
This factorial growth persists to the decay rate even after dividing by the $n!$ coming from the identical nature of the final states. That could be an indication that in the high multiplicity limit, the decay rate grows or that perturbation theory cannot be trusted for a high multiplicity computation. The total final state energy of the system at rest is:
\begin{align}
E=(2k+1)m  \, .
\end{align}
and the phase space in this case is zero (it is just a point). If we assume an infinitesimal sphere around $E$ and that the momentum dependence is constant in this region we get only an overall small term trying to combat the factorial growth, that we will call $V_{n}(E)$:
\begin{align}
\Gamma_{2k+1}(E) = (2k+1)! \left( \frac{\lambda}{48m^{2}} \right)^{2k} V_{2k+1}(E)   \, .
\end{align}

This is just a naive approximation if we want to know the phase space contribution we need to be able to go beyond the threshold in such a way that we can get results around a specific configuration. This is already potentially problematic for the perturbative unitarity of the theory. The square amplitude divided by the symmetry factor grows factorially and can, in principle, pass any unitary bounds of these processes:
\begin{align}
\frac{\left| \mathcal{M}\right|^{2}}{n!} \propto n!
\end{align}

\subsection{$\frac{\lambda\phi^{4}}{4!}$ in the Broken Phase}
The $\phi^{4}$ has another regime where we can explore this processes. In this phase the mass term is negative so it is convenient to use the definition:
\begin{align}
m^{2}=-\mu^{2}  \, .
\end{align}
The reflection symmetry is broken and the configuration that minimizes the potential from Eq.~(\ref{eqmov}) is no longer zero:
\begin{align} \label{min}
\phi_{min}^{2} = \frac{3!\mu^{2}}{\lambda}  \, .
\end{align}
If we want to find the tree level amplitude at threshold it is easier to work in the shifted field, where we do not have an expectation value~\cite{Brown-nov-92,Smith-apr-93}:
\begin{align} \label{shift}
\sigma_{tree}=\phi_{tree}-\phi_{\min}  \, .
\end{align}
Using this definition the equation of motion becames:
\begin{align} \label{eqbro}
(\Box + 2\mu^{2}) \sigma_{tree}(x) + \frac{\lambda \phi_{min}}{2!}\sigma_{tree}^{2}(x) + \frac{\lambda}{3!} \sigma_{tree}^{3}(x)=\rho(x)  \, .
\end{align}

The steps from Eq.~(\ref{eqbro}) to Eq.~(\ref{broken}) are essentially the same as detailed in the unbroken phase above. We need to solve this equation to find $\sigma$ as a functional of the source $\rho$. Choosing again the same exponential source Eq.~(\ref{salsa}), transforms this problem into finding the solution for the sourceless case with the boundary condition that:
\begin{align}
\sigma \rightarrow 0 \quad \mbox{as} \quad Im(t) \rightarrow \infty  \, .
\end{align}
Now, we look for a perturbative solution for the spatially homogeneous case of Eq.~(\ref{eqbro}). Doing that we find the perturbative series in terms of the unperturbed solution:
\begin{align}
z(t)= z_{0}e^{i \sqrt{2}\mu t}  \, ,
\end{align}
where the physical mass is now $\sqrt{2}\mu$ in the on-shell limit. It is easy to check that a solution for Eq.~(\ref{eqbro}), taking the $\rho \rightarrow 0$ limit but keeping $z_{0}$ finite is:
\begin{align} \label{broken}
\sigma(t)= \frac{z}{1-\frac{z}{2\phi_{min}}}  \, .
\end{align}
Keep in mind that $\phi_{min}$ has all the coupling dependence. We find such a solution performing a perturbative expansion, ressuming, and analytically continuing to the full complex $z(t)$ plane.

 To check that this is a solution is direct:
\begin{align}
\ddot{\sigma} = -2\mu^{2} \frac{\frac{z^{2}}{2\phi_{\min}}-z}{(1-\frac{z}{2\phi_{\min}})^{3}}  \, .
\end{align}
Plugging this in the equation of motion and simplifying the denominator we get:
\begin{align}
z^{3}(-\frac{\lambda}{12}+\frac{\mu^{2}}{2\phi_{min}^{2}})+z^{2}(-3\frac{\mu^{2}}{\phi_{\min}}+\frac{\lambda \phi_{\min}}{2}) + z(-2\mu^{2}+2\mu^{2})=0  \, ,
\end{align}
where we already used that $\omega^{2} \rightarrow  2\mu^{2}$ and set $\rho_{0} \rightarrow 0$. The l.h.s vanishes using Eq.~(\ref{min}).

Now that we have this solution, we can compute the tree level threshold amplitude for the broken phase, following the same logic as already explained in the unbroken case:
\begin{align} \label{ampbroken1}
\mathcal{A}_{B}(1 \rightarrow n) = \left( \pdv{z} \right)^{n} \sigma[z] \Bigg|_{z=0}  \, .
\end{align}
As before we find an series expansion and pick up the nth-term to get:
\begin{align}
\mathcal{A}_{B}(1 \rightarrow n) = n! \left( \frac{\lambda}{24\mu^{2}} \right)^{(n-1)/2}  \, .
\end{align}
We can see that the factorial growth is still present in this phase. The difference is that now we can have even final states. We also get a factorially growing decay rate:
\begin{align}
\Gamma_{B}(1 \rightarrow n) = n! \left( \frac{\lambda}{24\mu^{2}} \right)^{(n-1)} V_{n}(E)  \, .
\end{align}

As highlighted before this can be a potential danger for the unitarity of the theory. If these processes start to dominate with factorial power, then the cross section for a process of few particles going to $n$ start to grows as well. This growth is inconsistent with the perturbative unitarity of the theory. There are a few possible explanations for this feature and ways to save this growth. We will continue working within the perturbative approach, at threshold, and see if the factorial growth persists after quantum corrections, at the one-loop level. Later on, we explore going beyond threshold such that the assumption that $V_{n}(E)$ is constant can be checked and obtain a better decay rate expression.

\section{One-Loop Amplitude at Threshold} \label{loop}
Until now, we saw that the tree level amplitude at the threshold for the $\phi^{4}$ theory displays a factorial growth already in the first term of the series. We expect that these series that appear in Quantum Field Theory to be divergent. However, it is not the usual case when the first term of the series is already large. This, in principle, could mean that we cannot even trust the first term of the series as a good approximation. If we want to understand this better, we need to compute quantum corrections for this theory to see if somehow these factorial growths get tamed. It is possible to implement loop corrections in this formalism if we expand the operators in a $\hbar$ expansion, where the first term corresponds to the tree level. In this expansion, we have to be careful because $\hbar$ has dimension, and so far have been using units such that:
\begin{align}
\hbar = 1  \, .
\end{align}
To get around this, we can introduce a deformation parameter $\epsilon$ where $\hbar$ would appear, such that the limit $\epsilon \rightarrow 0$ is the classical limit that $\hbar \rightarrow 0$.

The expansion of the field operator is:
\begin{align}
\hat{\phi}(x)= \phi_{0} \hat{I} + \sqrt{\epsilon} \hat{\phi}_{1/2}+\epsilon \hat{\phi}_{1} + \mathcal{O}(\epsilon^{3/2})  \, .
\end{align}

The expansion is in $\sqrt{\epsilon}$ because we are working at the level of the equation of motion. We will see from the calculation that the one-loop correction appears in the $\epsilon$ term using this convention. With this definition, we can find the expectation value of the field order by order in $\epsilon$, and then the amplitude can be computed up to that same order by an appropriate functional derivative. Now, let us specialize in the cases that we have considered above to see how quantum corrections modify them.

\subsection{$\frac{\lambda\phi^{4}}{4!}$ in the Unbroken Phase}
The first case of interest is the $\phi^{4}$ theory in the unbroken phase~\cite{loop1}, the equation of motion in terms of the field operator, before expanding in $\epsilon$ is:
\begin{align}
\ev{T\left[ (\Box + m^{2})\hat{\phi}+ \frac{\lambda}{3!} \hat{\phi}^{3}\right]}{\Omega}_{\rho}=0  \, .
\end{align}
We will use the notation:
\begin{align}
\ev{\hat{\phi}_{i}}{\Omega}_{\rho} = \phi_{i}[\rho]  \, .
\end{align}
Expanding the field operator up to $\mathcal{O}(\epsilon^{3/2})$ the equation of motion is\footnote{The source that appear in the equation is the classical external one. It appears after we bring the d'Alembert operator out of the expectation value, passing trough the time ordering.}
\begin{align} \label{loopexp}
\left[ (\Box+m^{2})\phi_{0}(x) +\frac{\lambda}{3!}\phi_{0}^{3}(x)-\rho(x) \right] +
\end{align}
\begin{align} \nonumber
+ \sqrt{\epsilon}\left[ \left( \Box+m^{2} + \frac{\lambda}{2}\phi_{0}^{2}(x) \right) \phi_{1/2}(x) \right] + 
\end{align}
\begin{align} \nonumber
+\epsilon \left[  \left( \Box+m^{2} + \frac{\lambda}{2}\phi_{0}^{2}(x) \right) \phi_{1}(x) + \frac{\lambda \phi_{0}}{2} \ev{T\left(\hat{\phi}_{1/2}(x)\hat{\phi}_{1/2}(x)\right)}{\Omega}_{\rho} \right]+ \mathcal{O}(\epsilon^{3/2}) =0  \, .
\end{align}

Now we do the threshold limit. This means that  $\phi_{0}$ is as calculated before, Eq.~(\ref{unbroken}):
\begin{align} \label{unbroken1}
\phi_{0} = \frac{z(t)}{1-\frac{\lambda}{48m^{2}}z(t)^{2}}  \, .
\end{align} 
The equation of order $\sqrt{\epsilon}$ defines the two-point function that appears at order $\epsilon$ in Eq.~(\ref{loopexp}). It is the zero mode equation for the differential operator:
\begin{align} \label{diamante}
\diamondsuit=\Box+m^{2} + \frac{\lambda}{2}\phi_{0}^{2}(t)  \, ,
\end{align}
where the tree level and one-loop fields are at threshold, and the $\phi_{1/2}$ is allowed to have spatial dependence that appears in $\diamondsuit$ and its Green function. We will see that the boundary conditions kill all zero modes of $\diamondsuit$, so this equation give us that $\phi_{1/2}$ is zero as expected. This is a common feature since we are treating things at the level of the equation of motion. The order $\epsilon$ is what we are interested in and gives us the one-loop contribution for the amplitude. In Eq.~(\ref{loopexp}), there is a contribution of the two-point Green function of $\hat{\phi}_{1/2}$. The operator that we need to invert to find this Green function is Eq.~(\ref{diamante}), and in the end, set $x'= x$. This Green function will be taken at the same point so we can expect divergences to appear. These divergences are familiar from more standard Quantum Field Theory arguments.

 Using $\phi_{0}$ we can write the operator, Eq.~(\ref{diamante}), as:
\begin{align} \label{carvao}
\diamondsuit=  \Box+m^{2} + \frac{\lambda}{2}\left( \frac{z(t)}{1-\frac{\lambda}{48m^{2}}z^{2}} \right)^{2}   \, .
\end{align}
From the start we have a problem, this operator is not Hermitian because $z(t)$ is complex. This makes our job a little harder. To deal with this fact, we proceed as proposed in~\cite{loop1}. Going to Euclidian time, we can make this problem simpler. However, in Euclidean time we have a pole on the countour of integration. To adress this, we can do a shift in the Euclidian time variable to avoid the pole and then analytically continue the solution to the whole complex Euclidean plane. It is convenient to work in terms of~\cite{loop1}:
\begin{align}
u(\tau) = e^{m\tau}  \, ,
\end{align}
where $u(\tau)$ is defined as:
\begin{align} \label{rot1}
-\frac{\lambda}{48m^{2}} z(t)^{2} = u(\tau)^{2} \, ,
\end{align}
and the new Euclidian time coordinate is:
\begin{align}
\tau = i t + i \frac{i \pi}{2m} + \frac{\ln(\frac{\lambda z_{0}^{2}}{48m^{2}})}{2m}  \, .
\end{align}
The tree level solution takes the form:
\begin{align}
\phi_{0}(t) = i \sqrt{\frac{48m^{2}}{\lambda}}  \frac{u(\tau)}{1+u(\tau)^{2}} = i \sqrt{\frac{48m^{2}}{\lambda}} \frac{\sech(m\tau)}{2}  \, .
\end{align}

Then the operator to be inverted, Eq.~(\ref{carvao}), reads:
\begin{align}
\diamondsuit = \left( -\partial_{\tau}^{2} - \nabla^{2} + m^{2} - 6m^{2}\sech^{2}(m\tau) \right)  \, .
\end{align}
Doing a partial Fourier transform only in the spatial section we can write the Green function of this operator as:
\begin{align} \label{fourier}
G(\vec{x},\vec{x}' ; \tau, \tau') = \int \frac{\dd[3]{k}}{(2\pi)^{3}} G_{k}(\tau,\tau') e^{i \vec{k}.(\vec{x}-\vec{x}')} \, ,
\end{align}
the Green function satisfying:
\begin{align}
\diamondsuit(\vec{x},\tau) G(\vec{x},\vec{x}' ; \tau, \tau') = \delta^{3}(\vec{x}-\vec{x}')\delta(\tau-\tau')  \, .
\end{align}
Using Eq.~(\ref{fourier}) we focus on the following Green function:
\begin{align}
\left( -\partial_{\tau}^{2} + \theta^{2} - 6m^{2}\sech^{2}(m\tau) \right)G_{\theta}(\tau,\tau') =\delta(\tau -\tau')  \, ,
\end{align}
where we defined $\theta^{2}=\vec{k}^{2}+m^{2}$. From $G_{\theta}$ we can then find the full Green functions by doing momenta integrals using Eq.~(\ref{fourier}). We are doing this computation in $(3+1)D$,  but the generalization to other dimensions is straightforward only changing the numbers of $k$ integrals. In fact, the Green function that appears in Eq.~(\ref{loopexp}) is evaluated at coincident space-time points, therfore, we have:
\begin{align} \label{samepoint}
\ev{T\left( \hat{\phi}_{1/2}(x)\hat{\phi}_{1/2}(x)\right)}{\Omega}_{\rho} = \int \frac{\dd[3]{k}}{(2\pi)^{3}}  G_{\theta}(\tau,\tau)  \, .
\end{align}

Surprisingly this Green function is very similar to a known quantum mechanical potential (Poschl-Teller Potential)~\cite{pol}. We will transform this problem into that quantum mechanical problem and then show how to solve this potential exactly. After that, we continue with the solution to find the Green function and in the end, the loop correction. To look for this Green function, we search for two regular solutions to the homogeneous equation, one regular at $\tau \rightarrow \infty$ the other at $\tau \rightarrow -\infty$. With both solutions $f_{-}(\tau)$, $f_{+}(\tau)$ and with the Wronskian $\mathcal{W}$:
\begin{align} \label{wronk}
\mathcal{W} = f_{+}(\tau)f_{-}(\tau)' - f_{+}(\tau)'f_{-}(\tau)  \, ,
\end{align}
we then construct the Green function:
\begin{align} \label{gf}
G_{\theta}(\tau,\tau') = \frac{ f_{+}(\tau)f_{-}(\tau')}{\mathcal{W}}  \qquad \textnormal{for} \qquad \tau > \tau'  \, ,
\end{align}
\begin{align}
G_{\theta}(\tau,\tau') = \frac{ f_{+}(\tau')f_{-}(\tau)}{\mathcal{W}}  \qquad \textnormal{for} \qquad \tau' > \tau  \, .
\end{align}

Both solutions can be written in a Schrodinger like form:
\begin{align}
\left( -\partial_{\tau}^{2} - 6m^{2} \sech^{2}(m\tau) \right) f(\tau) = -\theta^{2} f(\tau)  \, ,
\end{align}
where we identify $\tilde{E}_{\theta}=-\theta^{2}$ and the Hamiltonian being the operator on the left. We can solve this only using algebra. To facilitate we can work using coordinates without dimension by doing a change of variables, $x=m \tau$:
\begin{align}
\hat{H} =  \left( -\partial_{x}^{2} - 6 \sech^{2}(x) \right)  \, .
\end{align}
The dimensionless energy is defined as:
\begin{align}
E_{\theta}=-\frac{\theta^{2}}{m^{2}}  \, ,
\end{align}
such that we need to solve the eigenvalue problem:
\begin{align} \label{prob1}
\hat{H}f(x)=E_{\theta}f(x)  \, .
\end{align}
It turns out that to find the solution for this Hamiltonian we need to generalize it to:
\begin{align} \label{prob}
\hat{H}_{l} = p^{2} - l(l+1) \sech^{2}(x)  \, .
\end{align}
Our case is $l=2$. Now we will do a little sidetrack to solve this eigenvalue problem because even though this is an exactly solvable quantum mechanical system, it is not so trivial to find the solution.

\subsection{Eigenvalues and Eigenfunctions of the Poschl-Teller Potential}
This section can be skipped without affecting the core of the thesis. Here we want to solve the quantum mechanical problem,  Eq.~(\ref{prob}):
\begin{align}
\hat{H}_{l} \ket{p,l} = E_{l} \ket{p,l}  \, .
\end{align}
We want to find the eigenfunctions for the special case of $l=2$. For each $l$ we have a quantum system with eigenvalues $E_{l}$. The spectrum that we are interested in is the continuum band of the Poschl-Teller potential:
\begin{align} \label{banana}
V(x)= -l(l+1)\sech^{2}(x) \, .
\end{align}
The esiest one is $l=0$, the system is the free particle, and the energy does not depend on $l$ trivially:
\begin{align}
\hat{H}_{0} \ket{p,0}= p^{2}\ket{p,0}  \, .
\end{align}

The special propriety of this system is that it belongs to a class of factorizable potentials. This feature is reminiscent from the supersymmetric version of this potential~\cite{pt1}:
\begin{align}
V_{s}(x)= -l(l+1)\sech^{2}(x) + l^{2} \, .
\end{align}
Because we are interested in the continuum spectrum of Eq.~(\ref{banana}), we will not introduce Supersymmetric Quantum Mechanics~\cite{susy1,susy2}. Nevertheless, Supersymmetry is the basis of why this process work and why this potential is solvable. The factorization of the Hamiltonian  Eq.~(\ref{prob}) can be archived using the following operators:
\begin{align} \label{caca}
a_{l} = p - i l \tanh(x)  \, ,
\end{align}
\begin{align} \label{coco}
a_{l}^{\dagger} = p + i l \tanh(x)  \, .
\end{align} 
They are choosen in this form to satisfy:
\begin{align}
\comm{a_{l}^{\dagger}}{a_{l}} = 2l \sech^{2}(x)  \,  ,
\end{align}

\begin{figure}[h!]
\centering
\includegraphics[width=15cm]{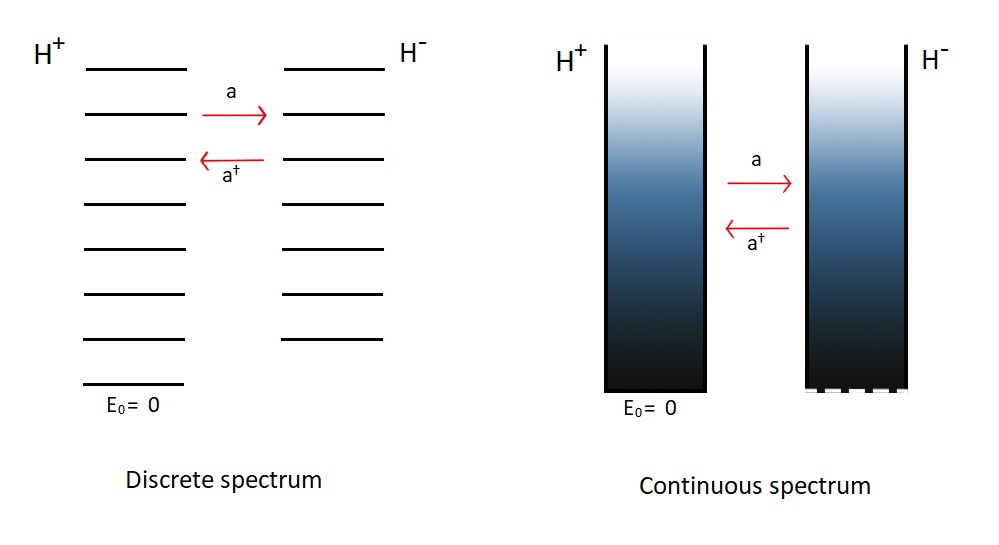}
\caption{Relation between the two parterns Hamiltonians..}
\label{supa}
\end{figure}

With the operators Eq.~(\ref{caca}) and Eq.~(\ref{coco}) we can construct the initial Hamiltonian Eq.~(\ref{prob}):
\begin{align} \label{caraca}
H_{l}^{+} \equiv a_{l}^{\dagger}a_{l}= H_{l}+l^{2}  \,  .
\end{align}
The object in the l.h.s of Eq.~(\ref{caraca}) is the Hamiltonian for the supersymmetric description of this potential. We can define its parter Hamiltonian exchanging the order of the operators:
\begin{align} \label{caracafermionico}
H_{l}^{-} \equiv a_{l}a_{l}^{\dagger}=H_{l-1}-l^{2}  \, .
\end{align}
In this case both supersymmetric partners are related only by a change in constants inside the Hamiltonian. This class of systems is called Shape Invariant Potentials~\cite{shape1}, and this plays a pivot role in making this potential solvable. With the definitions of  Eq.~(\ref{caraca}) and Eq.~(\ref{caracafermionico}) we can see that, if we have an eigenstate of $H^{+}$ with contiuum spectrum:
\begin{align}
H_{l}^{+} \ket{E^{+},l}=E^{+}_{l} \ket{E^{+},l}  \, ,
\end{align} 
there exists  an eigenstate of $H^{-}_{l}$ with the same eigenvalue, except for the ground state:
\begin{align} \label{mesalva}
a_{l}H_{l}^{+} \ket{E^{+},l}= H_{l}^{-} \left( a_{l} \ket{E^{+},l}  \right) =E^{+}_{l} \left( a_{l} \ket{E^{+},l} \right)  \,  .
\end{align} 
This occurs the other way around as well. This is a consequence of the supersymmetry on the system, both spectrums are related by the operators  $a_{l}$ and $a_{l}^{\dagger}$ as it is represented in Figure~\ref{supa}.

Because this potential is shape invariant, using  Eq.~(\ref{caraca}) and Eq.~(\ref{caracafermionico}) we can see that the spectrum does not depend on $l$. The shape invariance tells us that a change of $l$ relates $H^{+}$ and $H^{-}$.  This implies that the eigenvalues will be the same as the free case, $l=0$:
\begin{align}
E_{l} = p^{2}  \, .
\end{align}
Now, to construct the different eigenestates we can use Eq.~(\ref{mesalva}) to build up all the states of $H_{l}$. For $l=0$ we have:
\begin{align}
\bra{x}\ket{p,0}=e^{i p x}  \, ,
\end{align} 
and for l=1:
\begin{align}
\ket{p,1}=a^{\dagger}_{1}\ket{p,0}  \, .
\end{align}
Just to see that indeed this is a eigenstate of $H_{1}$ we can check directly:
\begin{align}
H_{1} \ket{p,1} = (\eta_{1}-1)a^{\dagger}_{1} \ket{p,0} = a_{1}^{\dagger} \xi_{1} \ket{p,0} - a_{1}^{\dagger} \ket{p,0} = \\
= (p^{2}+1)a_{1}^{\dagger}\ket{p,0} - a_{1}^{\dagger}\ket{p,0} = p^{2} (a_{1}^{\dagger}\ket{p,0}=p^{2} \ket{p,1}  \, .
\end{align}
The wave function for the $l=1$ case is:
\begin{align}
\bra{x}\ket{p,1} = \mel{x}{a_{1}^{\dagger}}{p,0}= \mel{x}{\hat{p}+i \tanh(\hat{x})}{p,0} = \left( p+i \tanh(x) \right) e^{ipx}  \, .
\end{align}
Finally, the case of interest is the $l=2$ case:
\begin{align}
\bra{x}\ket{p,2} = \mel{x}{a^{\dagger}_{2}}{p,1}=\left( 1+p^{2}+3ip \tanh(x)-3\tanh^{2}(x)\right) e^{ipx}  \, .
\end{align}

Using this wave function, we can extend this solution to all complex $p$ plane to get both functions to construct the Green function Eq.~(\ref{gf}):
\begin{align}
p =\pm \frac{\theta}{m}  \, .
\end{align}
Because we need regular solutions at $\pm \infty$, both $p$'s will be used in this contruction:
\begin{align} \label{sol1}
f_{-}(x)= \left( 1-\frac{\theta^{2}}{m^{2}} +3\frac{\theta}{m}\tanh(x) -3\tanh^{2}(x) \right)e^{\frac{\theta x}{m}}  \, ,
\end{align}
\begin{align} \label{sol2}
f_{+}(x)= \left(1-\frac{\theta^{2}}{m^{2}} -3\frac{\theta}{m}\tanh(x) -3\tanh^{2}(x)\right) e^{-\frac{\theta x}{m}}  \, .
\end{align}

\subsection{Back to $\frac{\lambda\phi^{4}}{4!}$ in the Unbroken Phase}

Now that we have both solutions Eq.~(\ref{sol1}) and Eq.~(\ref{sol2}) to construct the Green function of Eq.~(\ref{gf}) we just need to write them in terms of the variables that we are using:
\begin{align}
u(x) = e^{x}  \, ,
\end{align}
\begin{align}
f_{-}(x) = \frac{u^{4}(-\frac{\theta^{2}}{m^{2}}+3\frac{\theta}{m}-2) + u^{2}(-2\frac{\theta^{2}}{m^{2}}+8)-\frac{\theta^{2}}{m^{2}}-3\frac{\theta}{m}-2}{(1+u^{2})^{2}} u^{\frac{\theta}{m}}  \, ,
\end{align}
\begin{align}
f_{+}(x) = \frac{u^{4}(-\frac{\theta^{2}}{m^{2}}-3\frac{\theta}{m}-2) + u^{2}(-2\frac{\theta^{2}}{m^{2}}+8)-\frac{\theta^{2}}{m^{2}}+3\frac{\theta}{m}-2}{(1+u^{2})^{2}} u^{-\frac{\theta}{m}}  \, .
\end{align}
Having these two solutions we can calculate the Wronskian, Eq.~(\ref{wronk}):
\begin{align}
\mathcal{W} = 2\frac{\theta}{m^{4}}(\theta^{2}-m^{2})(\theta^{2}-4m^{2})  \, .
\end{align}

With this, we can construct the equal time Green function:
\begin{align}
G_{\theta}(\tau,\tau) = \frac{f_{+}(m\tau)f_{-}(m\tau)}{\mathcal{W}}=
\end{align}
\begin{align} \nonumber
= \frac{1}{2\theta (1+u^{2})^{4}} \left(  (1+u^{8}) + (u^{6}+u^{2})\frac{4(\theta^{2}+2m^{2})}{(\theta^{2}-m^{2})} + u^{4} \frac{6(\theta^{4}-m^{2}\theta^{2}+12m^{4})}{(\theta^{2}-m^{2})(\theta^{2}-4m^{2})} \right)  \, .
\end{align}
The Green function in this form is not so useful. We can use partial fraction expansion in the $u$ variable to separate the different kind of contributions to the Green function. Doing so it is straightforward to see that we have a finite part and a divergent part when doing the Fourier integral in Eq.~(\ref{fourier}). We can separate both parts in the following way that will facilitate the renormalization latter:
\begin{align}
G_{\theta}(\tau,\tau) =  G_{\theta}(\tau,\tau)^{div}  + G_{\theta}(\tau,\tau)^{fin}  \, ,
\end{align}
\begin{align}
G_{\theta}(\tau,\tau)^{div} = \frac{1}{2\theta} + \frac{6m^{2}u^{2}}{(1+u^{2})^{2}}\frac{1}{\theta^{3}}  \, ,
\end{align}
\begin{align}
G_{\theta}(\tau,\tau)^{fin} = \frac{6m^{4}}{\theta^{3}(\theta^{2}-m^{2})} \left(  \frac{u^{2}+u^{6}}{(1+u^{2})^{4}} \right)+ \frac{6m^{4}u^{4}}{(1+u^{2})^{4}}\left( \frac{14\theta^{2}-8m^{2}}{\theta^{3}(\theta^{2}-m^{2})(\theta^{2}-4m^{2})} \right)  \, .
\end{align}

Now we are ready to integrate this to get the full Green function that enters into the one-loop equation:
\begin{align}
G(\vec{x},\vec{x};\tau,\tau) = \int \frac{\dd[3]{k}}{(2\pi)^{3}} G_{\theta}(\tau,\tau)  \, ,
\end{align}
with $\theta^{2}= \vec{k}^{2}+m^{2}$. The divergent part of the two-point function is:
\begin{align}
G^{div}(\vec{x},\vec{x};\tau,\tau) =  \frac{1}{2}\mathcal{I}_{1} + \frac{6m^{2}u^{2}}{(1+u^{2})^{2}}\mathcal{I}_{2}  \, ,
\end{align}
where $\mathcal{I}_{1}$ is quadratically divergent and $\mathcal{I}_{2}$ has an logarithmic divergence:
\begin{align}
\mathcal{I}_{1} = \int \frac{\dd[3]{k}}{(2\pi)^{3}} \frac{1}{(\vec{k}^{2}+m^{2})^{1/2}}  \, ,
\end{align} 
\begin{align}
\mathcal{I}_{2} = \int \frac{\dd[3]{k}}{(2\pi)^{3}} \frac{1}{(\vec{k}^{2}+m^{2})^{3/2}}  \, .
\end{align} 
To get the finite part we need to do two integrals:
\begin{align} \label{int1}
\mathcal{I}_{f1} =  \int \frac{\dd[3]{k}}{(2\pi)^{3}}  \frac{6m^{4}}{\theta^{3}(\theta^{2}-m^{2})}  \, ,
\end{align}
\begin{align} \label{int2}
\mathcal{I}_{f2}=  \int \frac{\dd[3]{k}}{(2\pi)^{3}} \frac{6m^{4}(14\theta^{2}-8m^{2})}{\theta^{3}(\theta^{2}-m^{2})(\theta^{2}-4m^{2})}  \, .
\end{align}

The angular part of these integrals is trivial to solve, and the only difficult part is the radial integration. One important thing to remember is that the fields that generate such propagators have fixed boundary conditions to give the right vacuum projection. The boundary conditions fix the prescription to pass through the poles, and this means that the denominator in Eq.~(\ref{int1}) and Eq.~(\ref{int2}) has a $-m^{2}+i\epsilon$ term. This information is important only for the poles inside the domain of integration. The next step to solve these integrals is to change to a dimensionless variable:
\begin{align}
\theta = \omega m   \, .
\end{align}
Doing this transformation it is straightforward to find the solution for these integrals:
\begin{align}
\mathcal{I}_{f1} = \frac{3m^{2}}{\pi^{2}} \int_{0}^{\infty} \dd{\omega} \frac{1}{\omega^{2}(\omega^{2}-1)^{1/2}}= \frac{3m^{2}}{\pi^{2}}    \, ,
\end{align}
\begin{align}
\mathcal{I}_{f2}= \lim_{\epsilon \rightarrow 0}  \frac{3m^{2}}{\pi^{2}}   \int_{0}^{\infty} \dd{\omega} \frac{(14\omega^{2}-8)}{\omega^{2}(\omega^{2}-1)^{1/2}(\omega^{2}-4+i\epsilon)}= 
\end{align}
\begin{align} \nonumber
=\lim_{\epsilon \rightarrow 0}   \frac{6m^{2}}{\pi^{2}} \left[ \frac{4i \sqrt{-3+i\epsilon}\sqrt{4i+\epsilon}+\sqrt[4]{-1}(24-7i\epsilon)\sinh^{-1}(\sqrt[4]{-1}\sqrt{4i+\epsilon})}{\sqrt{i\epsilon-3} (\epsilon +4i)^{3/2}}  \right]  \, .
\end{align}
Doing the $\epsilon$ limit we get:
\begin{align}
\mathcal{I}_{f2}=  \frac{6m^{2}}{\pi^{2}} +\frac{3im^{2} \sqrt{3}}{\pi} - \frac{3m^{2}\sqrt{3}}{\pi^{2}} \ln(\frac{2+\sqrt{3}}{2-\sqrt{3}})  \, .
\end{align}

Thus, the full two-point function, Eq.~(\ref{samepoint}), is:
\begin{align}
G(\vec{x},\vec{x};\tau,\tau) =  \frac{1}{2}\mathcal{I}_{1} + \frac{6m^{2}u^{2}}{(1+u^{2})^{2}}\left( \mathcal{I}_{2}+\frac{1}{2\pi^{2}} \right) - \frac{6m^{2}u^{4}}{(1+u^{2})^{4}} F  \, ,
\end{align}
where we use the notation from~\cite{loop1}:
\begin{align}
F = \frac{\sqrt{3}}{2\pi^{2}} \left( \ln(\frac{2+\sqrt{3}}{2-\sqrt{3}}) - i \pi \right)  \, .
\end{align}

An important point to make is that the $\epsilon$ limit is evaluated from the right such that these expressions make sense.

 This result is in accordance with~\cite{loop1}. Now, we need to renormalize our theory, we could do dimensional regularization to extract only the divergent part from the integrals using minimal subtraction. However, it is simpler to just absorb the whole divergent part, using the same scheme as in~\cite{loop1}. To visualize better the renormalization we can re-write the equation of motion using:
\begin{align}
\lambda \rightarrow \lambda_{R} + \delta \lambda  \, ,
\end{align}
\begin{align}
m^{2} \rightarrow m^{2}_{R} + \delta m^{2}  \, ,
\end{align}
where the corrections start at order $\epsilon$, being a power series in this variable. To see that these two constants absorb the divergences, we insert back the two-point function in the equation of motion Eq.~(\ref{loopexp}) to get:
\begin{align} \label{qqtofazendo}
\left[ (-\partial_{\tau}^{2}+m^{2})\phi_{0}(t) +\frac{\lambda}{3!}\phi_{0}^{3}(t)-\rho(t) \right] + 
\end{align}
\begin{align} \nonumber
+ \sqrt{\epsilon} \left[ \left( \Box+m^{2} + \frac{\lambda}{2}\phi_{0}^{2}(x) \right) \phi_{1/2}(x) \right]+
\end{align}
\begin{align}
+\epsilon \left[  \left( -\partial_{\tau}^{2}+m^{2} + \frac{\lambda}{2}\phi_{0}^{2}(x) \right) \phi_{1}(x) + \frac{\lambda \phi_{0}}{2} \left( \frac{1}{2}\mathcal{I}_{1} + \frac{6m^{2}u^{2}}{(1+u^{2})^{2}}(\mathcal{I}_{2}+\frac{1}{2\pi^{2}}) - \frac{6m^{2}u^{4}}{(1+u^{2})^{4}} F \right) \right]+ \mathcal{O}(\epsilon^{3/2}) =0  \, .
\end{align}
Using the definition of the tree level solution, Eq.~(\ref{unbroken}), the divergent part has the distinctive forms:
\begin{align}
\phi_{0} \left( \frac{\lambda m \mathcal{I}_{1}}{4} \right)  \, ,
\end{align}
\begin{align}
\phi_{0}^{3} \left( -\frac{3\lambda^{2}}{48}(\mathcal{I}_{2}+\frac{1}{2\pi^{2}}) \right)  \, .
\end{align}

So we can see that the redefinition of mass and coupling could absorb these divergences. We write these constants like:
\begin{align}
m^{2} = m_{R}^{2} +\epsilon \delta m^{2}_{1} + \mathcal{O}(\epsilon^{2})  \, ,
\end{align}
\begin{align}
\lambda = \lambda_{R} +\epsilon \delta \lambda_{1}+ \mathcal{O}(\epsilon^{2})  \, .
\end{align}
Using this definition in the equation of motion, Eq.~(\ref{qqtofazendo}), we get:
\begin{align}
\left[ (-\partial_{\tau}^{2}+m^{2}_{R})\phi_{0}(t) +\frac{\lambda_{R}}{3!}\phi_{0}^{3}(t)-\rho(t) \right] + 
\end{align}
\begin{align} \nonumber
+ \sqrt{\epsilon} \left[ \left( \Box+m^{2}_{R} + \frac{\lambda_{R}}{2}\phi_{0}^{2}(t)\right) \phi_{1/2}(x) \right]+
\end{align}
\begin{align} \nonumber
+\epsilon \left[  \left( -\partial_{\tau}^{2}+m^{2}_{R} + \frac{\lambda_{R}}{2}\phi_{0}^{2}(\tau) \right) \phi_{1}(\tau) + \phi_{0} \left( \delta m^{2}_{1} + \frac{\lambda_{R}I_{1}}{4} \right) + \phi_{0}^{3} \left( \frac{\delta\lambda_{1}}{3!}-\frac{\lambda_{R}^{2}}{16}(I_{2}+\frac{1}{2\pi^{2}}) \right) - \frac{6m^{2}_{R}u^{4}}{(1+u^{2})^{4}} F \right]+
\end{align}
\begin{align} \nonumber
+ \mathcal{O}(\epsilon^{3/2}) =0  \, .
\end{align}
The obvious choice for the counter terms are:
\begin{align}
\delta m^{2}_{1} =- \frac{\lambda_{R}\mathcal{I}_{1}}{4}  \, ,
\end{align}
\begin{align}
\frac{\delta\lambda_{1}}{3!} =\frac{\lambda_{R}^{2}}{16}(\mathcal{I}_{2}+\frac{1}{2\pi^{2}})  \, .
\end{align}

Choosing this renormalization we can proceed to solve the one-loop generator of the amplitudes. The tree level solution stays the same as before, except for the replacement of bare for renormalized quantities. The one-loop equation simplifies to:
\begin{align}
\left( -\partial_{\tau}^{2}+m^{2}_{R} + \frac{\lambda_{R}}{2}\phi_{0}^{2}(\tau) \right) \phi_{1}(\tau) = \frac{6m_{R}^{2}u^{4}}{(1+u^{2})^{4}}F \frac{\lambda_{R}\phi_{0}}{2}  \, .
\end{align}
Writing the equation in terms of $u(\tau)=e^{m_{R}\tau}$ we get:
\begin{align}\label{qqto}
\left( -\partial_{\tau}^{2}+m^{2}_{R} -\frac{24m_{R}^{2}u^{2}}{(1+u^{2})^{2}} \right) \phi_{1}(\tau) = 12 i m_{R}^{3} F \sqrt{3\lambda_{R}} \frac{u^{5}}{(1+u^{2})^{5}}  \, .
\end{align}
From the form of the right side of Eq.~(\ref{qqto}) we can try to look for solutions of the form:
\begin{align}
\phi_{1}(\tau) = \alpha \frac{u^{5}}{(1+u^{2})^{3}}  \, ,
\end{align}
and confirm by direct computation that this is a solution if:
\begin{align}
\alpha = -\frac{iF \lambda_{R}}{8} \sqrt{\frac{48m_{R}^{2}}{\lambda_{R}}}  \, .
\end{align}

Having this solution we can analytically continue back to real time using the relation Eq.~(\ref{rot1}), using the renormalized mass inside $z(t)$. This gives us:
\begin{align}
\phi_{1}(t) =- \frac{F \lambda_{R}}{8} \left( \frac{\lambda_{R}}{48m_{R}^{2}} \right)^{2}  \frac{z^{5}}{(1-\frac{\lambda_{R}}{48m_{R}^{2}}z^{2})^{3}}  \, .
\end{align}
Then the full contribution for the generator of the amplitudes at one-loop is:
\begin{align}
\phi_{0}(t)+\phi_{1}(t)=\phi_{0+1}(t) = \frac{z}{(1-\frac{\lambda_{R}}{48m_{R}^{2}}z^{2})} \left[ 1 - \epsilon \frac{F\lambda_{R}}{8} \frac{(\frac{\lambda_{R}}{48m_{R}^{2}})^{2}z^{4}}{(1-\frac{\lambda_{R}}{48m_{R}^{2}}z^{2})^{2}} \right]  \, .
\end{align}
To calculate the amplitude from this solution we just need to differentiate $n$ times  it with respect to $z(t)$, following Eq.~(\ref{amp1}):
\begin{align} \label{ampun}
\mathcal{A}_{1loop} (1 \rightarrow 2k+1) = (2k+1)! \left( \frac{\lambda_{R}}{48m_{R}^{2}} \right)^{k} \left[ 1-\frac{F \lambda_{R} k(k-1)}{16} \right]  \, ,
\end{align}
where we set $\epsilon$ to one. This expression was obtained first in~\cite{loop1}.

 Here we can see that the factorial growth persists. The next correction only makes things worse, and this can be seen as an indication that we are using the wrong approximation for this regime. We expect that in a large $k$ approximation of the amplitude, the corrections to be of order $1/k$. In this case, the true object that needs to be small such that this approximation is useful is $\lambda_{R}k^{2}$. In the regime where this is small, this expression is well defined. It is not known how much we can trust this expression outside this regime even though we arrived at it only using that $\lambda_{R}$ is small. This is because, for a large $n$, the loop correction will be larger than the tree level one, signaling that the approximation is not good. We discuss this further in the context of a simpler toy model at the end of chapter~\ref{c3}. For now, let us see if this behavior is the same in the broken phase of this theory. 

\subsection{$\frac{\lambda\phi^{4}}{4!}$ in the Broken Phase}
In the case of broken reflection symmetry we need to be more careful in the renormalization because the shift done at tree level in Eq.~(\ref{shift}) is not the appropriate one. We use the shift in the variables done before Eq.~(\ref{shift}), and when we get to renormalization, we will do the appropriate adjustments as in~\cite{Smith-apr-93}. Aside from this, everything else is very similar to the unbroken case. We expand the $\hat{\sigma}$ operator:
\begin{align}
\hat{\sigma}(x) = \sigma_{0} \hat{I} + \sqrt{\epsilon} \hat{\sigma}_{1/2}(x) +\epsilon \hat{\sigma}_{1}(x) + \mathcal{O}(\epsilon^{3/2})  \, .
\end{align}
The equation of motion, Eq.~(\ref{eqbro}), using this expansion is:
\begin{align}
\left[ (\Box + M^{2})\sigma_{0}(x) + \frac{\lambda \phi_{min}}{2} \sigma_{0}^{2}(x) + \frac{\lambda}{3!}\sigma_{0}^{3}(x) - \rho(x) \right] + 
\end{align}
\begin{align} \nonumber
+ \sqrt{\epsilon} \left[ \left( \Box + M^{2} + \lambda\phi_{min} \sigma_{0}(x)+ \frac{\lambda}{2}\sigma_{0}^{2} \right)\sigma_{1/2}(x) \right]+ 
\end{align}
\begin{align}
+ \epsilon \left[ \left( \Box + M^{2} + \lambda\phi_{min} \sigma_{0}(x)+ \frac{\lambda}{2}\sigma_{0}^{2} \right) \sigma_{1}(x) + \frac{\lambda}{2}(\phi_{\min}+\sigma_{0})\ev{T\left( \hat{\sigma}_{1/2}(x)\hat{\sigma}_{1/2}(x) \right)}{\Omega}_{\rho} \right] + \mathcal{O}(\epsilon^{3/2}) =0  \, ,
\end{align}
where $\phi_{min}^{2}= \frac{3M^{2}}{\lambda}$ and $M= \sqrt{2}\mu$ is the mass of the excitation.  We are interested in solving the one-loop contribution at threshold. Just like before the tree level is already solved when we ask for homogeneous solution and the next order only give the information about the two-point function that enters in the one-loop equation. The tree level solution is the one that we calculated before:
\begin{align}
\sigma_{0}(t)= \frac{z(t)}{1- \frac{z(t)}{2\phi_{\min}}}  \, .
\end{align}

The operator that we need to invert to find the two-point function now is:
\begin{align}
\diamondsuit_{b} = \Box + M^{2} +\lambda\phi_{min} \sigma_{0} + \frac{\lambda\sigma_{0}^{2}}{2}  \, .
\end{align}
Using the solution $ \sigma_{0}(t)$ we get to the same problem of non-hermiticity of the operator. Doing the same step as before, we perform a Wick rotation taking care of the pole in the Euclidean line:
\begin{align}
-\frac{z(t)}{2\phi_{min}} = u(\tau) = e^{M\tau}  \, ,
\end{align}
\begin{align}
\tau = it + i \frac{\pi}{M} +\frac{1}{M} \ln(\frac{z_{0}}{2\phi_{min}})  \, .
\end{align}
Doing this change of variables, the operator that we want to invert becomes:
\begin{align}
\diamondsuit_{b} = -\partial_{\tau}^{2} -\nabla^{2} + M^{2} -2\phi_{min}^{2} \lambda \frac{u}{(1+u)^{2}}  \, .
\end{align}
We can re-write the last term in a familiar form:
\begin{align}
\diamondsuit_{b} = -\partial_{\tau}^{2} -\nabla^{2} + M^{2} - \frac{3M^{2}}{2} \sech^{2}(\frac{M\tau}{2})  \, .
\end{align}

Now the steps are very similar, we do a partial Fourier transform in the spatial part. The remaining operator that we need to invert is almost what we had before:
\begin{align}
\left( -\partial_{\tau}^{2} +\theta^{2} - \frac{3M^{2}}{2} \sech^{2}(\frac{M\tau}{2})\right) G_{\theta}(\tau,\tau') = \delta(\tau-\tau')  \, ,
\end{align}
where $\theta^{2}=\vec{k}^{2}+M^{2}$. Doing a change of variables to an dimensionless one we can cast the functions in a known form:
\begin{align}
\frac{M \tau}{2}  =\xi  \, ,
\end{align}
\begin{align}
\left( -\frac{M^{2}}{4}\partial_{\xi}^{2} +\theta^{2}-\frac{3M^{2}}{2} \sech^{2}(\xi) \right) f(\xi)=0  \, .
\end{align}
This is almost the equation that we had before, Eq.~({\ref{prob1}), just re-scaling the $\theta$:
\begin{align}
\theta \rightarrow 2\theta  \, .
\end{align}
In terms of $u(\tau)$ the solution will change the power because of the definition of $\xi$:
\begin{align}
u^{2} \rightarrow u  \, .
\end{align}

We already solved these equations in the last section, so the same point Green function before integrating is:
\begin{align}
G_{\theta}(\tau,\tau) =  G_{\theta}(\tau,\tau)^{div} + G_{\theta}(\tau,\tau)^{fin}  \, ,
\end{align}
\begin{align}
 G_{\theta}(\tau,\tau)^{div} = \frac{1}{4\theta} + \frac{3M^{2}}{\theta^{3}} \frac{u}{(1+u)^{2}}  \, ,
\end{align}
\begin{align}
G_{\theta}(\tau,\tau)^{fin} = \frac{3M^{4}}{4\theta^{3}(4\theta^{2}-M^{2})}\frac{u}{(1+u)^{4}} + \frac{3M^{4}(56\theta^{2}-8M^{2})}{4\theta^{3}(4\theta^{2}-M^{2})(4\theta^{2}-4M^{2})}\frac{u^{2}}{(1+u)^{4}} + 
\end{align}
\begin{align} \nonumber
+\frac{3 M^{4}}{4 \theta^{3}(4\theta^{2}-M^{2})} \frac{u^{3}}{(1+u)^{4}}  \, .
\end{align}
The same definition of the divergent integrals $\mathcal{I}_{1}$ and $\mathcal{I}_{2}$ are used to write the divergent part of the two-point function as:
\begin{align}
G(\vec{x},\vec{x};\tau,\tau)^{div} = \frac{1}{4}\mathcal{I}_{1} +  \frac{u}{(1+u)^{2}} \frac{3M^{2} \mathcal{I}_{2}}{4}  \, .
\end{align} 
Now for the finite part, we need to solve the following two integrals:
\begin{align}
\mathcal{I}_{f3}= \int \frac{\dd[3]{k}}{(2\pi)^{3}} \frac{1}{\theta^{3}(4\theta^{2}-M^{2})}  \, ,
\end{align}
\begin{align}
\mathcal{I}_{f4}= \int \frac{\dd[3]{k}}{(2\pi)^{3}} \frac{56\theta^{2}-8M^{2}}{\theta^{3}(4\theta^{2}-M^{2})(4\theta^{2}-4M^{2})}  \, .
\end{align}
This time there is no pole inside the domain of integration, so we don't need to worry about the Feynman prescription.  The finite part in terms of these integrals is:
\begin{align}
G(\vec{x},\vec{x};\tau,\tau)^{fin} = \frac{3M^{4}}{4} \left( \mathcal{I}_{f3} \frac{u +u^{3}}{(1+u)^{4}} + \mathcal{I}_{f2} \frac{u^{2}}{(1+u)^{4}} \right)  \, .
\end{align}
The integrals can be solved directly, in both cases we use the dimensionless coordinates $\theta=\omega M$:
\begin{align}
\mathcal{I}_{f3} = \frac{1}{M^{2}} \left( \frac{6 - \sqrt{3}\pi}{12\pi^{2}} \right)  \, ,
\end{align}
\begin{align}
\mathcal{I}_{f4} = \frac{1}{M^{2}} \left( \frac{6 + \sqrt{3} \pi}{6\pi^{2}} \right)  \, .
\end{align}
Using this information we can construct the two-point function. We just write in an appropriate form to interpret after during renormalization:
\begin{align}
G(\vec{x},\vec{x};\tau,\tau) = \frac{1}{4} \mathcal{I}_{1} + \frac{u}{(1+u)^{2}} \left( \frac{3M^{2}}{4} \mathcal{I}_{2} + \frac{3 M^{2}}{8\pi^{2}} - \frac{M^{2} \sqrt{3}}{16\pi} \right) + \frac{M^{2} \sqrt{3}}{4\pi} \frac{u^{2}}{(1+u)^{4}}  \, .
\end{align} 

We need to deal with the divergent part now. We can absorb them in the renormalization of the constants:
\begin{align}
M^{2} = M_{R}^{2} +\epsilon \delta M^{2}_{1} + \mathcal{O}(\epsilon^{2})  \, ,
\end{align}
\begin{align}
\lambda = \lambda_{R} + \epsilon \delta \lambda_{1} + \mathcal{O}(\epsilon^{2})  \, .
\end{align}
To do the renormalization properly let us focus only in the contribution for the one-loop equation, that is in the form:
\begin{align} \label{nacaba}
\frac{\lambda}{2} (\phi_{min}+\sigma_{0}) G(\vec{x},\vec{x};\tau,\tau)  \, .
\end{align}
As we said before it is better to work in the unshifted field because this shift receives quantum corrections. Using the definition $\frac{\phi_{0}}{\phi_{\min}}= \gamma$ we can write Eq.~(\ref{nacaba}) in general as:
\begin{align}
\frac{\lambda \phi_{min}}{2} \gamma \left[ A - \frac{B}{4}(\gamma^{2}-1) +\frac{C}{16}(\gamma^{2}-1)^{2} \right]  \, ,
\end{align} 
where the constants are:
\begin{align}
A= \frac{\mathcal{I}_{1}}{4}  \, ,
\end{align}
\begin{align}
B =  \frac{3M^{2}}{4} \mathcal{I}_{2} + \frac{3 M^{2}}{8\pi^{2}} - \frac{M^{2} \sqrt{3}}{16\pi}  \, ,
\end{align}
\begin{align}
C =  \frac{M^{2} \sqrt{3}}{4\pi}   \, .
\end{align}
Only $A$ and $B$ have divergences, the contribution for the one-loop equation is written as:
\begin{align}
\gamma \frac{\lambda \phi_{min}}{2} \left( A-\frac{B}{4} \right) - \frac{\lambda \phi_{min}}{2}\gamma^{3}B + \frac{\lambda \phi_{min}}{2} C \left( \frac{\gamma^{5}}{16}-\frac{\gamma^{3}}{8}+ \frac{\gamma}{16} \right)  \, .
\end{align}
The renormalization scheme of~\cite{Smith-apr-93} will absorb the two initial terms. Because we are using the unshifted field it is easy to read the renormalized mass and coupling. We just need to be careful because in the unshifted field the mass has the ``wrong'' sign:
\begin{align}
m^{2}= m_{R}^{2}+\epsilon \delta m_{1}^{2} + \mathcal{O}(\epsilon^{2})  \, ,
\end{align}
\begin{align}
\lambda= \lambda_{R} + \epsilon \delta \lambda_{1} + \mathcal{O}(\epsilon^{2})  \, .
\end{align}
Using this in the equation of motion, we see that:
\begin{align}
\delta m_{1}^{2} = \frac{\lambda_{R}}{8} \left(\mathcal{I}_{1} - M_{R}^{2}(\frac{3 \mathcal{I}_{2}}{4} + \frac{3}{8\pi^{2}} -\frac{\sqrt{3}}{16\pi}) \right)  \, ,
\end{align}
\begin{align}
\frac{\delta \lambda_{1}}{3!} = \frac{\lambda_{R}^{2}}{48} \left( \frac{3\mathcal{I}_{2}}{2} +\frac{3}{4\pi^{2}} - \frac{\sqrt{3}}{16\pi} \right)  \, .
\end{align}

In this renormalization scheme we have to solve now the equation for $\sigma_{1}$, where we shift by the renormalized couplings. The equation for the tree level stays the same and the one-loop became:
\begin{align}
\left( -\partial_{\tau}^{2} + M^{2}_{R} + \lambda_{R}\phi_{min}^{R} \sigma_{0}(\tau)+ \frac{\lambda_{R}}{2}\sigma_{0}^{2}(\tau) \right) \sigma_{1}(\tau) = -\frac{\lambda_{R} \phi_{min}^{R}M_{R}^{2}\sqrt{3}}{8\pi} \left( \frac{\gamma^{5}}{16}-\frac{\gamma^{3}}{8}+ \frac{\gamma}{16} \right)  \, .
\end{align}
We can write everything in terms of $u(\xi)$ where $\xi=M_{R}\tau$ is the dimensionless time coordinate. The equation using these variables is:
\begin{align} \label{eqeq}
 \left( \partial_{\xi}^{2} -1 + \frac{6u}{(1+u)} - \frac{6u^{2}}{(1+u)^{2}} \right)\sigma_{1}(\xi) = \frac{3M_{R}\sqrt{\lambda_{R}} u^{2}(1-u)}{8\pi(1+u)^{5}}  \, .
\end{align}
To solve this equation we look for an ansatz of the form:
\begin{align}
\sigma_{1} = \alpha \frac{u^{2}}{(1+u)^{3}}  \, .
\end{align}
Plugging this solution in Eq.~(\ref{eqeq}), it is immediate to see that $\alpha =\frac{M_{R}\sqrt{\lambda_{R}}}{8\pi}$. This means that the one-loop solution is:
\begin{align}
\sigma_{1} = \frac{M_{R}\sqrt{\lambda_{R}}}{8\pi} \frac{u^{2}}{(1+u)^{3}}  \, .
\end{align} 
The complete solution for the generator of the tree level and threshold amplitudes at one-loop is then:
\begin{align}
\sigma_{0}+\sigma_{1}=\sigma_{0+1}(\tau)= \frac{-2\phi_{\min}^{R} u}{(1+u)} \left( 1+ \epsilon\frac{\lambda_{R}^{3/2}}{96\pi M_{R}}\left(-2\phi_{min}^{R} \frac{u}{(1+u)^{2}}\right) \right)  \, .
\end{align}
With this solution we can analytically continue to the whole complex $\tau$ plane to get the answer in real time, in terms of $z(t)$:
\begin{align}
\sigma_{0+1}(t) = \frac{z}{1-\frac{z}{2\phi_{\min}^{R}}} \left(1 + \epsilon\frac{\lambda_{R}^{3/2}}{96\pi M_{R}} \frac{z}{(1-\frac{z}{2\phi_{min}^{R}})^{2}}\right)  \, .
\end{align}

Having this solution we can find the one-loop amplitude at threshold doing $n$ derivatives in $z(t)$  using Eq.~(\ref{ampbroken1}) (and setting $\epsilon$ to one):
\begin{align} \label{ampbro}
\mathcal{A}_{1loop}^{B}(1 \rightarrow n) = n! \left( \frac{1}{2\phi_{0}^{R}}\right)^{n-1}\left(1+\frac{\lambda_{R}\sqrt{3}}{96\pi} n(n-1)\right)  \, .
\end{align}

We can see that this contribution is real, different from the unbroken case Eq.~(\ref{ampun}). It has a different sign, and it still has factorial growth even in the first order. It is clear that the important coupling is, in this case, $\lambda n^{2}$. This is the object that needs to be small such that the approximation makes sense. This is similar to the case before, so in a high multiplicity limit, it is not so straightforward to recover information from these initial terms. This discussion will be continued at the end of chapter~\ref{c3}, were we discuss the range of validity of perturbation theory for high multiplicity calculation. If we want to understand this better, we need to try to recover momentum dependence in these amplitudes. Without this, we cannot say anything about the decay rate that is meaningful. It is worth to point out that all the results so far are exact in 0 spatial dimensions, giving just quantum mechanics. There is no phase space in this case, and we still have the factorial growth of these amplitudes. This could be problematic to the unitarity of the theory if this equation can be trusted in the region of interest.  Next, we show a possible way to understand why this factorial growth is happening in the perturbative regime. After that, we start to investigate how we can recover the momentum dependence doing some trivial cases and trying to generalize for high multiplicity. In the end,  we review some general results in the literature about this regime so we can start the exploration of the Higgsplosion proposal.

\subsection{Discussion About the Factorial Growth}

Usually, the series that we deal with in Quantum Mechanics and Quantum Field Theory are divergent. This divergence is not a problem because we know how to deal with this kind of series with different summation machines like Borel or Pad\'e~\cite{bender1,bender2}.  To understand this other kind of divergence in the amplitude, we need to understand why the series is divergent in the first place.
The analysis is done for $\lambda x^{4}$ in Quantum Mechanics, but we can generalize to the Field Theory case. We can see that there is a connection between the graphs with $N$ vertices with the coefficients of a given series:
\begin{align}
\sum_{\text{graphs}} \text{all graphs with $N$ verticies} = a_{N}  \, .
\end{align}
Now, if we have $N$ vertices in $x^{4}$ theory the structure would be something as is represented in Figure~(\ref{xx});

\begin{figure}[h!]
\centering
\includegraphics[width=10cm]{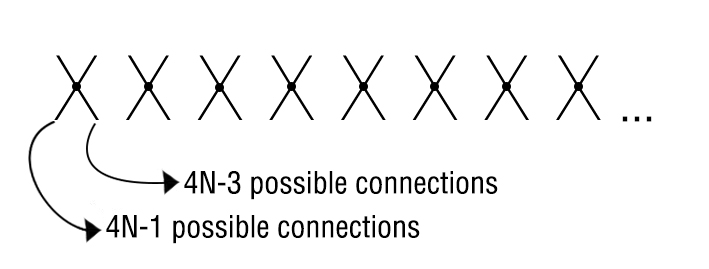}
\caption{Combinatorics for the $\phi^{4}$ interaction.}
\label{xx}
\end{figure}
We have to connect all these vertices. In a usual process, we have to connect some external lines, but they are only a finite amount so we can ignore in the large-$N$ limit. If we pick one of the lines, we have $(4N-1)$ possible connections, the next line we have $(4N-3)$. Going all the way we have:
\begin{align}
a_{N} \sim (4N-1)!!  \quad \text{as} \quad N \rightarrow \infty \, . 
\end{align}
In this case, we are over counting. There are different arrangements for the vertices that give the same result, $N!$. We still are overcounting because the lines on a vertex are identical, giving a factor of $4!$. In the end, we can write the coefficient of the series as:
\begin{align}
a_{N} \sim \frac{(4N-1)!!}{N! (24)^{N}}  \quad \text{as} \quad N \rightarrow \infty \, .
\end{align}

It could be that some of these graphs are disconnected and do not contribute to physical processes. In the counting of the graphs, the chance that this happens is small so we do not consider this. In the large-$N$ limit, it does not make a difference. Now, in this limit the coefficient of the series behaves like:
\begin{align}
a_{N} \sim N! C^{N} \quad \text{as} \quad N \rightarrow \infty  \, .
\end{align}
The factorial growth of the series indicates the divergent nature of it. This is just a heuristic argument, but it captures the spirit of what is happening. We could get some integrals that give a small contribution and change this picture. The divergence of the series occurs because, at small $N$, the small coefficients dominate. However, when going to large-$N$, the number of diagrams is so large that no matter how small the coefficient is, it will blow up. The behavior of a typical divergent series can be seen in the Figure~\ref{divergent}.

\begin{figure}[h!]
\centering
\includegraphics[width=15cm]{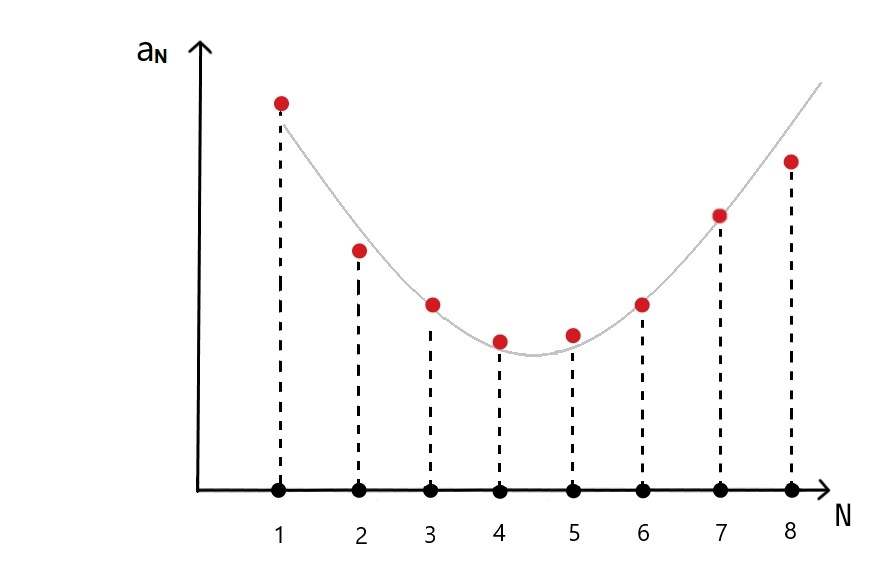}
\caption{Typical Behavior of a divergent series. The optimal truncation of a series like this would be when it hits the lowest point because of the definition of an asymptotic series.}
\label{divergent}
\end{figure}

Now, we can see why these amplitudes are growing factorially. We have too many diagrams for large multiplicity amplitudes already at tree level as it is shown in  Figure~\ref{fig:1to5} and  Figure~\ref{fig:1to7}. There is no small contribution that wins in the large-$n$ limit. The picture is significantly different in this case because there is never a region where the small coefficient dominates. This means that we still can extract information from the series, but any partial sum will give a bad approximation for the function. If we use summation machines to the loop corrections, we get the real behavior of these amplitudes. In this framework, we cannot trust the perturbative expression even at the tree level, but they have information about the actual function. In high multiplicity processes not only $\lambda$ matters but $n$ as well. This argument does not hold for different kinds of approximations like semiclassical calculation, but in those cases, it should be possible to do a similar analysis so we can understand the limitations of the results.

\section{Investigation of Beyond Threshold Amplitude}

So far, we only calculated threshold amplitudes. With this information only, we cannot reconstruct the decay rate because at the threshold the phase space is just a point. Nevertheless, this result is exact in 0 spatial dimensions, being the usual $x^{4}$ theory. If we want to see if the decay rate has an exponential growth at high multiplicity, we need to analyze if the phase space contribution has enough strength to combat the factorial growth of Eq.~(\ref{ampun}) and Eq.~(\ref{ampbro}). Going beyond the threshold at high multiplicity is an incredibly difficult task. There are too many momenta in the final state. To surpass this problem, we can try to construct these amplitudes in the near threshold limit where all final particles are non-relativistic. In this Section, we try to generalize Brown's method for beyond threshold amplitudes and discuss the difficulties of doing so. After that, we use Feynman rules for small multiplicity to try to understand what we should expect in the near threshold limit. Finally, in the end, we comment about some results in the literature of beyond threshold amplitude and what we can take from them.

\subsection{Naive Generalization of Brown's Methods and its Problems}

We saw that the Brown method~\cite{Brown-nov-92} of using the expectation value of the field with respect to a source works well for the threshold amplitude. If we want to generalize this, a first naive attempt is trying a source of the type:
\begin{align} \label{salsa2}
\rho(x_{\mu}) = \rho_{0} e^{ik_{\mu}x^{\mu}}  \, .
\end{align}
In the limit where $\vec{k} \rightarrow 0 $ we recover the solution of before. If we do this in the tree level solution, we need to solve the equation:
\begin{align}
(\Box + m^{2}) \phi_{0} + \frac{\lambda}{3!} \phi_{0}^{3} = \rho_{0} e^{ik_{\mu}x^{\mu}} \, ,
\end{align}
it turns out that we can solve this equation in the same way as the threshold case. The only difference is that the mass shell condition will be for the four-momentum. The solution is similar, because:
\begin{align}
\Box e^{n i k_{\mu}x^{\mu}} = -n^{2} k^{2} e^{n i k_{\mu} x^{\mu}}  \, .
\end{align}
Then, in the on-shell limit we get the same result as before:
\begin{align}
\phi_{0}(x)= \frac{z(x_{\mu})}{\left( 1-\frac{\lambda}{48m^{2}}z(x_{\mu})^{2}\right)}  \, .
\end{align}
were $z(x_{\mu})$ is a similar expression compared to Eq.~(\ref{zdete}):
\begin{align}
z(x_{\mu})= z_{0}e^{i k_{\mu} x^{\mu}}  \, ,
\end{align}
\begin{align}
z_{0} = -\frac{\rho_{0}}{k^{2}-m^{2}}  \, .
\end{align}

This is strange, because in the LSZ reduction formula we get something like:
\begin{align}
\fdv{\phi_{0}[\rho]}{\rho(x_{i})}= \frac{1}{p^{2}_{i}-m^{2}} \delta^{4}(x-x_{i}) \pdv{\phi_{0}}{z}  \, .
\end{align}
The problem is that, in the end, we set $\rho$ to zero and all the momentum dependence vanishes, obtaining a threshold result again. Even though this is a solution of the equation of motion with a space-time dependent source, this is not enough to find beyond threshold amplitudes. It seems that this source, Eq.~(\ref{salsa2}) can only excite one frequency mode, not having enough information about the field to construct beyond threshold amplitudes:
\begin{align}
\int \dd[4]{x} \phi_{0}(x)\rho(x) \propto \phi_{0}(k)  \, .
\end{align}

The alternative to this is to find a more comprehensive source that we can solve that has the right threshold limit and has at least some non-relativistic generalization. It turns out that this is a hard task because of the non-linearity and for now we cannot advance further. Before proceeding to the next part, it is worth pointing out one thing. The limit of $z=0$ in the LSZ reduction formula may appear strange and in this case, be responsible for vanishing the momentum dependence. If we do this computation with caution, we see that in the double limit of $k^{2} \rightarrow m^{2}$ and $\rho \rightarrow 0$ we are left with a constant $z_{0}$ term, so it is possible to try out instead $z=z_{0}e^{ik_{\mu}x^{\mu}}$. However, if we do all the work before going to mass shell the expression for $z(x)$ is finite, and we can take the limit for the source going to zero there, taking $z_{0}$ to zero first. The order of these limits is essential, and in the LSZ reduction formula, the $\rho$ going to zero is the first one, so this should not alter the results.

\subsection{Tree Level Investigation of Beyond Threshold ($1 \to 3$)}
If we want to recover the momenta dependence of a high multiplicity amplitude, it is worth to calculate simpler cases. The region of interest is with all external particles in a non-relativistic limit. Here we work the first non-trivial case of 1 particle going to 3 with non-relativistic momenta. This case is useful to check the threshold computation done in the previous Section, using Feynman diagrams, and see how is the shape of the first momentum correction. The process that we are interested in is represented in Figure~\ref{1t3}.

\begin{figure}[h!]
\centering
\includegraphics[width=8cm]{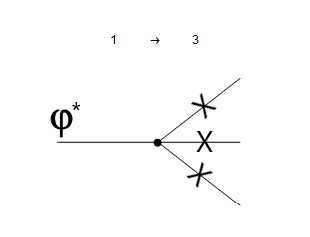}
\caption{Amputated amplitude that we are interested in is the $1 \to 3$ process. Generated with FeynArts~\cite{feyart}.}
\label{1t3}
\end{figure}
Here we have to remember that we do not remove the incoming leg, like we usually do. Then the amplitude is:
\begin{align} 
\mathcal{M}(1 \rightarrow 3) = (p^{2}-m^{2})\mathcal{A}(1 \rightarrow 3)  \, .
\end{align}
Feynman diagrams construct $\mathcal{M}(1 \rightarrow 3)$ and $p_{\mu}$ is the momentum of the incoming particle. Using the usual Feynman rule for the $\phi^{4}$ theory as is represented in Figure~\ref{fe}

\begin{figure}[h!]
\centering
\includegraphics[width=8cm]{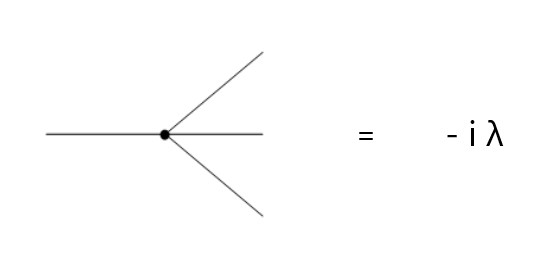}
\caption{Feynman rule for the $\phi^{4}$ case in the normalization that we are using. Generated with FeynArts~\cite{feyart}.}
\label{fe}
\end{figure}
We get the amplitude as:
\begin{align} \label{ampamp}
\mathcal{A}(1 \rightarrow 3) = \frac{\lambda}{p^{2}-m^{2}}  \, .
\end{align}

To investigate the non-relativistic limit of this expression we need to expand $p^{2}$ in the appropriate manner. Calling the outgoing momenta as $q_{i}$, in the non-relativistic limit we have $\abs{\vec{q}_{i}} \ll m $. The definition of the incoming momentum is:
\begin{align}
p = q_{1}+q_{2}+q_{3}  \, ,
\end{align}
where we write each external momentum as:
\begin{align}
q_{i} = (\gamma_{i} m, \vec{q}_{i})  \, .
\end{align}

In the non-relativistic limit we expand the first component as:
\begin{align}
q_{i} = (m + \frac{\vec{q}_{i}^{\hspace{0.1cm}2}}{2m} - \frac{\vec{q}_{i}^{\hspace{0.1cm}4}}{8m^{3}} + \dots, \vec{q}_{i})  \, ,
\end{align}
this means that the denominator of Eq.~(\ref{ampamp}) can be written as:
\begin{align}
(q_{1}+q_{2}+q_{3})^{2}-m^{2}= 2 \left( m^{2} + q_{1}\cdot q_{2} + q_{1}\cdot q_{3}+ q_{2}\cdot q_{3} \right)=
\end{align}
\begin{align} \nonumber
2 \Bigg( 4m^{2} + \vec{q}^{\hspace{0.1cm}2}_{1} +\vec{q}^{\hspace{0.1cm}2}_{2} +\vec{q}^{\hspace{0.1cm}2}_{3} - \vec{q}_{1}\vec{q}_{2} - \vec{q}_{1}\vec{q}_{3} - \vec{q}_{2}\vec{q}_{3} +
\end{align}
\begin{align} \nonumber
+\frac{1}{4m^{2}} ( \vec{q}^{\hspace{0.1cm}2}_{1}\vec{q}^{\hspace{0.1cm}2}_{2} +\vec{q}^{\hspace{0.1cm}2}_{1}\vec{q}^{\hspace{0.1cm}2}_{3}+\vec{q}^{\hspace{0.1cm}2}_{2}\vec{q}^{\hspace{0.1cm}2}_{3}-\vec{q}^{\hspace{0.1cm}4}_{1}-\vec{q}^{\hspace{0.1cm}4}_{2}-\vec{q}^{\hspace{0.1cm}4}_{3}) + \dots  \Bigg)  \, .
\end{align}
The interesting feature of this expression that remains in the high multiplicity cases is that at leading order we can write it in terms of the non-relativistic energy of the outgoing particles in a general frame:
\begin{align} \label{nonenergy}
E = \frac{1}{2m} \sum \vec{q}_{i}^{\hspace{0.1cm}2}  -\frac{1}{2mn} \left( \sum \vec{q}_{i} \right)^{2} = \frac{n-1}{2mn} \sum \vec{q}_{i}^{\hspace{0.1cm}2} - \frac{1}{nm} \sum_{i \neq j} \vec{q}_{i}\vec{q}_{j}  \, .
\end{align}
Using Eq.~(\ref{nonenergy}) the denominator of Eq.~(\ref{ampamp}) can be written as:
\begin{align} \label{caa}
(q_{1}+q_{2}+q_{3})^{2}-m^{2}= 8m^{2}+6mE +\frac{1}{2m^{2}} ( \vec{q}^{\hspace{0.1cm}2}_{1}\vec{q}^{\hspace{0.1cm}2}_{2} +\vec{q}^{\hspace{0.1cm}2}_{1}\vec{q}^{\hspace{0.1cm}2}_{3}+\vec{q}^{\hspace{0.1cm}2}_{2}\vec{q}^{\hspace{0.1cm}2}_{3}-\vec{q}^{\hspace{0.1cm}4}_{1}-\vec{q}^{\hspace{0.1cm}4}_{2}-\vec{q}^{\hspace{0.1cm}4}_{3}) + \dots   \, .
\end{align}

The first important thing to notice is that in the next order we cannot write the denominator in terms of only $E$. It is easy to see this because $E^2$ would contain odd powers of momenta and this does not appear in Eq.~(\ref{caa}). This feature remains when we increase the number of external particles. We can see that in the large $n$ limit this problem vanishes, leaving only even powers of momentum in Eq.~(\ref{nonenergy}). If we write the amplitude in a expansion of small spatial momenta (small E) we get:
\begin{align} \label{ampi}
\mathcal{A}(1 \rightarrow 3) =  \frac{\lambda}{8m^{2}} \Bigg( 1-\frac{3}{4} \frac{E}{m} + \frac{9}{8} \frac{E^{2}}{m^{2}} - 
\end{align}
\begin{align} \nonumber
-\frac{1}{8m^{4}}( -\vec{q}^{\hspace{0.1cm}3}_{1}\vec{q}_{2}-\vec{q}^{\hspace{0.1cm}3}_{1}\vec{q}_{3}-\vec{q}^{\hspace{0.1cm}3}_{3}\vec{q}_{1}-\vec{q}^{\hspace{0.1cm}3}_{2}\vec{q}_{1}-\vec{q}^{\hspace{0.1cm}3}_{2}\vec{q}_{3}-\vec{q}^{\hspace{0.1cm}3}_{3}\vec{q}_{2} + 2\vec{q}^{\hspace{0.1cm}2}_{1}\vec{q}^{\hspace{0.1cm}2}_{2}+2\vec{q}^{\hspace{0.1cm}2}_{1}\vec{q}^{\hspace{0.1cm}2}_{3}+2\vec{q}^{\hspace{0.1cm}2}_{2}\vec{q}^{\hspace{0.1cm}2}_{3} )+ \dots  \Bigg)   \, .
\end{align}
From this result, we can see that for small momenta the first term depends only on the non-relativistic energy. If we go to high orders in the momentum expansion, it is expected that we cannot describe the system only with $E$ and other functions will have an important play. From Eq.~(\ref{nonenergy}) we can see that these other objects are subdominant, and only the non-relativistic energy dominates. However, even with this expression, we cannot say much about the decay rate. If we are working in a limit where the kinetic energy is small, we are in a shell of the phase space as represented in Figure~\ref{deca}

\begin{figure}[h!]
\centering
\includegraphics[width=10cm]{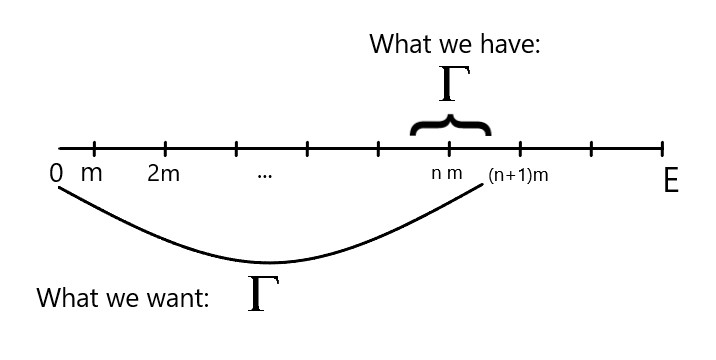}
\caption{Difference between the decay rate that we can get using a non-relativistic approximation to what it would be useful to have.}
\label{deca}
\end{figure}

This means that the decay rate can only be calculated for this shell that is smaller than  $2m$. For a good approximation of the decay rate, we expect to be able to calculate in a broader region. The limit of validity of this expression does not help us much in understanding the behavior of the decay rate, but it is a step in this direction. Next, we do the same calculation for the $1 \to 5$ case where we start to see a trend in behavior. One thing that we can take from this is that the threshold computation works, being the first term of Eq.~(\ref{amp1}).

\subsection{Tree Level Investigation of Beyond Threshold ($1\rightarrow 5$) and its Generalities}

Let us go one step beyond and analyze the process of $1\rightarrow 5$ particles in the same non-relativistic limit. Now the problem starts to appear because we have ten different combinations of external leg positions for the Feynman diagram, as is shown in Figure~\ref{1t5}. These are all the possible combinations that the internal propagator can have, considering the particles are identical.

\begin{figure}[h!]
\centering
\includegraphics[width=15cm]{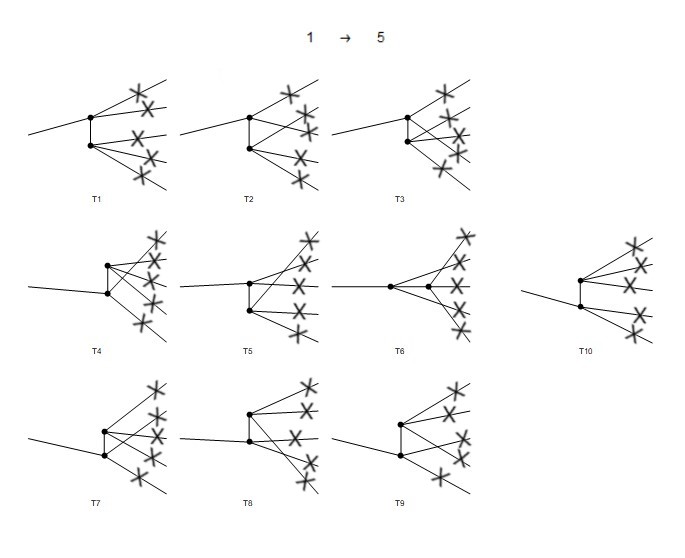}
\caption{Amputated amplitude that we are interested in for the $1\rightarrow 5$ process. Generated with FeynArts~\cite{feyart}.}
\label{1t5}
\end{figure}
 
 The amplitude can be written as:
\begin{align} \label{15}
\mathcal{A}(1 \rightarrow 5) =  \frac{1}{p^{2}-m^{2}}  \sum_{ijk} \frac{1}{(q_{i}+q_{j}+q_{k})^{2}-m^{2}} \lambda^{2}   \, ,
\end{align}
where $i,j,k$ are the sum of the following combinations written in the Table~\ref{tab:1}.
\begin{table}[h!]
\centering

\begin{tabular}{|l|l|l|}
\hline
i & j & k \\ \hline
1 & 2 & 3 \\ \hline
1 & 2 & 4 \\ \hline
1 & 2 & 5 \\ \hline
1 & 3 & 4 \\ \hline
1 & 3 & 5 \\ \hline
1 & 4 & 5 \\ \hline
2 & 3 & 4 \\ \hline
2 & 3 & 5 \\ \hline
2 & 4 & 5 \\ \hline
3 & 4 & 5 \\ \hline
\end{tabular}
\caption{Possible combinations for the sum in i, j and k.}
\label{tab:1}
\end{table}

The problem now is to do the non-relativistic limit, because we have two propagators.  One of them with all the momenta and other only with three at a time. Doing the non-relativistic limit just like before we can write the amplitude as:
\begin{align} \label{abosa}
\mathcal{A}(1 \rightarrow 5) = \left( \frac{\lambda^{2}}{2}\right)^{2} \frac{1}{\Delta_{1}+\Delta_{2} t^{2} + \Delta_{3}t^{4}} \sum_{ijk} \frac{1}{\Delta_{4}+\Delta_{5}^{ijk}t^{2} -\frac{1}{4m^{2}}\Delta_{6}^{ijk} t^{4}} \, .
\end{align}
In this expression $t$ is just a fictitious parameter to help expand this in small spatial momentum, it keeps track of the power of $\left|\vec{q}_{i}\right|$. In the end we expand Eq.~(\ref{abosa}) in $t$ and set $t=1$ to get the non-relativistic approximation. These deltas appears from doing the expansion on the denominators of Eq.~(\ref{15}), up to quartic order:
\begin{align}
\Delta_{1} =  12 m^{2} \, ,
\end{align} 
\begin{align}
\Delta_{2} =  2 ( \vec{q}_{1}^{\hspace{0.1cm}2}+\vec{q}_{2}^{\hspace{0.1cm}2}+\vec{q}_{3}^{\hspace{0.1cm}2}+\vec{q}_{4}^{\hspace{0.1cm}2}+\vec{q}_{5}^{\hspace{0.1cm}2}) - \vec{q}_{1}\vec{q}_{2} - \vec{q}_{1}\vec{q}_{3} - \vec{q}_{1}\vec{q}_{4} -
\end{align} 
\begin{align} \nonumber
 -\vec{q}_{1}\vec{q}_{5} - \vec{q}_{2}\vec{q}_{3} - \vec{q}_{2}\vec{q}_{4} - \vec{q}_{2}\vec{q}_{5} - \vec{q}_{3}\vec{q}_{4} - \vec{q}_{3}\vec{q}_{5} - \vec{q}_{4}\vec{q}_{5} \, ,
\end{align}
\begin{align}
\Delta_{3} = -\frac{1}{2m^{2}} ( \vec{q}_{1}^{\hspace{0.1cm}4}+\vec{q}_{2}^{\hspace{0.1cm}4}+\vec{q}_{3}^{\hspace{0.1cm}4}+\vec{q}_{4}^{\hspace{0.1cm}4}+\vec{q}_{5}^{\hspace{0.1cm}4})+\frac{1}{4m^{2}} ( \vec{q}_{1}^{\hspace{0.1cm}2}\vec{q}_{2}^{\hspace{0.1cm}2} +\vec{q}_{1}^{\hspace{0.1cm}2}\vec{q}_{3}^{\hspace{0.1cm}2}+\vec{q}_{1}^{\hspace{0.1cm}2}\vec{q}_{4}^{\hspace{0.1cm}2}+\vec{q}_{1}^{\hspace{0.1cm}2}\vec{q}_{5}^{\hspace{0.1cm}2}+
\end{align} 
\begin{align} \nonumber
+\vec{q}_{2}^{\hspace{0.1cm}2}\vec{q}_{3}^{\hspace{0.1cm}2}+\vec{q}_{2}^{\hspace{0.1cm}2}\vec{q}_{4}^{\hspace{0.1cm}2}+\vec{q}_{2}^{\hspace{0.1cm}2}\vec{q}_{5}^{\hspace{0.1cm}2}+\vec{q}_{3}^{\hspace{0.1cm}2}\vec{q}_{4}^{\hspace{0.1cm}2}+\vec{q}_{3}^{\hspace{0.1cm}2}\vec{q}_{5}^{\hspace{0.1cm}2}+\vec{q}_{4}^{\hspace{0.1cm}2}\vec{q}_{5}^{\hspace{0.1cm}2}) \, ,
\end{align}
\begin{align}
\Delta_{4} =4m^{2} \, ,
\end{align}
\begin{align}
\Delta_{5}^{ijk} = \vec{q}_{i}^{\hspace{0.1cm}2}+\vec{q}_{j}^{\hspace{0.1cm}2}+\vec{q}_{k}^{\hspace{0.1cm}2} -\vec{q}_{i}\vec{q}_{j}-\vec{q}_{i}\vec{q}_{k}-\vec{q}_{j}\vec{q}_{k} \, ,
\end{align}
\begin{align}
\Delta_{6}^{ijk} = \vec{q}_{i}^{\hspace{0.1cm}4} + \vec{q}_{j}^{\hspace{0.1cm}4} + \vec{q}_{k}^{\hspace{0.1cm}4} - \vec{q}_{i}^{\hspace{0.1cm}2}\vec{q}_{j}^{\hspace{0.1cm}2} - \vec{q}_{i}^{\hspace{0.1cm}2}\vec{q}_{k}^{\hspace{0.1cm}2} -\vec{q}_{j}^{\hspace{0.1cm}2}\vec{q}_{k}^{\hspace{0.1cm}2} \, .
\end{align}

With this information we can do the non-relativistic expansion of the amplitude:
\begin{align}
\mathcal{A}(1 \rightarrow 5) = \lambda^{2} \sum_{ijk} \left( \frac{1}{192m^{4}} - \frac{1}{2304m^{6}}(\Delta_{2}+3\Delta^{ijk}_{5}) + \dots \right) \, .
\end{align}
The first term gives the right threshold amplitude, the second term we will re-write in terms of the non-relativistic energy. Using the definition of the non-relativistic energy, Eq.~(\ref{nonenergy}) it is a direct but boring computation to show:
\begin{align}
\sum_{ijk} \Delta_{2} + 3\Delta_{5}^{ijk} = 95mE \, .
\end{align}
The amplitude up to first order is:
\begin{align}
\mathcal{A}(1 \rightarrow 5) = 5! \left( \frac{\lambda}{48m^{2}}\right)^{2}\left(1-\frac{19}{24} \frac{E}{m} + \dots \right) \, .
\end{align}

This result is in agreement with~\cite{Papadopoulos-nov-92,Libanov-jul-94}. This shows that indeed, the first correction beyond threshold comes only from the non-relativistic energy. Again, if we try to go to the next order, this ceases to be the case because of odd powers of momentum in the expression for the energy square that does not appear in the calculation. The next step in this investigation is to show and discuss some results coming from recursion relation and semiclassical computation. Mostly we will comment and try to interpret inside this framework. From these cases, we can take that the information beyond the threshold is tough to get.

\subsection{Recurrence Relations and General Results of Beyond Threshold Amplitudes at High Multiplicity}

The focus so far was to study these high multiplicity processes in the perturbative regime. We saw already why these amplitudes are growing factorially. The last piece of information on the perturbative regime comes from recursion relations. Here we highlight the history and main results of it. After that, we comment on some recent results coming from the semiclassical calculation. These results were essential to motivate the studies of high multiplicity processes after the initial wave of results that we covered so far.

The first improvement came in 1992 when E.N Argyres and Costa G. Papadopoulos used recurrence relations to find amplitudes with one momentum in the final state~\cite{Papadopoulos-nov-92}. The general form of the recursion relation is represented in Figure~\ref{rec}. They find the solution for general monomial interaction of the form:
\begin{align}
V(\phi) = \lambda_{m} \frac{\phi^{m}}{m!}
\end{align} 
\begin{figure}[h!]
\centering
\includegraphics[width=15cm]{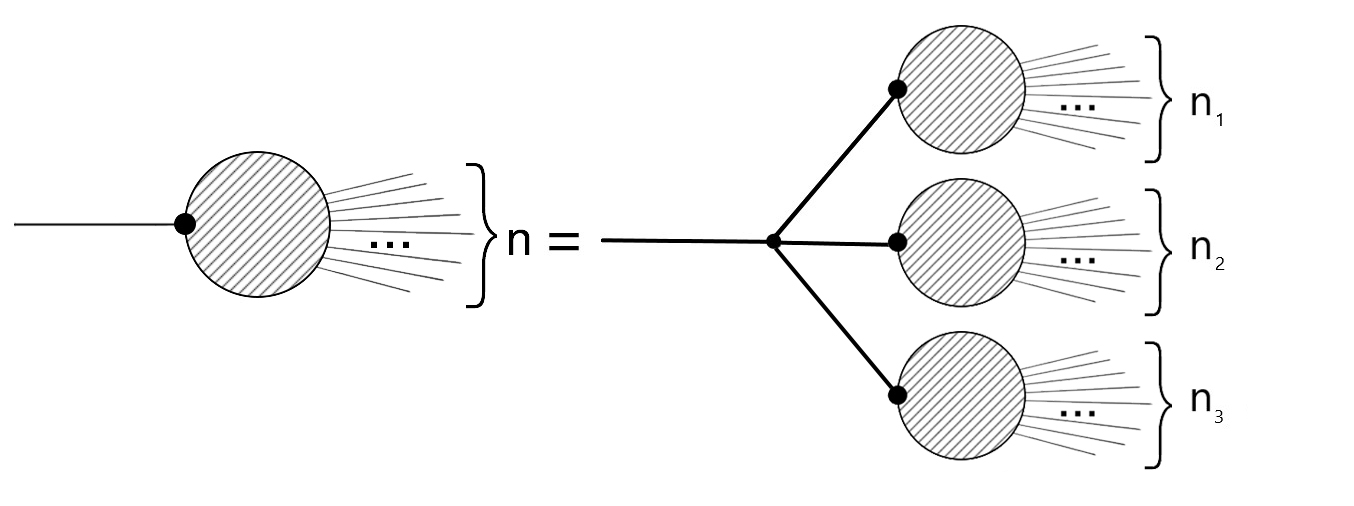}
\caption{Recursion relation for $\phi^{4}$ theory in the unbroken phase, $n=n_{1}+n_{2}+n_{3}$.}
\label{rec}
\end{figure}
The specialization for our $\phi^{4}$ case gave the result:
\begin{align}
\mathcal{A}(1 \rightarrow 2k+1) = (2k)!(2k)(2k+1+\omega) \left( \frac{\lambda_{4}}{48}\right)^{k} \left( k + 1 + \frac{2(3-\omega)}{3+\omega} k + \frac{(3-\omega)(1-\omega)}{(3+\omega)(5+\omega)}(k-1) \right)  \, ,
\end{align}
where the mass is set to one and $\omega=2E-1$ with $E$ being the energy of the final particle with nonzero spatial momentum.  The threshold result is recovered when $\omega=1$. We can find with these results $2 \rightarrow n$ processes, if we use a negative $\omega$, as it was shown to recover the threshold results:
\begin{align}
\mathcal{A}(2 \rightarrow n) = 0  \quad  n>4 \, .
\end{align}
In the same paper the authors calculate the case of broken phase with one particle in the final state with arbitrary momentum:
\begin{align}
\mathcal{A}^{B} (1 \rightarrow n ) = (n-1)! (n-1)(n+\omega) \left( \frac{\lambda_{4}}{12}\right)^{(n-1)/2} \left( n + 2(n-1)\frac{(1-\omega)}{(2+\omega)} - \frac{\omega (1-\omega)}{(2+\omega)(3+\omega)}\right) \, ,
\end{align}
and again the limit of $\omega =1$ checks out. These results seem to show that in the perturbative region, even momentum dependence cannot save the behavior of these amplitudes.

After these results, people realized that if we restrict ourselves to the quantum mechanical limit of this, we get the complete result. This motivated the study of transition amplitudes in the $x^{4}$ oscillator by C.A Diamantis, B.C Georgalas, A. B. Lahanas and E. Papantonopolus~\cite{anarmo}. They looked for bounds in the transition amplitudes generated by an external source. This would be a simplification of our case in the (1+0)D case. They introduced the concept of holy grail function $F(\lambda n)$ for the study of these amplitudes:
\begin{align}
\mathcal{A}(1 \rightarrow n) = \kappa e^{\frac{F(\lambda n)}{\lambda^{2}}} \, .
\end{align}
In~\cite{anarmo}, they showed that in the trusted region, these transitions never exponentially grow. However, this is not a definitive answer because we cannot go to all the parameter space. This result was significant to understand the unitarity of the quantum mechanical case, but better expressions for the amplitudes were needed.

The situation changed dramatically when M.V. Libanov, V.A. Rubanov, D. T. Son and S. V. Troitsky published two papers about the exponentiation of these amplitudes~\cite{Libanov-mar-95,Libanov-jul-94}. It was conjectured that the amplitudes at high multiplicity should have the special form:
\begin{align}
\mathcal{A}(1 \rightarrow n) \propto \sqrt{n!} e^{\frac{F(\lambda n,\epsilon)}{\lambda}} \, ,
\end{align}
inspired by the instantonic cross section. The behavior of the amplitude is completely determined by the holy grail function $F$. To show evidence of this conjecture they redid all the results so far in this formalism and wrote the corresponding holy grail function. For the $\frac{\lambda \phi^{4}}{4}$ case~\cite{Libanov-mar-95} at tree level and treashold:
\begin{align}
F_{unbroken} = \frac{\lambda n}{2} \ln(\frac{\lambda n}{8}) - \frac{\lambda n}{2} \, ,
\end{align}
\begin{align}
F_{broken} = \frac{\lambda n}{2} \ln(\frac{\lambda n}{2}) - \frac{\lambda n}{2} \, .
\end{align}
For the treshold amplitude at tree level of the O(2) model with interaction $\frac{\lambda (\phi_{1}^{2}+\phi_{2}^{2})^{2}}{4}$~\cite{Libanov-mar-95}:
\begin{align}
F_{unbroken} = \lambda n \left( \ln (\frac{\lambda n(\sqrt{m_{1}}+\sqrt{m_{2}})^{2}}{8(m_{1}+m_{2})}) -1 \right) \, ,
\end{align}
\begin{align}
F_{broken} = \lambda n \left( \ln (\frac{\lambda n(\sqrt{m_{1}}+\sqrt{2m_{2}})^{2}}{2(m_{1}+2m_{2})}) -1 \right) \, .
\end{align}
Then they introduced new results for the beyond threshold and loop corrections. They solved the recursion relations of Figure~\ref{rec} for the large $n$ limit using the fact that at leading order only the non-relativistic energy contribute. In this region they showed that for the unbroken $\frac{\lambda \phi^{4}}{4}$ case at tree level:
\begin{align}
\mathcal{A}(1 \rightarrow n) = n! \left( \frac{\lambda}{8} \right)^{(n-1)/2} e^{-\frac{5}{6}E} \, ,
\end{align}
where $m=1$ and $E$ is the non-relativistic kinetic energy of the final particles. Other important result was to show that at leading order the $n$-loop correction exponentiate to the form:
\begin{align}
\mathcal{A}(1 \rightarrow n) = \mathcal{A}_{tree}(1 \rightarrow n) e^{B\lambda n^{2}} \, .
\end{align}
In the second paper, they started to lay the ground for the generalization of the WKB method to produce semiclassical calculation in Quantum Field Theory. For the first time, we have a closed expression for the large multiplicity amplitudes with all the momentum dependence. All of these results were still perturbative, and they keep showing the same behavior.

The next year D.T Son~\cite{Son-may-95} showed how to generalize the WKB method to calculate semiclassical amplitudes in the regime:
\begin{align}
\lambda \rightarrow 0 \, , \quad n \rightarrow \infty  \, , \quad \lambda n = g = \text{cte} \, , \quad \epsilon = \text{cte} \, ,
\end{align}
where $\epsilon$ is the non-relativistic energy per particle per unit of mass in the final state:
\begin{align}
\epsilon = \frac{E-nm}{m}
\end{align}
This was one of the missing pieces to understand these processes. The question is now written in terms of the right coupling $g= \lambda n$.  That coupling has a t' Hooft like form similar to an large-N Yang-Mills expansion. Finding a expansion for small and large g, we can then see if these processes can grow exponentially. The amazing thing about this semiclassical calculation is that we can get the decay rate automatically, without the need to integrate the $n$ particle phase space. The semiclassical calculation only computes few $\rightarrow$ many processes, but it is expected that the exponential part is independent of the number of initial particles provided they are small. At the time only the limit $g \ll 1$ could be explored, it is not so useful at first glance:
\begin{align}\label{smallg}
F(g,\epsilon) = g \ln(\frac{g}{16}) - g + \frac{3g}{2}(\ln(\frac{\epsilon}{3\pi}) +1) - \frac{25}{12} g \epsilon + \frac{ \sqrt{3}}{4\pi} g^{2} \, .
\end{align}
This was still not enough because we need the expression in the limit of large $g$ to say anything sensible. It was, nevertheless, a big step in the right direction.

A decade passed and the new generation of results started to appear, first the generalization of recurrence relations for different particle content to approximate these results to the Standard Model by Valentin V. Khoze~\cite{Khoze-apr-14}. These results were still perturbative but showed the same behavior of the simpler scalar case. After that, the most important results for understanding high multiplicity processes came from Valetin V. Khoze again~\cite{Khoze-jun-18}. They used the D.T. Son method plus some new tricks to obtain a semiclassical amplitude in the right limit of $g \gg 1 $. This result only exists for $\phi^{4}$ in (1+3)D and broken phase but this is a revolutionary solution nonetheless:
\begin{align} \label{eq10}
\Gamma(\epsilon) \propto e^{n \left( \ln(\frac{g}{4}) +0.85\sqrt{g} -1 +  \frac{3}{2}(\ln(\frac{\epsilon}{3\pi}) +1) - \frac{25}{12}  \epsilon  \right)} \, .
\end{align} 
Eq.~(\ref{eq10}) is valid in the regime:
\begin{align}
\lambda \rightarrow 0 \, , \quad n \rightarrow \infty  \, ,\quad \lambda n = g \gg 1 , \quad  \epsilon \ll 1 \, .
\end{align}
This was the first result outside ordinary perturbation theory were the exponential growth appeared. The analysis of this result is done at the end of the next chapter.

Concluding, it is worth showing another impressive result coming from Joerg Jackel and Sebastian Schenk~\cite{Jaeckel-jun-18}. They did the full perturbative analysis for the quantum mechanical case. In there it is shown that the amplitude indeed exponentiate and after resuming the series there is no unitarity violation. This shows that in this case, we cannot trust the partial sum of the initial terms of the perturbative series, as we expected.

\chapter{Higgsplosion and Higgspersion}  \label{c3}

\pagenumbering{arabic}

\section{The Rise of the Higgsplosion}

\subsection{The Higgsplosion Mechanism}

It is know that in a Scalar Field Theory a high multiplicity amplitude could violate perturbative unitarity, making the theory inconsistent in this limit~\cite{Arkhipov-nov-82,Goldberg-may-90}. We did this computation and confirmed these results~\cite{Brown-nov-92,Voloshin-feb-92,loop1,Smith-apr-93}. In a concrete example the off-shell amplitude, like is shown in Figure~\ref{offshell}, for a $n$-particle decay grows factorially with $n$:
\begin{align}
\mathcal{A}(1 \rightarrow n ) \propto \lambda^{n} n! \, ,
\end{align}
such that the decay rate behave like:
\begin{align}
\Gamma(1 \rightarrow n) \propto \lambda^{n} n! V_{n}(E) \, .
\end{align}

\begin{figure}[h!]
\centering
\includegraphics[width=10cm]{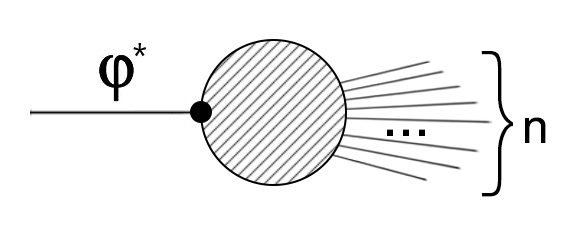}
\caption{Off-shell $1 \to n$ particles process.}
\label{offshell}
\end{figure}

Even after we remove one factor of $n!$ from the expression because the particles at the end are identical, the decay rate grows factorially. At the time that they found this result, they interpreted it as a sign that the perturbation theory became effectively strong coupled when $n > 1/\lambda $. The picture changed in 2017~\cite{Khoze-higgsplosion} when Valentin V. Khoze and Michael Spannowsky proposed that such behavior renders the cross-section of physical processes unitary, and even generate additional effects inside the theory. The central point of the proposal is to look for the full propagator of a virtual scalar in some intermediary process as is depicted in Figure~\ref{sper}. 

\begin{figure}[h!]
\centering
\includegraphics[width=10cm]{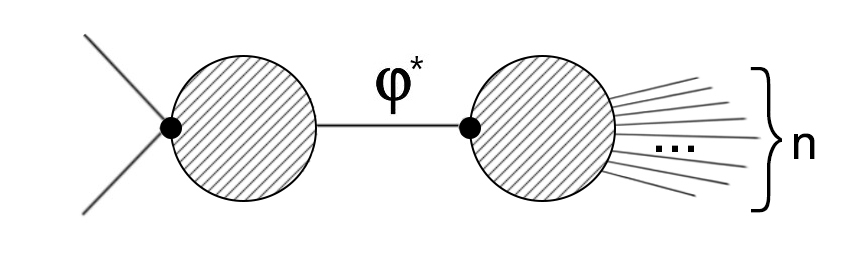}
\caption{Diagrammatic representation of the Higgspersion. The propagator between processes is the full propagator of the field $\phi$.}
\label{sper}
\end{figure}
In the Standard Model, this role would be played by the Higgs. The full propagator can be written in the form:
\begin{align}
\Delta_{H}(p) = \frac{i}{p^{2} - M^{2}(p^{2}) + i \frac{m}{Z_{\phi}}\Gamma(p^{2})} \, ,
\end{align}
where $M^{2}(p^{2})$ has the contribution from the bare mass and the real part of the 1PI function Eq.~(\ref{1pid}). The results of a factorially growing decay rate would mean that the propagator became strongly suppressed by $p^{2}$ factors. This blowing up of the decay rate is called Higgsplosion and the propagator suppression Higgspersion. This mechanism would save unitarity because, in an intermediary process, the suppression wins at large $p^{2}$ in a physical process.
\begin{align}
\sigma_{n} \propto \sqrt{s} \frac{\Gamma_{n}(s)}{s^{2} +(\frac{m}{Z_{\phi}})^{2}\Gamma^{2}(s)} \, .
\end{align}
Where $\Gamma(s)$ is the total offshell decay width for the scalar field. This remains finite, even when $\Gamma_{n} \rightarrow \infty$. The decay rate going to infinity is not problematic because this is not a real process, even though we can use it to construct real ones. The major consequence of this effect is the suppression of loops at high energy. In any process involving loops, we can trade the free propagator by the full propagator if we go to high enough orders as is exemplified in Figure~\ref{rego}.

\begin{figure}[h!]
\centering
\includegraphics[width=10cm]{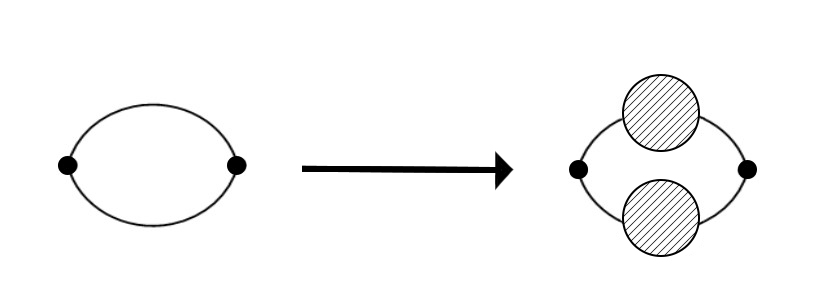}
\caption{Diagrammatic representation of changing the free propagator for the full propagator inside processes. The hashed circle represents the full propagator of the field $\phi$.}
\label{rego}
\end{figure}

 When considering the full propagator, the integral in the momentum will shutoff at the Higgsploding scale because of the exponential suppression of the propagator. This scale $E^{*}$ is a new dynamical scale of the theory that appears in the energy scale where the decay rate grows exponentially. The strong claims that are made because of this proposal are that the killing of the loops makes the couplings stop running. This means that the theory becomes UV finite and we are at a nontrivial fixed point at this scale~\cite{wilson}. There is strong evidence that this can happen at least in $\phi^{4}$ theory in the broken phase where we have semiclassical computations in a region where we can trust the decay rate, and it has the characteristic behavior~\cite{Khoze-jun-18}:
\begin{align} \label{eqimpor}
\Gamma(\epsilon) \propto e^{n \left( \ln(\frac{g}{4}) +0.85\sqrt{g} -1 +  \frac{3}{2}(\ln(\frac{\epsilon}{3\pi}) +1) - \frac{25}{12}  \epsilon  \right)} \, ,
\end{align} 
this expression is calculated in the limit $\lambda \rightarrow 0 $, $n \rightarrow \infty$, $\lambda n =g \gg 1$ and $\epsilon \rightarrow 0$. 

 In the first part of this thesis, we redid the perturbative computations of these amplitudes at and beyond threshold and commented some results coming from different methods. Now the focus is to understand what are the consequences if Higgsplosion is true. If the Higgsplosion does not occur, we need to understand why. To study this, we need to review what are the significant steps for the Higgsplosion to work and then analyze each step to see if there are any flaws or inconsistencies. 
  
The central player of the Higgsplosion mechanism is the exponential growth of the $n$-particle decay rate at some scale. As we argued, it is difficult to obtain these expressions using ordinary perturbation theory because we are stuck near the threshold of some point in energy space. The picture changed when semiclassical computation confirmed the results coming from perturbation theory, that the decay rate could indeed grow. This is by no means the end of the story as better decay rates expressions need to be found, and we do not fully understand what is the parameter region exactly where Eq.~(\ref{eqimpor}) can be trusted. If the Higgsploding mechanism exists in a Scalar Field Theory, we need to be able to compute the decay rate at high $p^{2}$ to see it. The next essential ingredient is the Optical Theorem and the Dyson resummed propagator. The Optical Theorem holds when we have a unitary theory~\cite{qft} and, as we argued in chapter~\ref{c1}, the propagator is valid even at the non-perturbative level. These are the principal players of this mechanism. If we want to know if this happens, we need to dig deeper into these points to see what we can extract from it.  It is very intriguing how simple steps can potentially generate a powerful new phenomenon in a Quantum Field Theory that can be accessible in principle perturbatively. 

The concept of Higgsplosion was presented and by itself it is straightforward to understand, now we turn our attention to the underlying assumptions to see if they are consistent in such a way that we can have this mechanism. To do this we review and introduce some potential consequences of Higgsplosion. After that, we review some of the modern criticism made in two papers~\cite{Belyaev-aug-18,Monin-aug-18} and discuss some results from the lattice~\cite{lattice1,lattice2}. Having done that, we will try to estimate $E^{*}$ for the case where we have the decay rate Eq.~(\ref{eqimpor}). Finally, we try to understand the role of the perturbation expansion, its validity and work out a toy model where the propagator decays exponentially.

\subsection{The Potential Power of the Higgsplosion}

The Higgsplosion mechanism could generate an interesting phenomena even when we have only scalar fields. We saw in the computation of the one-loop amplitude in the broken Eq.~(\ref{ampbro}) and unbroken Eq.~(\ref{ampun}) phase that we had to deal with divergences, for the unbroken phase:
\begin{align}
\delta m^{2}_{1} =- \frac{\lambda_{R}\mathcal{I}_{1}}{4} \, ,
\end{align}
\begin{align}
\frac{\delta\lambda_{1}}{3!} =\frac{\lambda_{R}^{2}}{16} \left( \mathcal{I}_{2}+\frac{1}{2\pi^{2}} \right) \, ,
\end{align}
and for the broken phase:
\begin{align}
\delta m_{1}^{2} = \frac{\lambda_{R}}{8}\left[ \mathcal{I}_{1} - M_{R}^{2}\left( \frac{3 \mathcal{I}_{2}}{4} + \frac{3}{8\pi^{2}} -\frac{\sqrt{3}}{16\pi}\right) \right] \, ,
\end{align}
\begin{align}
\frac{\delta \lambda_{1}}{3!} = \frac{\lambda_{R}^{2}}{48} \left[ \frac{3 \mathcal{I}_{2}}{2} +\frac{3}{4\pi^{2}} - \frac{\sqrt{3}}{16\pi} \right] \, .
\end{align}
These divergences were appearing because in the loop integral we integrated to arbitrary high momentum in the propagator. This changes when we consider the Higgsplosion mechanism. The full propagator can substitute the propagators inside the loop if we reorganize order by order the expansion like it is represented in the Figure~\ref{rego}. Making this substitution in the theory, it suddenly becomes clear that we should not integrate into all momenta. The Higgsplosion shuts down the propagator above the scale $E^{*}$, so we integrate up to a sphere of radius $E^{*}$. This means that the theory is finite, and we have a natural scale to measure our observables. Doing the running of the coupling, we see that it shut down when we hit the Higgsploding scale, entering the new phase in the theory as it is represented in Figure~\ref{fazer}.

\begin{figure}[h!]
\centering
\includegraphics[width=10cm]{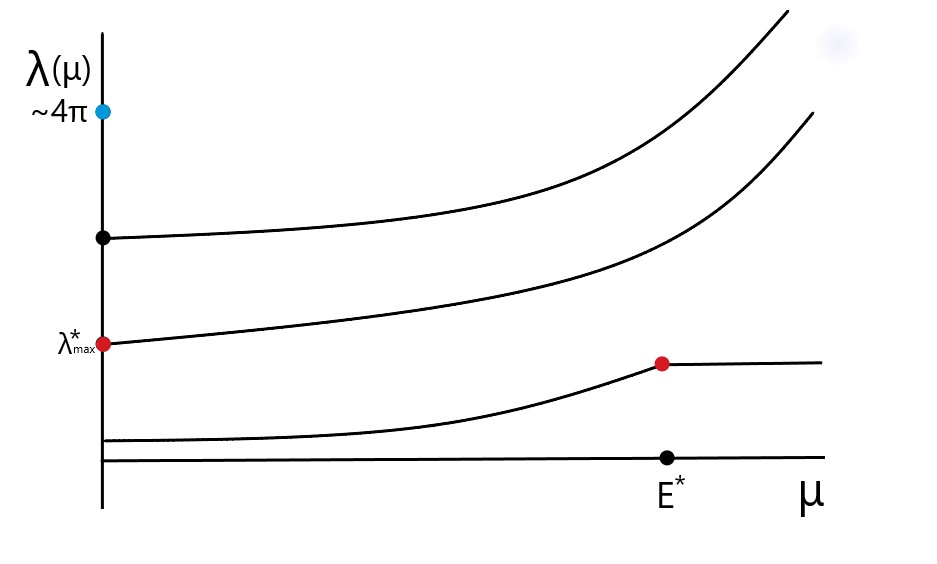}
\caption{Representation of the coupling flow in a $\phi^{4}$ theory with Higgsplosion. The $\lambda^{*}_{max}$ is the largest coupling for which we can trust the expressions indicating the Higgsplosion. It could be that even outside this region the theory Higgsplodes at some point, then it would have a barrier where all couplings go to constant independently of where it started.}
\label{fazer}
\end{figure}

 In $\phi^{4}$ the flow makes the coupling grow, this means that if we do not start at an ultra perturbative value, we will quickly move away from it. The expressions for the decay rate can only be trusted in this limit. If we start at a given $\lambda$ that is too big we do not know if it will Higgsplode. This new phase depends on the value of the coupling to be achieved. Passing through this value, we need to find an $n$ and $\epsilon$ that can generate the growing decay rate of Eq.~(\ref{eqimpor}). This means that for each coupling value exists a Higgsploding scale .
 
  The story could potentially be different if the Higgs of the Standard Model Higgsplodes, the coupling there is decreasing, signaling a potential vacuum decay. The theory, in principle, would always flow to the perturbative regime where we can trust the semiclassical calculation and would always Higgsplode provided it does not vacuum decay first. This is just a schematic picture, and further calculations are needed to understand the full flow of the coupling. The same behavior occurs for the mass, meaning that in a Higgsplosion mechanism the mass is of order $E^{*}$, not of the heavier particle that exists on the theory. This is shielding the scalar from receiving quantum correction above this scale and in principle could be used in the Standard Model to render the Higgs mass natural. The nature of this shielding needs to be understood if this mechanism happens in ordinary Quantum Field Theory, probably indicating an enhancement of symmetry in the system.
This feature can be interpreted as a UV fixed point that appears dynamically. We do not know if it will stay forever on it. The reorganization of degrees of freedom could generate new dynamics, and the running would need to be adequately analyzed. It could be that after the theory Higgsplodes, we cannot trust the calculations done with the Scalar Field because the theory ceases to create one-particle states as represented diagrammatically in Figure~\ref{aa}.

\begin{figure}[h!]
\centering
\includegraphics[width=15cm]{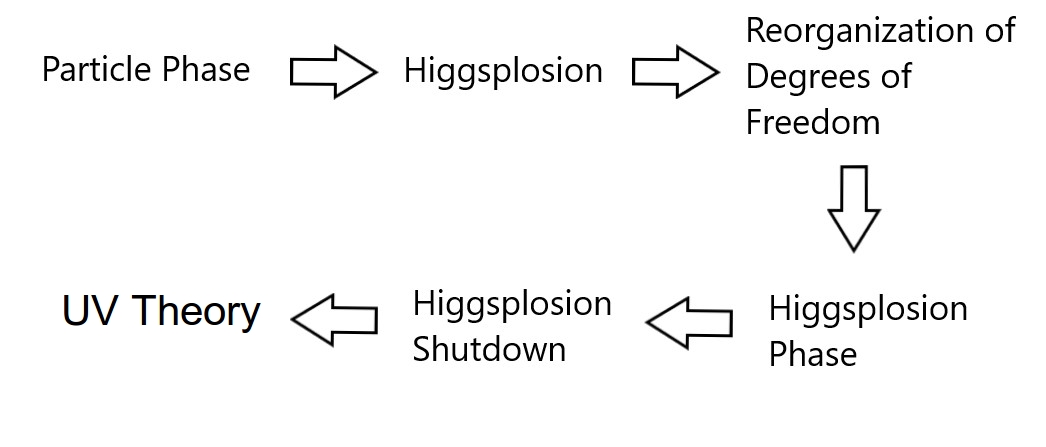}
\caption{Representation of the possible phases inside the scalar theory if Higgsplosion occurs, but it is shut down above a certain scale to preserve the proprieties that we expect from a normal Quantum Field Theory. It may be an indication that the UV completion of this theory is on a smaller scale than the Landau pole.}
\label{aa}
\end{figure}
 For the rest of the observables, the story stays pretty similar. The Higgspersion forces any process to be unitary and the coupling freezes. This feature is what makes Higgsplosion too good to be true. We can try to imagine what would happen if we implemented something like this in the Standard Model.

When we want to apply this to the Standard Model, we have to be real careful. There is no semiclassical result for the case of additional matter fields. We cannot trust the perturbative calculations to say that the Standard Model will Higgsplode, but we can try to extract some qualitative results. The first thing that we want to know is the scale $E^{*}$. We cannot make a reliable prediction for the Standard Model case, but if we trust naturality arguments for the Higgs mass parameter, the theory should Higgsplode at $10$ to $10^{3}$ TeV.  This is an assumption coming from outside the theory and  it can be that this is not the case. If the Higgs, in fact, Higgsplodes then any loop involving it would be rendered finite. This, however, does not apply to loops with other fields in the Standard Model. The running of the couplings would be modified above this scale, but there is no reason to freeze them. If, however, all particles Higgsplode as proposed in~\cite{Khoze-jun-17}, then the Standard Model would be UV finite. Perturbatively we can see that there is not much difference between the scalar and the fermion Higgsploding or even a spin one particle as it is represented at Figure~\ref{hpuniverse}.

\begin{figure}[h!]
\centering
\includegraphics[width=10cm]{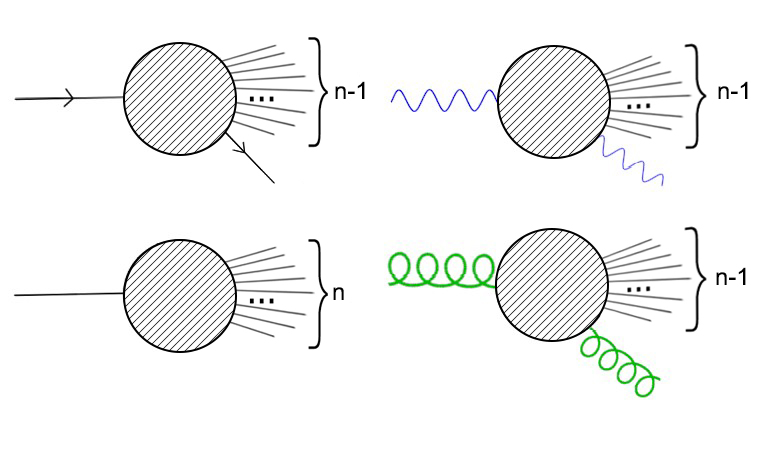}
\caption{Diagrammatic representation of different high multiplicity decays.}
\label{hpuniverse}
\end{figure}

However, the perturbative analysis cannot be trusted for high multiplicity computation, as we will discuss at the end of this chapter. The analysis done using the perturbative assumption was made in~\cite{Khoze-jun-17,precision}. While in this thesis, we are interested in the Standard Model application of this mechanism, we find it necessary to understand this proposal deeply before trying to use it.  

The focus shifts now to understand the problems with this mechanism and some useful discussions that ensued. We need to understand if the Higgsplosion can occur in the first place, and this is related to how much we can trust in the parameter space the semiclassical expression in Eq.~(\ref{eqimpor}).  In the end, we use some results coming from String Theory to study the exponential decay of a propagator in a toy model.

\section{Some Questions about Higgsplosion}

\subsection{Criticism From ``Problems with Higgsplosion"}
We saw the  evolution in power to compute the decay rate in a Scalar Field Theory across time. With modern techniques~\cite{Libanov-jul-94}, it was possible to write the decay rate in the exponential form:
\begin{align}
\Gamma(E,n) \propto e^{\frac{F(\lambda n, \epsilon)}{\lambda}} \, ,
\end{align}
For this expression to work, the conjecture is that at high multiplicity, only the non-relativistic energy contribute. The whole deal with the Higgsplosion framework is to compute this function $F$ in an appropriate regime and analyze its behavior. For small $g$ and $\epsilon$ it was found an expression for the holy grail function~\cite{Libanov-jul-94}, Eq.~(\ref{smallg}):
\begin{align}
F(g,\epsilon) = g \ln(\frac{g}{16}) - g + \frac{3g}{2}(\ln(\frac{\epsilon}{3\pi}) +1) - \frac{25}{12} g \epsilon + \frac{ \sqrt{3}}{4\pi} g^{2} \, ,
\end{align}
where we ignore terms like $g^{3}$,$g^{2}\epsilon$ and $g\epsilon^{2}$. We can see that as $g \rightarrow 0$, $F \rightarrow -\infty$, and the decay rate is suppressed. While the opposite is also true, increasing the coupling makes the decay rate grow exponentially.

 However, as it is pointed out in~\cite{Belyaev-aug-18} this expression is only valid in the limit:
\begin{align}
g \ll 1  \, ,\quad  \epsilon \ll 1 \, .
\end{align}
So we cannot trust this expression for couplings of order 1:
\begin{align}
g \approx 1  \, , \quad  n \approx \frac{1}{\lambda} \, .
\end{align}
In this range of validity (small g and $\epsilon$), the expression is always negative and Higgplosion would not occur. This changed when new methods for computing F were developed~\cite{Khoze-jun-18} and it was extended for the new region $g \gg 1$:
\begin{align}
F(g,\epsilon) = g \left( \ln(\frac{g}{4}) +0.85g^{1/2}-1  + \frac{3}{2}\left( \ln(\frac{\epsilon}{3\pi})+1 \right) - \frac{25}{12} \epsilon +\dots \right) \, .
\end{align}
We can see that for sufficiently large coupling and small $\epsilon$ this function increases. The detailed analysis of this result is made in section~\ref{s10}. This result is one of the most important for the Higgsplosion proposal. It shows that in an appropriate limit, the decay rate can indeed grow.

 However, as it is pointed out in~\cite{Belyaev-aug-18}, this semiclassical solution exists only for $\phi^{4}$ in the broken phase in (1+3)D. It is by no means a general result for an arbitrary scalar field, and the generalization for the Standard Model is not clear. It is argued in~\cite{Belyaev-aug-18} that the contributions of order $g^{2}\epsilon$ need to be taken into account so we can determine the validity range of $\epsilon$ in this function. Without these terms, we cannot determine what values of $\epsilon$ can be trusted in this approximation. Such mixed terms could prevent the exponential growth of the decay rate and by consequence prevent the Higgsplosion from happening.  The tricky part of this expression is that it has multiple parameters, and each of them plays a crucial role.

On the other hand, at least in the ultra-small limit for $\epsilon$, this expression can be trusted, and we can find a coupling that makes the decay rate grow. Going away from this coupling could kill this growth, so we need to be careful with the RG flow in this theory. If for instance, we have a small value of $\epsilon$ where the decay rate explodes, then we never probe larger values of $\epsilon$  because the theory would always decay before reaching this point. Nevertheless the need for a better expression of $F(g,\epsilon)$ as argued in~\cite{Belyaev-aug-18} is legitimate and something important to focus for the better understanding of the Higgsplosion claim and even Quantum Field Theory in general.

In the last part of the paper~\cite{Belyaev-aug-18}, they turn the attention to the full propagator. There it is argued that we cannot resum the propagator because the expression of the 1PI diagram is not convergent. That was already addressed, so we do not comment any further here. Lastly, it is pointed out in~\cite{Belyaev-aug-18} that unitarity needs to be restored somehow if Higgsplosion does not enter to play. The behavior of Eq.~(\ref{eqimpor}) shows some potential danger for unitarity violation. From their side, they argue that a better expression for $F$ would play the role of restoring unitarity. From the Higgplosion side, the Higgspersion plays that role in the theory.

\subsection{Criticism From "Inconsistencies with Higgsplosion"}

Another source of discussion about Higgsplosion arises with the paper from Monin~\cite{Monin-aug-18}. In this paper, he argued for the impossibility of the Higgsplosion mechanism inside local Quantum Field Theory. To understand it better, let us run by its arguments and then discuss a little about these results. We already saw the persistence in the exponential growth of the amplitude. Usually, we are talking about the decay rates but is worth pointing out that in the same way that the decay rate explodes the spectral density also explode:
\begin{align}
\rho (E,N) \propto  \epsilon^{\frac{3N}{2}} e^{\tilde{c} N^{3/2} \sqrt{\lambda}} \left( 1 + O(N\ln(N))\right) \, ,
\end{align}
with $\tilde{c}$ being some constant. This, in the context of Higgsplosion, is related to the exponential decay of the propagator in the UV. While the author argues that this behavior is inconsistent with local Quantum Field Theory, he points out that the natural cutoff for the theory is much smaller than the Landau pole if these expressions cease to be valid at some point. The persistence of the exponential behavior will happen if this is the case. 

Now, let us discuss a little why this behavior is unusual in an ordinary Quantum Field Theory. Normally, we have the spectral representation for the propagator (this could be any two insertions of a local operator):
\begin{align}
\Delta_{F}(p^{2}) = \int \dd{s} \frac{1}{p^{2}-s+i\epsilon} \rho(s) \, .
\end{align}
It is expected that this expression is divergent, and we need to do several subtractions to have a finite result. This procedure is related to the usual renormalization that adds $m$ independent parameters that need to be fixed by experiments. In this context, a non-renormalizable or non-local theory would have infinite subtractions and by consequence an infinite number of new parameters that would need to fix:
\begin{align}
\Delta_{F} (p^{2})= P_{m-1}(p^{2}) + p^{2m}\int \dd{s} \frac{1}{p^{2}-s+i\epsilon} \frac{\rho(s)}{s^{m}} \, .
\end{align}

The general form of the propagator allows us to extract what is the spectral density in terms of its imaginary part:
\begin{align}
-\frac{1}{\pi} \Im  \left(  \Delta_{F}(p^{2}) \right) = \rho (p^{2}) \, .
\end{align}

Up to this point, this is standard and very general Quantum Field Theory. What is done next is to assume what kind of distributions these operators are. We usually use tempered distributions, but they are very limiting as pointed out in~\cite{Khoze-sep-18}. This happens because we cannot treat a non-renormalizable theory with such distribution~\cite{Jaffe-jan-67}. As a consequence, restricting to tempered distribution would create a possible inconsistency with the Higgsplosion mechanism. These distributions only exist with a finite number of subtractions and by consequence cannot fall faster than an arbitrary polynomial. An exponential decay would be inconsistent with this choice. This is addressed in~\cite{Khoze-sep-18}, where they argue that Higgsplosion could exist within local Quantum Field Theory if we do not restrict to such distributions. They show that in principle, the Higgsplosion mechanism does not tell how fast the propagator falls apart from being exponential, and because of that, it can be consistent with local Quantum Field Theory. 

Something that we find strange in this process is the necessity of doing an infinite number of subtractions in the Higgsplosion scenario. The usual Scalar Field Theory in (1+3)D with potential $\phi^{4}$ is known to be perturbatively renormalizable and local, and somehow this is potentially being lost by this phenomena. If the theory became non-local above this scale, this would be strange behavior as well. It can happen that most of these parameters are not relevant, and we need to fix only a finite number of constants, and the theory then is local. That raises a flag about the possibility of such feature. Maybe such mechanism is shut down after some scale saving the behavior at high energies and not making the theory UV finite. In intermediary energies, the theory would enter in this new phase. We could see this new phase as a large number of field excitations behaving collectively. This reorganization of the degrees of freedom would happen at the Higgsplosion scale, but after that, the new degree of freedom will become relevant. This will generate by itself a new class of interactions that could regain the theory at the UV. In this picture, the Higgsplosion is a phase transition inside the theory. In this interpretation, the Higgsploding scale became the cutoff where a UV completion would enter. Such UV completion describes the dynamics of this reorganized degrees of freedom in the high energy limit. This is discussed further in the last section of this chapter.

Finally, we think that the discussion in the second appendix of~\cite{Monin-aug-18} is useful to understand better what is happening in the perturbative calculation. We present here what is the discussion and follow from it later  to dig deeper into this. In there, they propose a toy model, the (0+0)D Quantum Field Theory with the quartic interaction:
\begin{align} \label{integral0}
A(\lambda,N) = \int_{0}^{\infty} x^{2N} e^{-\frac{x^{2}}{2} - \lambda \frac{x^{4}}{4}} \, .
\end{align}
The nice thing about this integral is that we can solve it exactly to test different approximation schemes. The solution for this integral is:
\begin{align}
A(\lambda,N) = \lambda^{-(\frac{N}{2} + \frac{1}{4})} \Gamma(N+\frac{1}{2}) F_{1 , 1}\left( \frac{N}{2}+\frac{1}{4},\frac{1}{2}, \frac{1}{4\lambda} \right) \, ,
\end{align}
where $F_{1,1}$ is the Hypergeometric function. Doing a normal perturbation theory in the integral generate us a asymptotic series that is divergent in nature:
\begin{align} \label{epepep}
A(\lambda, N) \approx 2^{N-\frac{1}{2}} \Gamma(N + \frac{1}{2}) \left( 1 - \frac{\lambda}{4}(2N+3)(2N+1) + \dots \right) \, .
\end{align}
The interesting thing of this series is that not only $\lambda$ needs to be small, but in fact, $\lambda N $ needs to be small such that this expression is a good approximation. Because we are interested in the large $\lambda N $ limit we need a different kind of approximation. In this case, we can work out the proposal of~\cite{Monin-aug-18} and use a nontrivial saddle point to do the perturbation theory:
\begin{align}
x^{2N} = e^{2N \ln(x)}
\end{align}
Taking this into account, we can arrive at the expansion for $\lambda \rightarrow 0$, and $N \rightarrow \infty$:
\begin{align} \label{epep}
A(\lambda,N) \approx \frac{\sqrt{\pi}}{(8\lambda N)^{1/4}} e^{\frac{N}{2}(\ln(\frac{2N}{\lambda}) -1 - \sqrt{\frac{2}{\lambda N}})}(1 + O(\frac{1}{\sqrt{\lambda N}}))  \, .
\end{align}
This expression is valid as $\lambda N \gg 1$. Such expression is what we need to discuss the Higgsplosion. In this case, we can check that both Eq.~(\ref{epepep}) and Eq.~(\ref{epep}) are correct. Many parameters play important roles, and the balance between them determines the validity region of the expansion.

 Both papers discussing the problems~\cite{Belyaev-aug-18,Monin-aug-18} with Higgsplosion show the necessity to find a better expression for the beyond threshold amplitude and decay rate. We discuss in the next section the arguments coming from lattice simulations that could contest the Higgsplosion hypothesis. After that, we try to extimate the value of $E^{*}$ from Eq.~(\ref{eqimpor}).

\subsection{Lattice Results About Higgsplosion and Some Comments.}

The need for better analytic expression for the decay rate can be surpassed if we can extract information from lattice about this process. The main problem of this, ignoring the inherent problem of obtaining the result in 4D (triviality problem~\cite{trivi1,trivi2,trivi3}), is that the objects of interest are highly virtual. It is not possible to calculate the off-shell decay rate or the imaginary part of the inverse of the propagator. With this in mind, Yeo-Yie Charng searched for reasonable bounds in $\phi^{4}$ at $3D$ and $2D$ for the Higgsplosion~\cite{lattice1,lattice2}. Instead of looking for the decay rate, he looked for the cross section of two fermions going to $n$ scalars as it is represented in Figure~\ref{fer}.

\begin{figure}[h!]
\centering
\includegraphics[width=10cm]{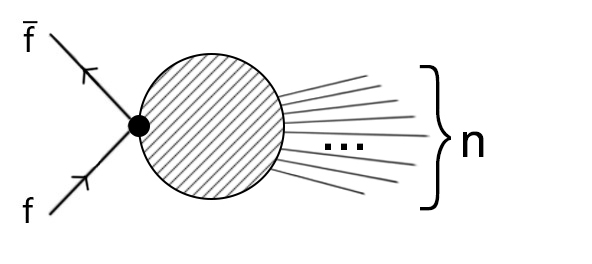}
\caption{Scattering of two fermions going to N scalars.}
\label{fer}
\end{figure}

Then, with the cross section $\sigma_{n}$ he defined the inclusive cross section $\sigma$:
\begin{align}
\sigma = \sum_{n} \sigma_{n} \, .
\end{align}
In this object we don't expect the factorial growth because the Higgspersion mechanism would dominate. Because the computation of $\sigma$ is hard he finds an inequality involving an object that it is easy to find in the lattice:
\begin{align}
\int \dd{s} \sigma(s) \leq (\frac{1}{Z'} -1) \, ,
\end{align}
where $Z'$ is defined as:
\begin{align}
\frac{\partial (G^{(2)})^{-1}}{\partial p^{2}} \vert_{p^{2}=0} = \frac{1}{Z'} \, .
\end{align}
This is a non-perturbative bound in nature. This bounds how large the integral of the total cross section can be. This is what we expect from unitarity, no growing cross sections at arbitrary energy. 

He proceeded to simulate $\phi^{4}$ to obtain the values of $Z'$ and see how much of a bound we have. For this object, in $3D$, the result that he gets with 85\% confidence level is:
\begin{align}
\int \dd{s} \sigma(s) \leq 0.026 \, .
\end{align}
So we expect the inclusive cross-section to be small. By the predictions of the Higgsplosion, we expect that this cross section to be exponentially suppressed at large $s$, making this possible. For the details of the simulation, it is recommended to read the original paper~\cite{lattice2}.

This result on its own does not say much about the possibility of Higgsplosion only that it is consistent, provided Higgspersion works as intended to restore unitarity. Maybe a more in-depth investigation from lattice could settle down any doubt about the possibility of such a mechanism. Calculating the imaginary part of the inverse of the off-shell propagator could do the trick. Until then the Higgsplosion mechanism checks out at least in the lattice.

\section{Analysis of the Decay Rate Expression, What we can say about $E^{*}$} \label{s10}

The focus of this section is trying to find a value for the Higgsplosion scale $E^{*}$.  This turns out to be a hard problem because we do not know in what parameter space the Eq.~(\ref{eqimpor}) is trustworthy. Specifically, we need to know how large $\epsilon$ can be. Before studying Eq.~(\ref{eqimpor}), let us look for the older result using a semiclassical approximation for the small $g$ limit, Eq.~(\ref{smallg}). Here we want to show that in this limit, we do not have any exponential growth. To do that, we can fix a coupling and plot different values of $\epsilon$, trying to find a value of $n$ that makes $F(g,n)$ positive. We will choose three values of the coupling $\lambda$: 0.1, 0.001, 0.0001. These values are inside the perturbative regime for $\lambda \rightarrow 0$. The problem of the small $g$ limit is that the maximum allowed value of $n$ is such that $g$ is at maximum of order one. For larger values, we cannot say anything, and a more in-depth investigation is necessary. These are the safer values to ensure that the expression is inside the domain of validity. With that in mind, we have the plots for the different couplings in Figures~\ref{smalll}.

\begin{figure}[h!]
  \centering
  \begin{subfigure}[b]{0.5\linewidth}
    \includegraphics[width=\linewidth]{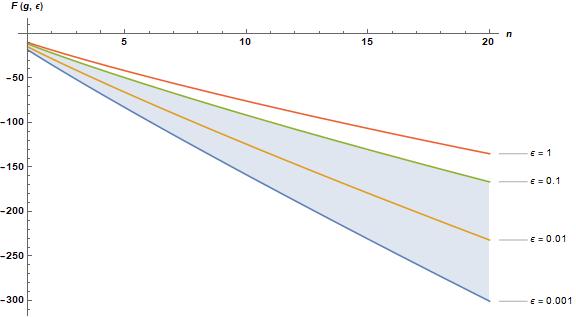}
	\caption{Plot of the different values of $\epsilon$ for the small coupling expression of the decay rate. The maximum allowed value of $n$ is such that $\lambda n = 1$, fixing $\lambda =0.1$.}
  \end{subfigure}
  \begin{subfigure}[b]{0.5\linewidth}
    \includegraphics[width=\linewidth]{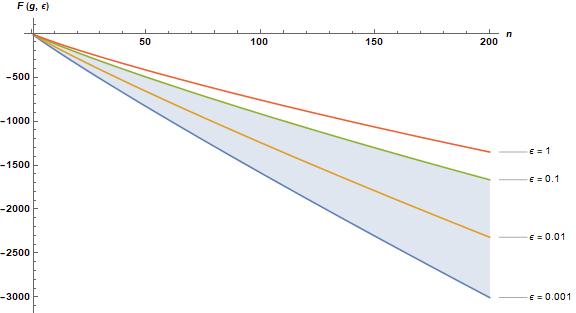}
 	\caption{Plot of the different values of $\epsilon$ for the small coupling expression of the decay rate. The maximum allowed value of $n$ is such that $\lambda n = 1$, fixing $\lambda=0.01$.}
  \end{subfigure}
  \begin{subfigure}[b]{0.5\linewidth}
    \includegraphics[width=\linewidth]{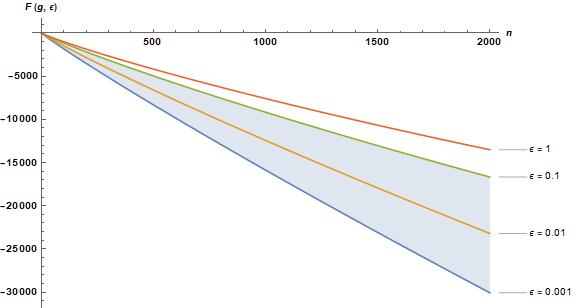}
	\caption{Plot of the different values of $\epsilon$ for the small coupling expression of the decay rate. The maximum allowed value of $n$ is such that $\lambda n = 1$, fixing $\lambda=0.001$.}
  \end{subfigure}
  \caption{The colored region is the trusted region between $\epsilon=0.1$ and $\epsilon= 0.001$, in reality, this region goes all the way to minus infinity the smaller the $\epsilon$. We chose to plot only the regions closer to zero.}
  \label{smalll}
\end{figure}
In these three plots, it is clear that the function $F(g,\epsilon)$ never grows. This indicates that there is no exponential growth and by consequence, no Higgsplosion. 

The story is different when we go for the strong $g$ limit. In this limit, we need to use Eq.~(\ref{eqimpor}) and use values of $n$ such that $\lambda n$ in the minimum is one. We can now plot the different values for $\epsilon$ in each choice of $\lambda$ just as before, represented in Figure~\ref{stronggg}

\begin{figure}[h!]
  \centering
  \begin{subfigure}[b]{0.5\linewidth}
    \includegraphics[width=\linewidth]{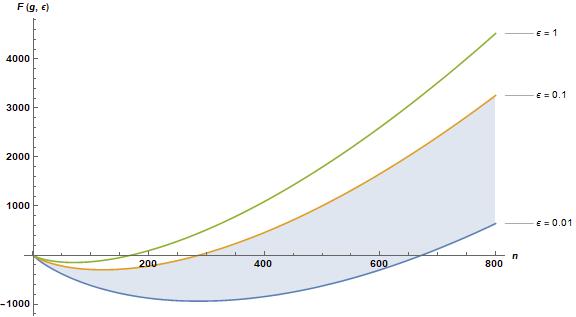}
	\caption{Plot of the different values of $\epsilon$ for the strong coupling expression of the decay rate. The minimum allowed value of $n$ is such that $\lambda n = 1$, fixing $\lambda=0.1$.}
  \end{subfigure}
  \begin{subfigure}[b]{0.5\linewidth}
    \includegraphics[width=\linewidth]{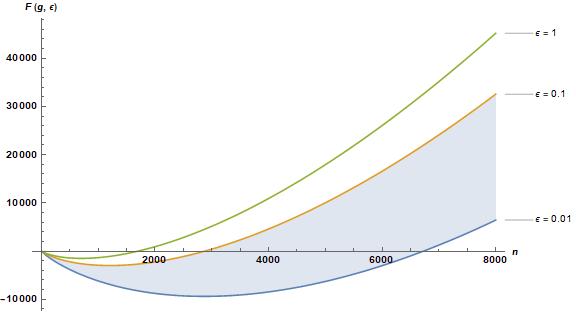}
 	\caption{Plot of the different values of $\epsilon$ for the strong coupling expression of the decay rate. The minimum allowed value of $n$ is such that $\lambda n = 1$, fixing $\lambda=0.01$.}
  \end{subfigure}
  \begin{subfigure}[b]{0.5\linewidth}
    \includegraphics[width=\linewidth]{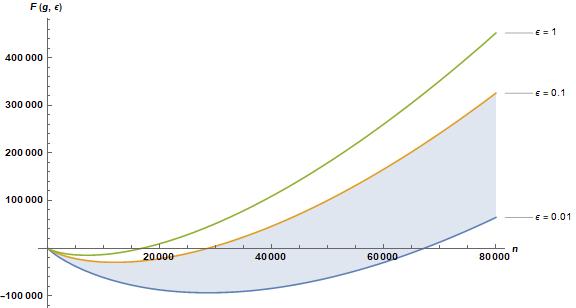}
	\caption{Plot of the different values of $\epsilon$ for the strong coupling expression of the decay rate. The minimum allowed value of $n$ is such that $\lambda n = 1$, fixing $\lambda=0.001$.}
  \end{subfigure}
  \caption{The colored region is the trusted region between $\epsilon=0.1$ and $\epsilon= 0.001$, in reality, this region goes all the way to minus infinity the smaller the $\epsilon$. We chose to plot only the regions closer to zero.}
  \label{stronggg}
\end{figure}

In the large $g$ region we can get exponential growth. There is always a point where the behavior shifts from exponential suppression to exponential growth. This point is what caraterizes the Higgsplosion scale $E^{*}$. For each $\lambda$ we have a different Higgsplosion scale. The precise determination of this scale is only possible if we know the maximum allowed value for $\epsilon$. We chose a region smaller than $\epsilon =0.1$ to find values but in principle this could be even smaller. The smaller the $\epsilon$ the larger the value of $E^{*}$. For instance, fixing $\lambda =0.1$ and $\epsilon=0.1$ the value of the Higgsploding scale is approximately:
\begin{align}
E^{*}\approx 300m.
\end{align}
While for $\lambda =0.01$ and $\epsilon=0.1$ the value of the Higgsploding scale is approximately:
\begin{align}
E^{*} \approx 3000m.
\end{align}
Finnaly for  for $\lambda =0.001$ and $\epsilon=0.1$ we get:
\begin{align}
E^{*} \approx 30000m.
\end{align}

This behavior is similar to what is predicted in~\cite{precision} for the general form of the Higgsploding scale:
\begin{align}
E^{*}= C \frac{m}{\lambda}
\end{align}
In our case, we are obtaining $C \approx 30$, because we fixed $\epsilon=0.1$. The value for this constant is close to one, respecting the naturality argument. These values come from the choice of $\epsilon$ and possibly larger choices give a more natural value for this constant. To obtain more precise values for the Higgsplosion scale, we need to understand the limitations of Eq.~(\ref{eqimpor}). In this example, if one obtains the next order in $\epsilon$ correction, this becomes an achievable task. Now that we more or less covered what we can say about the Higgsplosion scale let us enter the discussions of applicability of the perturbation theory and a new interpretation for the Higgsplosion mechanism using two toy models.

\section{0-Dimensional Case of Study.}

Let us try to understand better the approximations that we are doing. The case of study is a modification of the integral Eq.~(\ref{integral0}) (the modification is almost trivial as we just extend the domain of integration):
\begin{align}
Z[m,\lambda] = \int_{-\infty}^{\infty} \dd{x} e^{-m\frac{x^{2}}{2} - \lambda \frac{x^{4}}{4!}} \, .
\end{align}

Ultimately we are interested in calculating the expectation value of $2N$ ``fields":
\begin{align}
A[m,\lambda, N] = \int_{-\infty}^{\infty} \dd{x}  x^{2N} e^{-m\frac{x^{2}}{2} - \lambda \frac{x^{4}}{4!}} \, .
\end{align}
These are the ``$2N$-point Green functions" (moments) for a zero dimensional field theory. We could try to work with the connected amplitudes (cumulants), however, because we are interested in testing the approximation methods it is similar to work with both cases. We can relate $A[m,\lambda,N]$ and $Z[m,\lambda]$ by doing derivatives with respect to $m$.

  This case is useful because we can in fact solve the integral to compare with the approximations. The first approximation that we will do is small $\lambda$, for both functions. We want to see if this approximation of small $\lambda$ can be trusted in the high multiplicity limit. For the $Z[m,\lambda]$ the perturbative expansion can be found easily by expanding the exponential and exchanging the integration and summation:
\begin{align}
Z_{p}[m,\lambda] \thicksim  \sum_{0}^{\infty} \left( \left( \frac{2}{m} \right)^{1/2} \left( \frac{1}{6m^{2}}\right)^{n} \frac{\Gamma(2n+\frac{1}{2})}{n!} \right) \lambda^{n}  \quad \text{as} \quad \lambda \rightarrow 0 \, .
\end{align}
The expansion for the amplitude is analogous and we obtain:
\begin{align}
A_{p}[m,\lambda,N]   \thicksim  \sum_{0}^{\infty} \left( \left( \frac{2}{m}\right)^{\frac{1+2N}{2}} \left( \frac{1}{6m^{2}} \right)^{n} \frac{\Gamma(2n+N+\frac{1}{2})}{n!} \right) \lambda^{n}  \quad \text{as} \quad \lambda \rightarrow 0 \, .
\end{align}
Both expressions are valid when $\lambda$ is small compared to $m$.  Let us see if the partial sums of these series are a good approximation for the full result, that in this case we can compute:
\begin{align}
Z[m,\lambda] = \sqrt{\frac{3m}{\lambda}} e^{\frac{3m^{2}}{4\lambda}} K\left(\frac{1}{4},\frac{3m^{2}}{4\lambda}\right) \, ,
\end{align}
\begin{align}
A[m,\lambda,N] = 2^{-\frac{3}{4} + \frac{3N}{2}} 3^{\frac{1}{4}+\frac{N}{2}} \lambda^{-\frac{3}{4}-\frac{N}{2}} ( \sqrt{2\lambda} \Gamma(\frac{1}{4} + \frac{N}{2}) F_{11}\left( \frac{1}{4}+\frac{N}{2},\frac{1}{2},\frac{3m^{2}}{2\lambda}\right) - 
\end{align}
\begin{align} \nonumber
2\sqrt{3m^{2}}  \Gamma(\frac{3}{4} + \frac{N}{2}) F_{11}\left( \frac{3}{4}+\frac{N}{2},\frac{3}{2},\frac{3m^{2}}{2\lambda}\right)  )  \, ,
\end{align}
where $F_{11}[a,b,z]$ is the hypergeometric function and $K[a,z]$ is the modified Bessel function of the second kind. To test the approximations, we set $m=1$ and scan different scales for the coupling. The first one is the approximation for the partition function that is expected to be good for small $\lambda$ and some values are listed in the Table~\ref{tabela1}. For all the analysis done here we will use the notation $X^{(n)}$, meaning the partial sum up to the nth-term of the series of $X$.
\begin{table}[h!]
\centering
\begin{tabular}{|c|c|c|c|}
\hline
$\lambda$       & 0.01    & 0.1     & 1       \\ \hline
$Z_{ap}^{(1)}$ & 2.50976 & 2.5379 & 2.8199 \\ \hline
$Z_{ap}^{(2)}$  & 2.50978 & 2.5420 & 3.0484 \\ \hline
$Z_{ap}^{(3)}$  & 2.50978 & 2.5405 & 3.3625 \\ \hline
$Z_{ex}$        & 2.50352 & 2.4773 & 2.3033 \\ \hline
\end{tabular}
\caption{Comparison between the partial sum and the exact result for the partition function.}
\label{tabela1}
\end{table}

Considering the approximation for $Z[\lambda]$, we can see that going for larger values of $\lambda$, the partial sum starts to become worst. In this case, it is not so off from the exact value, let us see if this is the same for the amplitude $A[\lambda,N]$. It is important to remember that this analysis is only for the partial sum, and different summation machines could improve these results. When we introduce different values of $N$ the approximation starts to break down, signaling that the appropriate coupling is indeed $\lambda N=g$, as it is shown in the Tables~\ref{tabela2},~\ref{tabela3} and~\ref{tabela3}. For all the numerical values we set $m=1$.

\begin{table}[h!]
\centering
{\small
\begin{tabular}{|c|c|c|c|c|c|c|}
\hline
$g$                 & $0.1$     & $1$                    & $10$                     & $30$                     & $60$                      & $100$                     \\ \hline
$N$                 & $1$       & $10$                   & $100$                    & $300$                    & $600$                     & $1000$                    \\ \hline
$A_{ap}^{(0)}[0.1]$ & $2.50663$ & $1.641 \times 10^{9}$  & $1.671 \times 10^{187}$  & $5.0883 \times 10^{703}$ & $3.0313 \times 10^{1587}$ & $1.9279 \times 10^{2867}$ \\ \hline
$A_{ap}^{(1)}[0.1]$ & $2.55329$ & $4.944 \times 10^{9}$  & $2.8576 \times 10^{189}$ & $7.6885 \times 10^{706}$ & $1.8251 \times 10^{1591}$ & $3.2199 \times 10^{2871}$ \\ \hline
$A_{ap}^{(2)}[0.1]$ & $2.68385$ & $9.5886 \times 10^{9}$ & $2.5401 \times 10^{191}$ & $5.8860 \times 10^{709}$ & $5.3124 \times 10^{1594}$ & $2.699 \times 10^{2875}$  \\ \hline
$A_{ex}[0.1]$       & $2.36727$ & $3.6169 \times 10^{8}$ & $4.7914 \times 10^{161}$ & $2.1524 \times 10^{580}$ & $7.1316 \times 10^{1271}$ & $2.5246 \times 10^{2250}$ \\ \hline
\end{tabular}
}
\caption{Comparison between the partial sum and the exact result for the amplitude with $\lambda=0.1$.}
\label{tabela2}
\end{table}

\begin{table}[h!]
\centering
{\small
\begin{tabular}{|c|c|c|c|c|c|c|}
\hline
$g$                  & $0.01$    & $0.1$                  & $1$                       & $3$                       & $6$                       & $10$                      \\ \hline
$N$                  & $1$       & $10$                   & $100$                     & $300$                     & $600$                     & $1000$                    \\ \hline
$A_{ap}^{(0)}[0.01]$ & $2.5066$ & $1.6911 \times 10^{9}$ & $1.671 \times 10^{187}$   & $5.0883 \times 10^{703}$  & $3.0313 \times 10^{1587}$ & $1.9279 \times 10^{2867}$ \\ \hline
$A_{ap}^{(1)}[0.01]$ & $2.5222$  & $1.9714 \times 10^{9}$ & $3.008 \times 10^{188}$   & $7.7343 \times 10^{705}$ & $1.8278 \times 10^{1590}$ & $3.2216 \times 10^{2870}$ \\ \hline
$A_{ap}^{(2)}[0.01]$ & $2.5225$  & $2.0178 \times 10^{9}$ & $2.8123 \times 10^{189}$ & $5.9557 \times 10^{707}$ & $5.5977 \times 10^{1592}$ & $2.7024 \times 10^{2873}$ \\ \hline
$A_{ex}[0.01]$       & $2.4911$  & $1.3521 \times 10^{9}$ & $3.1363 \times 10^{181}$ & $5.2283 \times 10^{665}$ & $5.8780 \times 10^{1471}$ & $2.8990 \times 10^{2614}$ \\ \hline
\end{tabular}
}
\caption{Comparison between the partial sum and the exact result for the amplitude with $\lambda=0.01$.}
\label{tabela3}
\end{table}

\begin{table}[h!]
\centering
{\small
\begin{tabular}{|c|c|c|c|c|c|c|}
\hline
$g$                   & $0.001$   & $0.01$                 & $0.1$                    & $0.3$                     & $0.6$                     & $1$                       \\ \hline
$N$                   & $1$       & $10$                   & $100$                    & $300$                     & $600$                     & $1000$                    \\ \hline
$A_{ap}^{(0)}[0.001]$ & $2.50663$ & $1.6911 \times 10^{9}$ & $1.671 \times 10^{187}$  & $5.0883 \times 10^{703}$  & $3.0313 \times 10^{1587}$ & $1.9279 \times 10^{2867}$ \\ \hline
$A_{ap}^{(1)}[0.001]$ & $2.50819$ & $1.6741 \times 10^{9}$ & $4.5119 \times 10^{187}$ & $8.1922 \times 10^{704}$ & $1.8551 \times 10^{1589}$ & $3.2389 \times 10^{2869}$ \\ \hline
$A_{ap}^{(2)}[0.001]$ & $2.50820$  & $1.6746 \times 10^{9}$ & $7.0239 \times 10^{187}$ & $6.6976 \times 10^{705}$  & $5.7149 \times 10^{1590}$ & $2.7316 \times 10^{2871}$ \\ \hline
$A_{ex}[0.001]$       & $2.50506$ & $1.6085 \times 10^{9}$ & $3.2239 \times 10^{186}$ & $5.2109 \times 10^{697}$  & $2.1500 \times 10^{1565}$ & $7.0608 \times 10^{2810}$ \\ \hline
\end{tabular}
}
\caption{Comparison between the partial sum and the exact result for the amplitude with $\lambda=0.001$.}
\label{tabela4}
\end{table}
It is easy to see that for $\lambda =0.1$ only the $N=1$ case is close to the exact result and going for larger $N$ only make things worse. While for $\lambda =0.001$ we can go up to $N=100$ and still get a close enough result. This shows that besides $\lambda$ being small, $g$ need to be controlled as well.  We can even do an extreme result:
\begin{align}
\lambda N = 0.0001  \, , \quad  \lambda = 0.000001  \, ,\quad N=100 \, .
\end{align}
For this case the exact answer and partial sum of the first term in the series are close:
\begin{align}
A_{ex} = 1.66816 \times 10^{187} \, ,
\end{align}
\begin{align}
A_{ap}^{(1)} = 1.6994 \times 10^{187} \, .
\end{align}
Showing that the true perturbation parameter is $\lambda N= g$. 

We can make different approximations for this integral to explore different limits. The first thing that we can do is a strong coupling expansion. If we want a strong coupling expansion we need to reorganize the integral, for this we set $m=1$ again, and do the substitution:
\begin{align}
z= \lambda^{1/4} x \, .
\end{align} 
Following by a renaming of the coupling:
\begin{align}
c = \frac{1}{\sqrt{\lambda}} \, ,
\end{align}
such that the integral now is:
\begin{align}
A[c,N] = \sqrt{c} \int_{-\infty}^{\infty} \dd{z} c^{N} z^{2N} e^{- \frac{z^{4}}{4!} - c \frac{z^{2}}{2}} \, .
\end{align}

The procedure will be the same, we can solve it exactly and do an expansion for small $c$ that is equivalent to a large $\lambda$:
\begin{align}
A[c,N] = \frac{2}{2N+1} 6^{\frac{N}{2}+\frac{1}{4}} c^{N + \frac{1}{2}} \Gamma(\frac{3}{2}+N) U \left( \frac{1}{4}+\frac{N}{2},\frac{1}{2},\frac{3c^{2}}{2} \right)  \, ,
\end{align}
\begin{align}
A[c,N] \thicksim \sum_{0}^{\infty} \left( 2^{\frac{2n+6N-1}{4}} 3^{\frac{1+2N+2n}{4}} \Gamma(\frac{1}{4} + \frac{n+N}{2})(-1)^{n} \right)\frac{c^{n+N+\frac{1}{2}}}{n!} \quad \text{as} \quad c \rightarrow 0 \, .
\end{align}
We can test this approximation for different values of $c$ and $N$, shown in the Tables~\ref{tabela5},~\ref{tabela6} and~\ref{tabela7}.
\begin{table}[h!]
\centering
\begin{tabular}{|c|c|c|c|l|}
\hline
$\lambda N$                       & 10000                           & 100                           & 1                            & 0.01                     \\ \hline
$c$                               & 0.01                            & 0.1                           & 1                            & 10                       \\ \hline
$A_{ap}^{(1)}[1]$                 & 0.00652336                      & 0.172028                      & -5.39346                     & -3596                    \\ \hline
$A_{ap}^{(2)}[1]$                 & 0.00652635                      & 0.181483                      & 24.5033                      & 90945.6                  \\ \hline
$A_{ap}^{(3)}[1]$                 & 0.00652626                      & 0.1786318                     & -65.7756                     & $-2.76392 \times 10^{6}$ \\ \hline
\multicolumn{1}{|l|}{$A_{ex}[1]$} & \multicolumn{1}{l|}{0.00652484} & \multicolumn{1}{l|}{0.177318} & \multicolumn{1}{l|}{1.72872} & 2.49116                  \\ \hline
\end{tabular}
\caption{Comparison between the partial sum and the exact result for the amplitude with $N=1$ in the strong coupling limit.}
\label{tabela5}
\end{table}
\begin{table}[h!]
\centering
\begin{tabular}{|c|c|c|c|l|}
\hline
$\lambda$                          & 10000                                          & 100                               & 1                                           & 0.01                      \\ \hline
$c$                                & 0.01                                           & 0.1                               & 1                                           & 10                        \\ \hline
$A_{ap}^{(1)}[10]$                 & $2.9328 \times 10^{-13}$                      & $0.0044344$                      & $-1.3902 \times 10^{9}$                    & $-5.2795 \times 10^{20}$ \\ \hline
$A_{ap}^{(2)}[10]$                 & $2.9426 \times 10^{-13}$                       & $0.0075253$                      & $8.3837 \times 10^{9}$                     & $ 3.0380 \times 10^{22}$ \\ \hline
$A_{ap}^{(3)}[10]$                 & $2.9420 \times 10^{-13}$                      & $0.0056700$                      & $-5.0286 \times 10^{10}$                    & $-1.8249 \times 10^{24}$ \\ \hline
\multicolumn{1}{|l|}{$A_{ex}[10]$} & \multicolumn{1}{l|}{$2.9376 \times 10^{-13}$} & \multicolumn{1}{l|}{$0.0057133$} & \multicolumn{1}{l|}{$2.5129 \times 10^{6}$} & $1.1521 \times 10^{9}$   \\ \hline
\end{tabular}
\caption{Comparison between the partial sum and the exact result for the amplitude with $N=10$ in the strong coupling limit.}
\label{tabela6}
\end{table}
\begin{table}[h!]
\centering
\begin{tabular}{|c|c|c|c|l|}
\hline
$\lambda$                           & 10000                                          & 100                                            & 1                                              & 0.01                       \\ \hline
$c$                                 & 0.01                                           & 0.1                                            & 1                                              & 10                         \\ \hline
$A_{ap}^{(1)}[100]$                 & $1.51361 \times 10^{-69}$                      & $-4.23803 \times 10^{31}$                      & $-2.98781 \times 10^{133}$                     & $-9.96931 \times 10^{234}$ \\ \hline
$A_{ap}^{(2)}[100]$                 & $1.56881 \times 10^{-69}$                      & $1.32163 \times 10^{32}$                       & $5.22077 \times 10^{134}$                      & $1.73547 \times 10^{237}$  \\ \hline
$A_{ap}^{(3)}[100]$                 & $1.55915 \times 10^{-69}$                      & $-1.73165 \times 10^{32}$                      & $-9.13326 \times 10^{135}$                     & $-3.03593 \times 10^{239}$ \\ \hline
\multicolumn{1}{|l|}{$A_{ex}[100]$} & \multicolumn{1}{l|}{$1.53967 \times 10^{-69}$} & \multicolumn{1}{l|}{$ 1.03189 \times 10^{31}$} & \multicolumn{1}{l|}{$1.13733 \times 10^{125}$} & $3.13631 \times 10^{181}$  \\ \hline
\end{tabular}
\caption{Comparison between the partial sum and the exact result for the amplitude with $N=100$ in the strong coupling limit.}
\label{tabela7}
\end{table}

In this case, the approximation is good for small $c$ and moderate multiplicity as expected. In terms of $\lambda$ this would be the approximation for $\lambda \rightarrow \infty$. The approximation parameter, in this case, is $cN$ so we can see that at high multiplicity, we can still get a good approximation if we pick values of $cN$ that are small. Here we can explore a different regime where $\lambda$ is large and N large provided that $cN$ is small. The semiclassical limit is something like $\lambda \rightarrow 0 $ and $N \rightarrow \infty$ . If we want to mimic this behavior, we need to make a different approximation. Given these two cases, we can see that the perturbation parameter is not the naive one and $N$ mixes with him to say how good of approximation the partial sum of the series is. With this toy model, we can see that at high multiplicity, the small coupling expansion is useless for a partial sum. If we use Pad\'e in the coefficients, we will get a better approximation, but this is not the spirit for the interpretation of exponential growth of amplitudes at tree level. Using this example, we can see that it is not possible to draw this conclusion using the perturbation theory only. This leaves us with only the semiclassical calculation indicating the exponential growth of the decay rate. If we want to understand this approximation we can try to make a saddle point approximation of this integral as suggested in~\cite{Monin-aug-18} using the effective action:
\begin{align} \label{int10}
A[m,\lambda, N] = \int_{-\infty}^{\infty} \dd{x}  e^{-m\frac{x^{2}}{2} - \lambda \frac{x^{4}}{4!} + 2N \ln(x)} \, .
\end{align}

Because we want to make a saddle point approximation with $N$ as the large parameter we can change the variables to arrange the integral in the canonical form:
\begin{align}
x = \sqrt{N}z \, .
\end{align}
Then the Eq.~(\ref{int10}) becomes:
\begin{align}
A[m,\lambda , N ] = \sqrt{N} e^{N\ln(N)} \int_{-\infty}^{\infty} \dd{z} e^{N W(z)} \, ,
\end{align}
where $W(z)$ is the function that we will find the saddles to expand:
\begin{align}
W(z) = -m \frac{z^{2}}{2} - g \frac{z^{4}}{4!} +2 \log(z) \, .
\end{align}
In this action, we used the definition of the t' Hooft like coupling $g$. There is a final step that we need to do before starting the approximation. The integral is symmetric under parity that means that the saddle point only will pick one side for the approximation, because of that we can write the integral as two copies of only one of the sides. This guarantees that we are making the right approximation for that integral:
\begin{align} \label{int20}
A[m,\lambda,N] = 2 \sqrt{N}e^{N\ln(N)} \int_{0}^{\infty} \dd{z} e^{N W(z)} \, .
\end{align}

Because we will do a numerical check on this approximation we  use the system where $m=1$ so we can compare with the results that we got so far. Now we need to look for saddle points of $W(z)$:
\begin{align}
W(z)' = -z - g \frac{z^{3}}{3!} + \frac{2}{z} =0 \, .
\end{align}
This equation has four solutions, we pick only the positive $z_{0}$ solutions and then choose the one where the function has a maximum:
\begin{align}
W''(z_{0}) = -1 - g \frac{z^{2}_{0}}{2} - \frac{2}{z_{0}^{2}} < 0  \, .
\end{align}
Using these conditions we get the saddle point that we will use to expand the integral around:
\begin{align}
z_{0} = \sqrt{\frac{-3 + \sqrt{9 + 12g}}{g}} \, .
\end{align}

Proceeding with the saddle point approximation we do a change of variables in Eq.~(\ref{int20}):
\begin{align}
z= z_{0} + \frac{y}{\sqrt{N}} \, .
\end{align}
Doing the expansion around the saddle point of the term in the exponent gives:
\begin{align}
N W(z) = NW(z_{0}) + \frac{y^{2}}{2} W''(z_{0}) + \frac{y^{3}}{6\sqrt{N}} W'''(z_{0}) + \dots \, .
\end{align}
Since we are making an large $N$ expansion, we can expand the exponential in powers of $\frac{1}{\sqrt{N}}$:
\begin{align}
e^{N W(z)} = e^{N W(z_{0})}  e^{\frac{y^{2}}{2}W''(z_{0})} (1 + \frac{y^{3}}{6\sqrt{N}}W'''(z_{0}) + \frac{y^{4}W''''(z_{0}) + 3 y^{6}(W'''(z_{0}))^{2}}{72N} + \dots \, .
\end{align}
In the large $N$ limit we can extend the domain of integration. This can be done because the integral is localized. The expansion for Eq.~(\ref{int10}) becomes:
\begin{align}
A[\lambda,N] = 2 e^{N(W(z_{0})+\ln(N))} \int_{-\infty}^{\infty} \dd{y} e^{\frac{y^{2}}{2}W''(z_{0})} \left(1 + \frac{y^{3}}{6\sqrt{N}}W'''(z_{0}) + \frac{y^{4}W''''(z_{0}) + 3 y^{6}(W'''(z_{0}))^{2}}{72N} + \dots \right)  \, .
\end{align}
Doing the integral we get the saddle point approximation for this integral. Here we will just retain terms up to $N^{-1}$ but the rest of the terms can be esily generated following the steps shown:
\begin{align}
A[\lambda,N] = 2 e^{N(W(z_{0})+\ln(N))} \sqrt{\frac{2\pi}{-W''(z_{0})}} \left( 1 + \frac{W''''(z_{0})}{24(-W''(z_{0}))^{2}N} + \frac{5 (W'''(z_{0}))^{2}}{8(-W''(z_{0}))^{3}N} + \dots \right) \, .
\end{align}
The expression for the different derivatives of $W$ in the point $z_{0}$ are:
\begin{align}
-W(z_{0}) = \frac{-3 + 2g + \sqrt{9 +12g} - 4g \ln(\frac{-3 +\sqrt{9+12g}}{g})}{4g} \, ,
\end{align} 
\begin{align}
-W''(z_{0}) = \frac{6 + 8g - 2 \sqrt{9+12g}}{\sqrt{9 + 12g}-3} \, ,
\end{align}
\begin{align}
W'''(z_{0}) = \frac{-8g + 6( \sqrt{9+12g-3}}{g(\frac{-3+ \sqrt{9+12g}}{g})^{3/2}} \, ,
\end{align}
\begin{align}
W''''(z_{0} )= g \left( 1 - \frac{12g}{(-3+ \sqrt{9+12g})^{2}} \right) \, .
\end{align}

Let us start to investigate how good is this approximation in the strong coupling semiclassical limit:
\begin{align}
N \rightarrow \infty  \, , \quad \lambda \rightarrow 0  \, , \quad g \rightarrow \infty
\end{align}
For this, let us construct the comparison for the following values of the coupling, $\lambda = 0.1 ,0.01 , 0.001$.  We scan this approximation in the large $g$ limit, that in principle, is ideal to study the Higgsplosion mechanism. The values for these couplings are in the Tables~\ref{tabela8},~\ref{tabela9}, and~\ref{tabela10}.
\begin{table}[h!]
\centering
\begin{tabular}{|c|c|c|c|}
\hline
$N$                                 & $100$                                          & $1000$                                            & $10000$                                            \\ \hline
$g$                                 & $10$                                           & $100$                                             & $1000$                                             \\ \hline
$A_{ap}^{(0)}[0.1]$                 & $4.7939 \times 10^{161}$                      & $1.1376 \times 10^{125}$                         & $4.6363 \times 10^{79}$                           \\ \hline
$A_{ap}^{(1)}[0.1]$                 & $4.7888 \times 10^{161}$                      & $1.115 \times 10^{125}$                           & $3.9322 \times 10^{79}$                           \\ \hline
\multicolumn{1}{|l|}{$A_{ex}[0.1]$} & \multicolumn{1}{l|}{$4.7914 \times 10^{161}$} & \multicolumn{1}{l|}{$2.5246 \times 10^{2250}$} & \multicolumn{1}{l|}{$8.3371 \times 10^{25318}$} \\ \hline
\end{tabular}
\caption{Comparison between the partial sum and the exact result for the amplitude with $\lambda=0.1$ in the strong coupling limit of $g$.}
\label{tabela8}
\end{table}
\begin{table}[h!]
\centering
\begin{tabular}{|c|c|c|c|}
\hline
$N$                                  & $1000$                                            & $10000$                                             & $100000$                                            \\ \hline
$g$                                  & $10$                                              & $100$                                               & $1000 $                                             \\ \hline
$A_{ap}^{(0)}[0.01]$                 & $2.89919 \times 10^{2614}$                       & $2.5247 \times 10^{2250}$                        & $ 5.5863 \times 10^{1798}$                       \\ \hline
$A_{ap}^{(1)}[0.01]$                 & $2.89888 \times 10^{2614}$                       & $2.5197 \times 10^{2250}$                        & $5.5015 \times 10^{1798}$                          \\ \hline
\multicolumn{1}{|l|}{$A_{ex}[0.01]$} & \multicolumn{1}{l|}{$2.67897 \times 10^{2614}$} & \multicolumn{1}{l|}{$1.7379 \times 10^{29044}$} & \multicolumn{1}{l|}{$4.3090 \times 10^{293359}$} \\ \hline
\end{tabular}
\caption{Comparison between the partial sum and the exact result for the amplitude with $\lambda=0.01$ in the strong coupling limit of $g$.}
\label{tabela9}
\end{table}
\begin{table}[h!]
\centering
\begin{tabular}{|c|c|c|c|}
\hline
$N$                                   & $10000$                                             & $100000$                                              & $1000000$                                              \\ \hline
$g$                                   & $10$                                                & $100$                                                 & $1000$                                               \\ \hline
$A_{ap}^{(0)}[0.001]$                 & $1.89717 \times 10^{36142}$                       & $7.33370 \times 10^{32503}$                         & $3.6033 \times 10^{27989}$                           \\ \hline
$A_{ap}^{(1)}[0.001]$                 & $1.89715 \times 10^{36142}$                         & $7.31224 \times 10^{32503}$                        & $3.5978 \times 10^{27989}$                           \\ \hline
\multicolumn{1}{|l|}{$A_{ex}[0.001]$} & \multicolumn{1}{l|}{$1.993025 \times 10^{30940}$} & \multicolumn{1}{l|}{$2.83770 \times 10^{312318}$} & \multicolumn{1}{l|}{$9.7164 \times 10^{3126099}$} \\ \hline
\end{tabular}
\caption{Comparison between the partial sum and the exact result for the amplitude with $\lambda=0.001$ in the strong coupling limit of $g$.}
\label{tabela10}
\end{table}

In these results, the huge numbers are appearing because we are not working only with the connected amplitudes. We can see that the saddle point is a good approximation for $g=10$ and $N=100$ and $1000$. Any other value the approximation is not useful for the partial sum.  This indicates that this saddle point approximation is trustworthy in the $g \approx 10 $ limit and deviates for larger values. With this result, we obtain three different regions of investigation. Ordinary perturbation theory for high multiplicity works when $\lambda N = g$ is small.  Going for strong perturbation theory in the high multiplicity case, it works for $\frac{N}{\sqrt{\lambda}}$ small, and this means that we can explore the high $\lambda$ limit. Finally, the saddle point approximation works best when $N$ is large, $\lambda$ small and $g$ of order ten.  If the semiclassical calculation for the Decay rate follows this behavior, this could be bad news because we want to explore larger values of $g$.

It is worth to remember that this is just a toy model, and the real limitation of the Eq.~(\ref{eqimpor}) cannot be extracted only from this. We can take from this that perturbation theory is not useful for partial sums in the high multiplicity limit, only in the ultra-perturbative limit. This is not the end of the story, and a useful approximation for large $g$ is still needed. Despite that for moderate values of $g$ and $N$, the semiclassical approximation can be trusted, and this could be enough to realize the Higgsplosion Mechanism. The next section, we introduce the last toy model and a new interpretation of the Higgsplosion mechanism.

\section{Exponential Decay of the Propagator, String Inspired Toy Model}

The appearance of an exponentially suppressed propagator is an unusual feature of a Quantum Field Theory. Most of the known cases have some form of non-locality in it. This does not mean that we cannot use these models to describe some physical system. Generally in these theories there is a scale $\Lambda$ where the non-locality becomes dominant. Below such scale, the system is causal and behaves like a local theory. If we want to understand the Higgsplosion mechanism near the Higgsplosion scale $E^{*}$ we could try to use some toy model that mimics this exponential behavior. 

It turns out that there is a physical system where an exponential decay of the propagator occurs. This system appears in String Field Theory~\cite{string3}, and we do not need to understand the String side fully to use it. Understanding String Field Theory would be a thesis on its own. We show the origins for such an action just for the sake of consistency, but given the theory, we can treat like any other Quantum Field Theory. This feature of exponential decay of propagators is a common occurrence in String Theory and String Field Theory~\cite{uv}.

The physical system is the Tachyon condensation in open strings~\cite{string1,string2,string3}. If an open string is attached to an unstable $D$-brane then there is a tachyon in the spectrum. This tachyonic vacuum will then describe a vacuum where the unstable $D$-brane decays and disappears. This feature is similar to what is proposed in this Thesis about the Higgsplosion entering a new phase because all the single particle states are gone. Using String Field Theory, it was possible to show that indeed there was no particle excitation in this vacuum configuration. If we try to describe this process using local fields, in the level 0\footnote{Level expansion is the expansion in the different excitations of the String Field. This is justified because the couplings to the higher excitations decrease exponentially. The level 0 is just the tachyon, and level 1 is the tachyon plus a massless vector boson, and so on.} sector of the open bosonic String Field Theory we have the following action:
\begin{align}
S_{tachyon} = \int \dd[26]{x} \left[ \frac{1}{2} \phi e^{-\frac{\Box}{M^{2}}}(\Box + \mu^{2})\phi - \frac{g}{3} \phi^{3} \right] \, ,
\end{align}
where $\phi$ is the tachyonic field with mass $m^{2}=-\mu^{2}$ that describe the instability of the $D$-brane. The exponential factor with the D'Alambertian operator gives the non-locality of this action, and the factor $M^{2}$ is the non-local scale. This action is studied in detail in~\cite{string1}. The propagator has the exponential decay feature:
\begin{align}
\Delta_{F}^{-1} = e^{\frac{p^{2}}{M^{2}}}(p^{2}-\mu^{2}) \, .
\end{align}

Here we study a toy model for the broken phase of the scalar field that has this exponential suppression. The action that we propose to work with is:
\begin{align}
S = \int \dd[4]{x} \left[ \frac{1}{2} \phi e^{-\frac{\Box}{(E^{*})^{2}}}(\Box - m^{2})\phi - \frac{\lambda}{4!} \phi^{4} \right] \, ,
\end{align}
in the broken phase where $m^{2}=-\mu^{2}$. The minimum configuration is the same as the local version because the field solution is constant:
\begin{align}
\phi_{\min}^{2} = \frac{3! \mu^{2}}{\lambda} \, .
\end{align}
Doing a shift in the field to remove the expectation value:
\begin{align}
\sigma = \phi - \phi_{\min} \, .
\end{align}
We can write the action as:
\begin{align}
S = \int \dd[4]{x} \left[ \frac{1}{2} \sigma \left(  e^{-\frac{\Box}{(E^{*})^{2}}}(\Box + \mu^{2}) -3\mu^{2} \right)\sigma + \mathcal{L}_{int} \right] \, .
\end{align}
We will ignore the interaction part because we want to analyze the spectrum of the free theory. The propagator in this phase is similar to the one obtained before:
\begin{align}
\Delta_{F}^{-1} = e^{\frac{p^{2}}{(E^{*})^{2}}}(-p^{2}+\mu^{2})-3\mu^{2} \, .
\end{align}

We want to look for the pole structure of this propagator. If we have more than one pole this will mean the appearance of a ghost in the spectrum, and the theory will not be valid above the mass of the ghost. Doing a change of variables to find these poles:
\begin{align}
-k^{2} = x \mu^{2}  \, ,
\end{align}
the equation that we need to solve is:
\begin{align} \label{eqeq1}
P(\nu,x)=e^{-\nu x}(x+1)-3 =0 \, .
\end{align}
In this equation we have $\nu = \frac{\mu^{2}}{(E^{*})^{2}}$ and the mass of the excitation is:
\begin{align}
m^{2} = x \mu^{2} \, .
\end{align}

It is possible to see the behavior of this function varying $\nu$ in the Figure~\ref{fig}. It has a distinct feature similar to the tachyonic case. There is a region where there is no pole ($\nu \gtrsim 0.14$). This region there is no particle in the spectrum, and we can interpret as the theory have Higgsploded. If we want to describe the theory above this scale, we would need to change the degrees of freedom. It has one special point where there is one particle in the spectrum ($\nu_{crit} \approx 0.14$). Finally, there exists a region where we have a particle and a ghost, the ghost being near infinity and getting closer as the coupling approaches the transition ($0<\nu \lesssim 0.14$). 
\begin{figure}[h!]
\centering
\includegraphics[width=15cm]{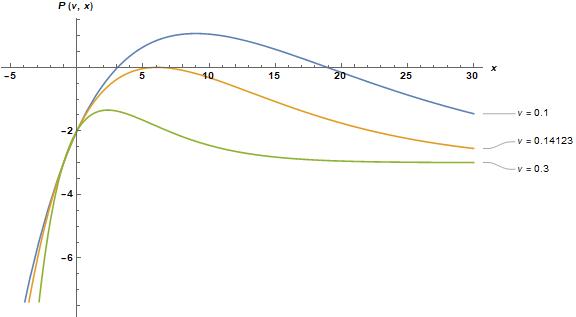}
\caption{Plot of Eq~(\ref{eqeq1}) for different scales $\nu =0.1, 0.14123$ and $0.3$.}
\label{fig}
\end{figure}

This is very similar to the spectrum found in~\cite{string1} that indicates the decay of the $D$-brane. In the Higgsplosion scenario, this could indicate that the system ``decays" and we need another description above that scale. This is not as strong as the UV finiteness but is a remarkable feature nonetheless. That could explain the similarities from a non-local theory pointed out in~\cite{Monin-aug-18}. We can see the values of $E^{*}$ given that $\mu$ is of the order of the weak scale demanding that the theory has no particles:
\begin{align}
E^{*} \approx 1.3 \, \text{TeV} \quad \text{for} \quad \mu = 500 \, \text{GeV} \, ,
\end{align}
\begin{align}
E^{*} \approx 0.66 \, \text{TeV} \quad \text{for} \quad \mu = 250 \,  \text{GeV} \, ,
\end{align}
\begin{align}
E^{*} \approx 2.6 \, \text{TeV} \quad \text{for} \quad \mu = 1 \, \text{TeV} \, .
\end{align}
These scales are too low, but this is just a toy model for the behavior proposed in the Higgsplosion mechanism. If we obtain a better expression for the off-shell propagator, we can try to model inside a non-local theory of this kind to investigate more deeply the dynamics near this scale.  This example shows how strange is the exponential decay of a propagator, and this possible interpretation of theory ``decay". In scales much lower than $E^{*}$ we can treat it using the local Lagrangian without a problem. As we get closer to $E^{*}$ we need to use this Lagrangian to discuss the physics. Finally, passing this scale, we would need another description in terms of the remaining degrees of freedom of the theory. The non-local features emerging near the Higgsplosion scale can be interpreted as the $n$ particles behaving collectively to form a structure. The scalar mass is shielded in this interpretation despite the absence of UV finitness because of the $n$ particle structure formed, in a similar way as a composite scalar is protected from the UV.
\chapter{Conclusion}
\pagenumbering{arabic}
In this thesis, we have examined the Higgsplosion mechanism. The Higgsplosion mechanism appears naturally when we study high multiplicity processes. This mechanism has the potential to generate exciting features inside a Quantum Field Theory. To understand this we need to study the results that lead to it.

Firstly, we have studied the systematic construction of high multiplicity amplitudes in a perturbative setup. In particular, we investigated the $\phi^{4}$ theory up to one-loop at treashold.  The analysis of these amplitudes showed the factorial growth behavior that is a danger to unitarity.  Even quantum corrections could not tame this behavior, indicating that ordinary perturbation theory is not suitable for these processes. It was then presented some beyond threshold results that showed the same behavior, but were perturbative. In this setup, we cannot say if the expressions are blowing because the approximation is terrible or because it is an essential feature of the theory. 

Next, we introduce the semiclassical result, Eq.~(\ref{eqimpor}), that shows this factorial behavior outside ordinary perturbation theory.  This kind of approximation was tested in the zero-dimensional toy model. We show then that usual perturbation theory and strong perturbation theory have a limited domain of validity. The semiclassical like expansion probed a different region than both of the approximations, showing an overall consistency with the result.  However, a better understanding of the limitations of Eq.~(\ref{eqimpor}) is still needed. 

Then, we introduced the Higgsplosion mechanism. We reviewed the original formulation of the Higgsplosion and showed that, at least using the semiclassical calculation, it could be possible to have in (1+3)D broken $\phi^{4}$. The extensions of the Higgsplosion mechanism depend on how you read the results. Using the original interpretation of UV finiteness renders the theory finite and the coupling stops running. The generalization for other models depends yet on a semiclassical approximation with different matter contents.  

The interpretation proposed in this thesis for the Higgsplosion mechanism is that the theory reorganizes the degrees of freedom at the Higgsploding scale. After this reorganization, another theory takes place in the UV. This is similar to the D-brane decay from tachyon condensation in string theory. The apparent non-locality can be seen in this setup because the $n$-scalars are behaving collectively in a finite size structure. After the change in the description, it is expected that the theory develops different kinds of interactions, and the flow to the UV is not necessarily finite.

This new interpretation needs to be tested, and one way is to prove that Higgsplosion occurs at least in the broken $\phi^{4}$ theory in (1+3)D. The state of the lattice now does not bring anything new to the table as we discussed, but the potential is there. Lattice investigation of this system could be fruitful because of its non-perturbative nature.  

The application for the Standard Model is yet far to be realized. It is not clear, however, if this mechanism occurs in this system. After this is resolved, then the generalization for different matter seems to be immediate. The Higgs would be the portal for this new phenomenon in the Standard Model.

\begin{appendix}

\end{appendix}

\bibliography{bibliography} 

\begin{thebibliography}{10}

\bibitem{higgs1}
S.~Chatrchyan {\em et~al.}, ``{Observation of a new boson at a mass of 125 GeV
  with the CMS experiment at the LHC},'' {\em Phys. Lett.}, vol.~B716,
  pp.~30--61, 2012.

\bibitem{higgs2}
D.~Buttazzo, {\em {Implications of the discovery of a Higgs boson with a mass
  of 125 GeV}}.
\newblock PhD thesis, Pisa, Scuola Normale Superiore, 2014.

\bibitem{dm1}
G.~Bertone, D.~Hooper, and J.~Silk, ``{Particle dark matter: Evidence,
  candidates and constraints},'' {\em Phys. Rept.}, vol.~405, pp.~279--390,
  2005.

\bibitem{dm2}
V.~Trimble, ``{Existence and Nature of Dark Matter in the Universe},'' {\em
  Ann. Rev. Astron. Astrophys.}, vol.~25, pp.~425--472, 1987.

\bibitem{DE}
R.~Durrer, ``{What do we really know about Dark Energy?},'' {\em Phil. Trans.
  Roy. Soc. Lond.}, vol.~A369, pp.~5102--5114, 2011.
\newblock [J. Cosmol.15,6065(2011)].

\bibitem{gravity1}
S.~Bose, A.~Mazumdar, G.~W. Morley, H.~Ulbricht, M.~Toro\ifmmode~\check{s}\else
  \v{s}\fi{}, M.~Paternostro, A.~A. Geraci, P.~F. Barker, M.~S. Kim, and
  G.~Milburn, ``Spin entanglement witness for quantum gravity,'' {\em Phys.
  Rev. Lett.}, vol.~119, p.~240401, Dec 2017.

\bibitem{gravity2}
C.~Marletto and V.~Vedral, ``{Gravitationally-induced entanglement between two
  massive particles is sufficient evidence of quantum effects in gravity},''
  {\em Phys. Rev. Lett.}, vol.~119, no.~24, p.~240402, 2017.

\bibitem{gravity3}
J.~Nemirovsky, E.~Cohen, and I.~Kaminer, ``{Spin-Spacetime Censorship},'' 2018.

\bibitem{gravity4}
S.~{Weinberg} and E.~{Witten}, ``{Limits on massless particles},'' {\em Physics
  Letters B}, vol.~96, pp.~59--62, Oct. 1980.

\bibitem{nat1}
G.~F. Giudice, ``{Naturally Speaking: The Naturalness Criterion and Physics at
  the LHC},'' pp.~155--178, 2008.

\bibitem{nat2}
G.~'t~Hooft, ``{Naturalness, chiral symmetry, and spontaneous chiral symmetry
  breaking},'' {\em NATO Sci. Ser. B}, vol.~59, pp.~135--157, 1980.

\bibitem{susy1}
G.~E. Stedman, ``{Simple Supersymmetry. 2. Factorization Method in Quantum
  Mechanics},'' {\em Eur. J. Phys.}, vol.~6, p.~225, 1985.

\bibitem{susy2}
F.~Cooper, A.~Khare, and U.~Sukhatme, ``{Supersymmetry and quantum
  mechanics},'' {\em Phys. Rept.}, vol.~251, pp.~267--385, 1995.

\bibitem{susy3}
A.~Arbey, M.~Battaglia, A.~Djouadi, F.~Mahmoudi, and J.~Quevillon,
  ``{Implications of a 125 GeV Higgs for supersymmetric models},'' {\em Phys.
  Lett.}, vol.~B708, pp.~162--169, 2012.

\bibitem{susy4}
H.~Baer and X.~Tata, {\em {Weak scale supersymmetry: From superfields to
  scattering events}}.
\newblock Cambridge University Press, 2006.

\bibitem{comp1}
G.~F. Giudice, ``{Naturalness after LHC8},'' {\em PoS}, vol.~EPS-HEP2013,
  p.~163, 2013.

\bibitem{comp2}
R.~Contino, ``{The Higgs as a Composite Nambu-Goldstone Boson},'' in {\em
  {Physics of the large and the small, TASI 09, proceedings of the Theoretical
  Advanced Study Institute in Elementary Particle Physics, Boulder, Colorado,
  USA, 1-26 June 2009}}, pp.~235--306, 2011.

\bibitem{comp3}
G.~Panico and A.~Wulzer, ``{The Composite Nambu-Goldstone Higgs},'' {\em Lect.
  Notes Phys.}, vol.~913, pp.~pp.1--316, 2016.

\bibitem{comp4}
M.~Redi and A.~Tesi, ``{Implications of a Light Higgs in Composite Models},''
  {\em JHEP}, vol.~10, p.~166, 2012.

\bibitem{comp5}
O.~Witzel, ``{Review on Composite Higgs Models},'' in {\em {36th International
  Symposium on Lattice Field Theory (Lattice 2018) East Lansing, MI, United
  States, July 22-28, 2018}}, 2019.

\bibitem{Khoze-higgsplosion}
V.~V. Khoze and M.~Spannowsky, ``{Higgsplosion: Solving the Hierarchy Problem
  via rapid decays of heavy states into multiple Higgs bosons},'' {\em Nucl.
  Phys.}, vol.~B926, pp.~95--111, 2018.

\bibitem{Khoze-jun-17}
V.~V. Khoze and M.~Spannowsky, ``{Higgsploding universe},'' {\em Phys. Rev.},
  vol.~D96, no.~7, p.~075042, 2017.

\bibitem{Brown-nov-92}
L.~S. Brown, ``Summing tree graphs at threshold,'' {\em Phys. Rev. D}, vol.~46,
  pp.~R4125--R4127, Nov 1992.

\bibitem{Voloshin-apr-93}
M.~B.~Voloshin, ``Zeros of tree-level amplitudes at multiboson thresholds,''
  {\em Physical review D: Particles and fields}, vol.~47, pp.~2573--2577, 04
  1993.

\bibitem{Goldberg-may-90}
H.~Goldberg, ``{Breakdown of perturbation theory at tree level in theories with
  scalars},'' {\em Phys. Lett.}, vol.~B246, pp.~445--450, 1990.

\bibitem{Son-may-95}
D.~Son, ``Semiclassical approach for multiparticle production in scalar
  theories,'' {\em Nuclear Physics B}, vol.~477, pp.~378--406, 06 1995.

\bibitem{qft}
M.~D. Schwartz, {\em {Quantum Field Theory and the Standard Model}}.
\newblock Cambridge University Press, 2014.

\bibitem{cumu}
G.-C. Rota and J.~Shen, ``On the combinatorics of cumulants,'' {\em Journal of
  Combinatorial Theory, Series A}, vol.~91, no.~1, pp.~283 -- 304, 2000.

\bibitem{Khoze-sep-18}
V.~V. Khoze and M.~Spannowsky, ``{Consistency of Higgsplosion in Localizable
  QFT},'' {\em Phys. Lett.}, vol.~B790, pp.~466--474, 2019.

\bibitem{Jaffe-jan-67}
A.~M. Jaffe, ``{HIGH-ENERGY BEHAVIOR IN QUANTUM FIELD THEORY. I. STRICTLY
  LOCALIZABLE FIELDS},'' {\em Phys. Rev.}, vol.~158, pp.~1454--1461, 1967.

\bibitem{feyart}
T.~Hahn, ``{Generating Feynman diagrams and amplitudes with FeynArts 3},'' {\em
  Comput. Phys. Commun.}, vol.~140, pp.~418--431, 2001.

\bibitem{Smith-apr-93}
B.~H. Smith, ``Summing one-loop graphs in a theory with b symmetry,'' {\em
  Phys. Rev. D}, vol.~47, pp.~3518--3520, Apr 1993.

\bibitem{loop1}
M.~B. Voloshin, ``{Summing one loop graphs at multiparticle threshold},'' {\em
  Phys. Rev.}, vol.~D47, pp.~R357--R361, 1993.

\bibitem{pol}
A.~O. Barut, A.~Inomata, and R.~Wilson, ``Algebraic treatment of second
  poschl-teller, morse-rosen and eckart equations,'' {\em Journal of Physics A:
  Mathematical and General}, vol.~20, pp.~4083--4096, sep 1987.

\bibitem{pt1}
S.-A. Yahiaoui, S.~Hattou, and M.~Bentaiba, ``Generalized morse and
  posch-teller potentials: The connection via schrodinger equation,'' {\em
  Annals of Physics}, vol.~322, no.~11, pp.~2733 -- 2744, 2007.

\bibitem{shape1}
R.~K. Bhaduri, J.~Sakhr, D.~W.~L. Sprung, R.~Dutt, and A.~Suzuki, ``{Shape
  invariant potentials in SUSY quantum mechanics and periodic orbit theory},''
  {\em J. Phys.}, vol.~A38, p.~L183, 2005.

\bibitem{bender1}
C.~Bender and S.~A.~Orszag, {\em Advanced Mathematical Methods for Scientists
  and Engineers: Asymptotic Methods and Perturbation Theory}, vol.~1.
\newblock 01 1999.

\bibitem{bender2}
C.~M. Bender and C.~Heissenberg, ``{Convergent and Divergent Series in
  Physics},'' in {\em {22th Saalburg Summer School on Foundations and New
  Methods in Theoretical Physics Wolfersdorf, Thuringia, Germany, September
  5-16, 2016}}, 2017.

\bibitem{Papadopoulos-nov-92}
E.~N~Argyres, C.~Papadopoulos, and R.~H.P.~Kleiss, ``Multiscalar production
  amplitudes beyond threshold,'' {\em Nuclear Physics B}, vol.~395, 11 1992.

\bibitem{Libanov-jul-94}
M.~V. Libanov, V.~A. Rubakov, D.~T. Son, and S.~V. Troitsky, ``{Exponentiation
  of multiparticle amplitudes in scalar theories},'' {\em Phys. Rev.},
  vol.~D50, pp.~7553--7569, 1994.

\bibitem{anarmo}
G.~A. Diamandis, B.~C. Georgalas, A.~B. Lahanas, and E.~Papantonopoulos, ``{On
  the high-energy suppression of transition amplitudes in the driven anharmonic
  oscillator},'' {\em Phys. Lett.}, vol.~B306, pp.~319--326, 1993.

\bibitem{Libanov-mar-95}
M.~V. Libanov, D.~T. Son, and S.~V. Troitsky, ``{Exponentiation of
  multiparticle amplitudes in scalar theories. 2. Universality of the
  exponent},'' {\em Phys. Rev.}, vol.~D52, pp.~3679--3687, 1995.

\bibitem{Khoze-apr-14}
V.~V. Khoze, ``{Multiparticle Higgs and Vector Boson Amplitudes at
  Threshold},'' {\em JHEP}, vol.~07, p.~008, 2014.

\bibitem{Khoze-jun-18}
V.~V. Khoze, ``{Semiclassical computation of quantum effects in multiparticle
  production at large lambda n},'' 2018.

\bibitem{Jaeckel-jun-18}
J.~Jaeckel and S.~Schenk, ``{Exploring High Multiplicity Amplitudes in Quantum
  Mechanics},'' {\em Phys. Rev.}, vol.~D98, no.~9, p.~096007, 2018.

\bibitem{Arkhipov-nov-82}
A.~A. Arkhipov, ``Unitarity bounds for multiparticle interaction amplitudes,''
  {\em Theoretical and Mathematical Physics}, vol.~53, pp.~1067--1084, Nov
  1982.

\bibitem{Voloshin-feb-92}
M.~Voloshin, ``Multi-particle amplitudes at zero energy and momentum in scalar
  theory,'' {\em Nuclear Physics B}, vol.~383, no.~1, pp.~233 -- 248, 1992.

\bibitem{wilson}
K.~G. Wilson and J.~B. Kogut, ``{The Renormalization group and the epsilon
  expansion},'' {\em Phys. Rept.}, vol.~12, pp.~75--199, 1974.

\bibitem{Belyaev-aug-18}
A.~Belyaev, F.~Bezrukov, C.~Shepherd, and D.~Ross, ``{Problems with
  Higgsplosion},'' {\em Phys. Rev.}, vol.~D98, no.~11, p.~113001, 2018.

\bibitem{Monin-aug-18}
A.~Monin, ``{Inconsistencies of higgsplosion},'' 2018.

\bibitem{lattice1}
Y.-Y. Charng, ``Nonperturbative lattice simulation of high multiplicity cross
  section bound in $\phi^{4}_{3}$ on beowulf supercomputer,'' 11 2001.

\bibitem{lattice2}
Y.~Y. Charng and R.~S. Willey, ``{Nonperturbative bound on high multiplicity
  cross-sections in $\phi^{4}$ in three-dimensions from lattice simulation},''
  {\em Phys. Rev.}, vol.~D65, p.~105018, 2002.

\bibitem{precision}
V.~V. Khoze, J.~Reiness, M.~Spannowsky, and P.~Waite, ``{Precision measurements
  for the Higgsploding Standard Model},'' 2017.

\bibitem{trivi1}
P.~{Weisz} and U.~{Wolff}, ``{Triviality of $\phi^{4}$ theory: Small volume
  expansion and new data},'' {\em Nuclear Physics B}, vol.~846, pp.~316--337,
  May 2011.

\bibitem{trivi2}
M.~Aizenman, ``{Proof of the Triviality of $\phi^{4}$ in D-Dimensions Field
  Theory and Some Mean Field Features of Ising Models for D$>$4},'' {\em Phys.
  Rev. Lett.}, vol.~47, pp.~1--4, 1981.

\bibitem{trivi3}
J.~Frohlich, ``{On the Triviality of Lambda ($\phi^{4}$) in D-Dimensions
  Theories and the Approach to the Critical Point in D $>$ Four-Dimensions},''
  {\em Nucl. Phys.}, vol.~B200, pp.~281--296, 1982.

\bibitem{string3}
Y.~Okawa, ``{Analytic methods in open string field theory},'' {\em Prog. Theor.
  Phys.}, vol.~128, pp.~1001--1060, 2012.

\bibitem{uv}
S.~Abel and N.~A. Dondi, ``{UV Completion on the Worldline},'' 2019.

\bibitem{string1}
M.~N. Hashi, H.~Isono, T.~Noumi, G.~Shiu, and P.~Soler, ``{Higgs Mechanism in
  Nonlocal Field Theories},'' {\em JHEP}, vol.~08, p.~064, 2018.

\bibitem{string2}
K.~Ohmori, {\em {A Review on tachyon condensation in open string field
  theories}}.
\newblock PhD thesis, Tokyo U., 2001.

\end{thebibliography}
\bibliographystyle{ieeetr}

\end{document}